%% file: Analysis_of_the_Semiclassical_Solution_of_CDT.tex
\documentclass[a4paper,12pt,notitlepage]{report}
\usepackage[english]{babel}
\usepackage[latin1]{inputenc}
\usepackage{amsmath}
\usepackage{graphics}
\usepackage[hyperfootnotes=false]{hyperref}
\hypersetup{pdftitle={Analysis of the Semiclassical Solution of CDT},pdfauthor={Tomasz Trze\'{s}niewski},pdfkeywords={quantum gravity, causal dynamical triangulations}}
\usepackage[all]{hypcap}

\title{Analysis of the Semiclassical Solution of CDT}
\author{Tomasz Trze\'{s}niewski}
\date{}
\begin{document}
\begin{center}
\LARGE\textbf{Analysis of the Semiclassical Solution of CDT\footnote{Master Thesis, supervised by prof. dr hab. J. Jurkiewicz and defended in June 2010.}} \\ 
\vspace{0.75cm}
\large Tomasz Trze\'{s}niewski \\ 
\vspace{0.5cm}
\normalsize\textit{Marian Smoluchowski Institute of Physics, \\ 
Jagiellonian University, Krak\'{o}w, Poland} \\ 
\vspace{0.25cm}
\texttt{e-mail:\ t.trzesniewski@uj.edu.pl}
\vspace{0.75cm}
\end{center}
\begin{abstract}
\noindent Causal dynamical triangulations (CDT) constitute a background independent, nonperturbative approach to quantum gravity, in which the gravitational path integral is approximated by the weighted sum over causally well-behaving simplicial manifolds i.e. causal triangulations. This thesis is an analysis of the data from the Monte Carlo computer simulations of CDT in $3+1$ dimensions. It is confirmed here that there exist the semiclassical limit of CDT for so-called $(4,1)$ (or equivalent $(1,4)$) simplices, being a discrete version of the mini-superspace model. Next, the form of the corresponding discrete action is investigated. Furthermore, it is demonstrated that the effective, semiclassical solution works also after the inclusion of remaining $(3,2)$ and $(2,3)$ simplices, treated collectively. A specific form of the resulting extended discrete action is examined and a transition from the broader framework to the former narrower one is shown.
\end{abstract}
\tableofcontents
\chapter*{Introduction}
\addcontentsline{toc}{chapter}{Introduction}
\vspace{-0.15cm}
In spite of numerous efforts, based on the most diverse assumptions, quantisation of gravitational field remains an elusive task. Regarding our standard approach to quantum field theory, perturbative nonrenormalisability is considered the fundamental obstacle here. Therefore attempts to formulate a nonperturbative theory of quantum gravity have gained a lot of interest. Probably the most intuitive path in this direction is founded on Feynman path integral defined for the geometry of spacetime (see Sec.~1.1). \\ 
\indent One of the approaches to quantum gravity with such a starting point is known by the name of causal dynamical triangulations (CDT). It is a modification (Sec.~1.4) of earlier Euclidean dynamical triangulations (EDT) (Sec.~1.2, 1.3) and consists in replacing the gravitational path integral with a weighted sum over piecewise linear manifolds assembled from simplices (i.e. triangulations), every of which must preserve causality and admit a global proper time foliation. Such an approach is background independent, as it does not favour any particular geometry. \\ 
\indent CDT can be naturally Wick rotated and becomes then a statistical system of random geometry. As a result, it is suited to be investigated numerically using Monte Carlo methods, in particular in $3+1$ dimensions (Sec.~2.1). The corresponding program has been written by J. Jurkiewicz and A. G\"{o}rlich and improved only recently by J. Gizbert-Studnicki, who also set up the latest simulations. Hence one should refer to \cite{JGS} for more details concerning numerics. The following thesis is an analysis of the obtained data and its point of departure is confirmation of the existence of a physically relevant semiclassical limit of CDT, being the discretised mini-superspace model (Sec.~2.2). \\ 
\indent In subsequent chapters I examine this effective, semiclassical solution of CDT employing two kinds of observables: average distributions of the number of simplices and the analogical (inverse) covariance matrices. Particularly, I seek in the form of the discrete effective action. In Chapter~3 I consider only so-called $(4,1)$ simplices, extend the description to $(\mathit{3,2})$ simplices in Chapter~4 and sketch out the possibility of separate inclusion of $(3,2)$ and $(2,3)$ simplices in Appendix~A. The whole analysis has been conducted with Mathematica and all graphs have been plotted with gnuplot.
\chapter{CDT model of quantum gravity}
\section{Path integral for gravitation}
\begin{figure}[ht]
\centering
\scalebox{0.16}[0.16]{\includegraphics{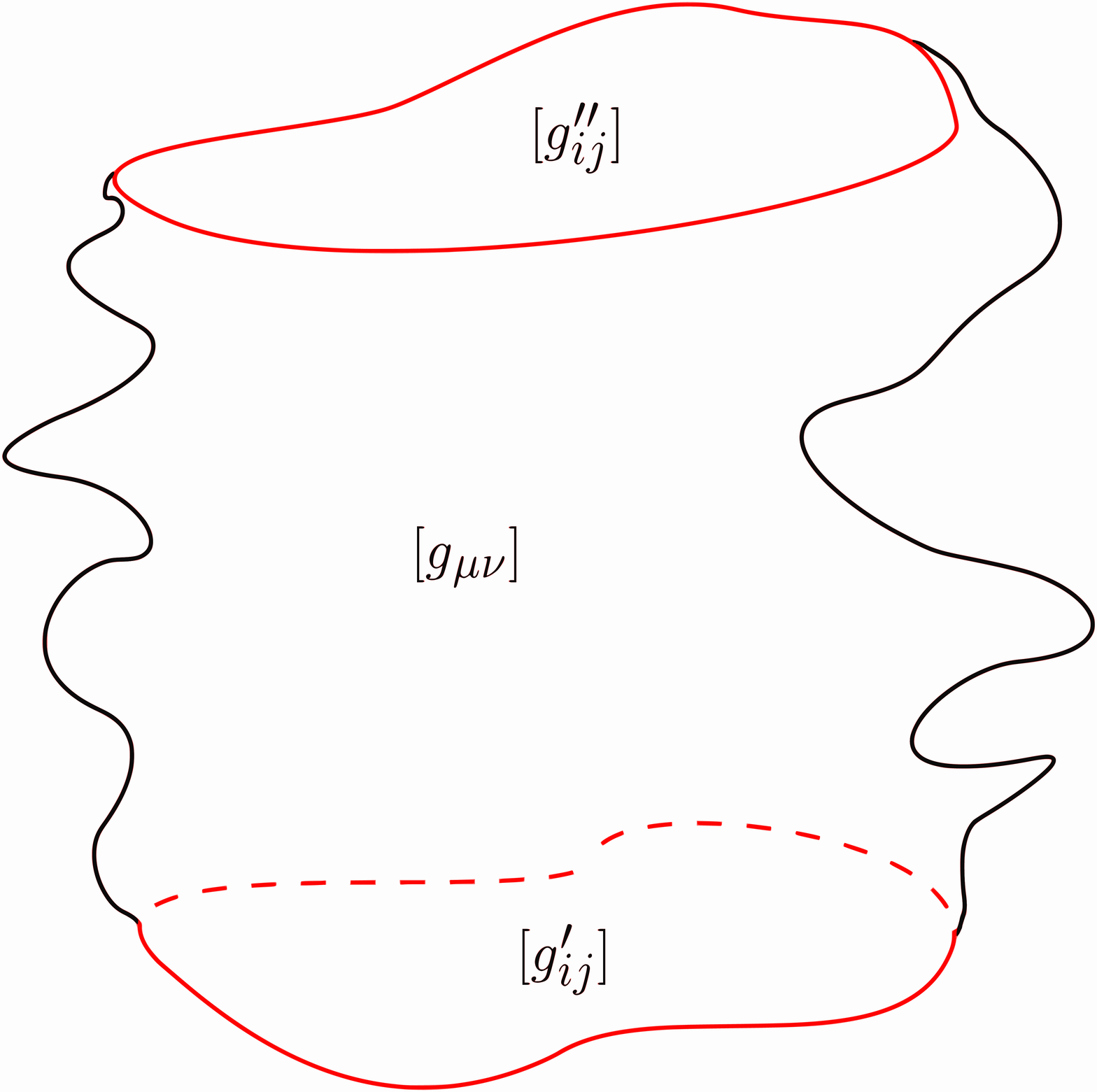}}
\caption{A quantum transition between an initial and final spatial geometry via quantum superposition of intermediate geometries described by the gravitational path integral (\ref{eq:13})\label{fig:11}}
\end{figure}

Classical field theory of gravitation, known as general relativity is specified by the Einstein-Hilbert action, which for a differentiable manifold $M$ of $d\equiv D+1$ dimensions has the form
\begin{align}\label{eq:11}
S_{EH}[g_{\mu\nu}]=\frac{1}{16\pi G}\int_{M}\!\!d^{d}x\ \sqrt{\vert\det g\vert}\,(R-2\Lambda) \\ \bigg(+\frac{1}{8\pi G}\int_{\partial M}\!\!d^{d-1}y\ \sqrt{\det h}\,K\bigg), \nonumber
\end{align}
where $g$ is a metric $2$-form field, $R$ a Ricci scalar, $G$ denotes the gravitational constant and $\Lambda$ the cosmological constant while the integral in brackets, appearing only when $M$ has boundary, is the Gibbons-Hawking-York term with an induced metric $h$ and trace of an extrinsic curvature $K$. \\ 
\indent Formally, one defines Feynman path integral for gravitational field, which actually is a grand canonical partition function for quantum gravity, as
\begin{align}\label{eq:12}
\mathcal{Z}(G,\Lambda):=\int_{\mathrm{Geom}(M)}\!\!\mathcal{D}[g_{\mu\nu}]\ \exp(i\, S_{EH}[g_{\mu\nu}]),
\end{align}
where $g_{\mu\nu}\equiv[g_{\mu\nu}]\in\mathrm{Geom}(M)$ denotes the equivalence class of metrics with respect to diffeomorphisms -- a geometry. In particular, the partition function for manifold with spatial boundaries at proper times $t^{\prime}$, $t^{\prime\prime}$,
\begin{align}\label{eq:13}
G_{G,\Lambda}(g^{\prime\prime}_{ij},g^{\prime}_{ij};t^{\prime\prime},t^{\prime}):=
\int_{\mathrm{Geom}(M)}\!\!\mathcal{D}[g_{\mu\nu}]\ \exp(i\, S_{EH}[g_{\mu\nu}]),
\end{align}
where $g^{\prime}_{ij}$, $g^{\prime\prime}_{ij}$ are induced boundary geometries, is called the propagator by analogy with Feynman path integral for a single particle and can be regarded\nopagebreak[4] as a transition amplitude between an initial and final spatial geometry.\nopagebreak[4] (Which is illustrated in Fig.~\ref{fig:11}.)
 \\ 
\indent Unfortunately, there is a number of mathematical problems associated with (\ref{eq:12}), (\ref{eq:13}), let alone the physical interpretation (cf. \cite{cure}, \cite{hist}). First and foremost, as always in quantum field theory one has to take care of convergence of a functional integral. This can not be circumvented here by applying a perturbative expansion, therefore a suitable (non-violating diffeomorphism invariance) regularisation and renormalisation of the path integral itself is necessary. Secondly, since there is no absolute parametrisation of the space of geometries $\mathrm{Geom}(M)$ we do not know what we are integrating over. Especially, how the diffeomorphism-covariant (because of the integral having to be diffeomorphism-invariant) measure $\mathcal{D}[g_{\mu\nu}]$ should be defined. As a result, one has to start with gauge-covariant (i.e. diffeomorphism-covariant in this case) fields and fix the gauge afterwards but this gives rise to Faddeev-Popov determinants, which are exceptionally difficult to be computed nonperturbatively. Another issue is the very existence of the measure. Thirdly, for any choice of variables the action (\ref{eq:11}) is quadratic in fields, which hinders evaluation of integrals in (\ref{eq:12}), (\ref{eq:13}). Furthermore, integrands of those are complex and therefore, a kind of Wick rotation i.e. an analytic continuation such that $t\mapsto i\,t$ is required. Due to the dependence of all metric components on time as well as non gauge-invariant nature of the above straightforward Wick rotation it is not so obvious what it should look, however. Were all these problems tackled, for $d\geq 3$ there is yet another, known as the conformal factor problem. Briefly speaking, it consists in unboundness (from below) of the action (\ref{eq:11}), triggered by the conformal mode of metric. This makes the Wick-rotated path integral still potentially divergent. Going beyond this section, in the context of causal dynamical triangulations the issue was addressed and probably resolved \cite{cure} (see also \cite{hist}). \\ 
\indent Last but not least, in the partition function (\ref{eq:12}) (as well as in (\ref{eq:13})) the topology of spacetime $M$ is \textit{fixed}. In principle, one might consider an extension of (\ref{eq:12}) including a sum over topologies
\begin{align}\label{eq:14}
\mathcal{Z}(G,\Lambda):=\sum_{M}\int_{\mathrm{Geom}(M)}\!\!\mathcal{D}[g_{\mu\nu}]\ \exp(i\, S_{EH}[g_{\mu\nu}]).
\end{align}
Although this makes the situation even more complicated, it is argued by some (see for instance \cite{top}) to be necessary from a point of view of the theory of quantum gravity. Nevertheless, there is still no experimental/observational evidence for the possibility of topology fluctuations which would arise then. In any case, this issue will be left out here.
\section{Method of dynamical triangulations}
There are several approaches \cite{appr} to fix the first of the above mentioned problems with the gravitational path integral i.e. finding a suitable regularisation and renormalisation. (Which -- hopefully -- should also help to resolve the other issues.) Inspiration has been coming especially from lattice gauge theory and its successes in quantum chromodynamics. In this way so-called \textit{quantum} Regge calculus appeared \cite{will}, originating from Regge calculus \cite{regge} -- a method of approximating solutions of classical gravitational theory with piecewise linear manifolds. It is characterised by the fact that one introduces a lattice regularisation in the form of a discretisation of geometries and the path integral is over discretised geometries only. It ought to be noted that in contrast with (classical) Regge calculus it is not an arbitrary continuous manifold that is to be approximated here but the integral over all such manifolds. Hence, discretisation of geometries does not need to be arbitrarily fine. For example, a quite limited number of lattice plaquettes can be considered. Nevertheless, it should be taken to infinity along with a plaquette size taken to zero in the limit of a presumable renormalised -- i.e. continuum -- theory. \\ 
\indent Dynamical triangulations (DT) constitute a distinct variant of such a scheme (see \cite{appr}). They are characterised by the integration over piecewise linear manifolds assembled from polytopes (a generalisation of polyhedra) which have fixed lengths of the edges, functioning as a gauge-invariant ultraviolet cutoff. Polytopes are traditionally regarded as being intrisically Minkowskian (i.e. in a sense flat -- which makes manifolds piecewise \textit{linear}), although this assumption may require a change for \textit{causal} dynamical triangulations. (Generally, there is no extrinsic geometry, since manifolds assembled from polytopes are not embedded anywhere -- the same as in the case of continuous ones.) \\ 
\indent More explicitly, in the method of dynamical triangulations the continuous partition function (\ref{eq:12}) is replaced with the discrete
\begin{align}\label{eq:15}
\mathcal{Z}(k,\lambda):=\sum_{\mathcal{T}}\frac{1}{C_{\mathcal{T}}} \exp(i\, S_{R}[\mathcal{T}]),\ C_{\mathcal{T}}\equiv|\mathrm{Aut}(\mathcal{T})|,
\end{align}
where $k$, $\lambda$ denote coupling constants of the regularised theory, corresponding to the inverse of gravitational constant $G$ and cosmological constant $\Lambda$ (divided by $G$, to be precise), respectively (cf. (\ref{eq:19}) here below); the sum, to which the integral reduces, goes over all "polytopialisations" (e.g. triangulations, cubulations) $\mathcal{T}$ of a fixed topology, weighted by the exponential of the regularised action $S_{R}[\mathcal{T}]$ and normalised by the factor $C_{\mathcal{T}}$ (rank of the automorphism group of $\mathcal{T}$). The latter replaces the measure $\mathcal{D}[g_{\mu\nu}]$ and does not allow for overcounting of geometries, which would violate diffeomorphism invariance. Such a description of geometry is stripped of the use of coordinates, though in principle a coordinate system might be assigned to each polytope. This is absolutely unnecessary, however, because \textit{all} geometric properties are contained in the connectivity of a polytopialisation (the set of neighbouring relations between polytopes), as it will be discussed below. Furthermore, thus we obtain a gauge-invariant parametrisation of the regularised $\mathrm{Geom}(M)$, a desire for which was expressed in the previous section. \\ 
\begin{figure}[hb]
\centering
\scalebox{0.22}[0.22]{\includegraphics{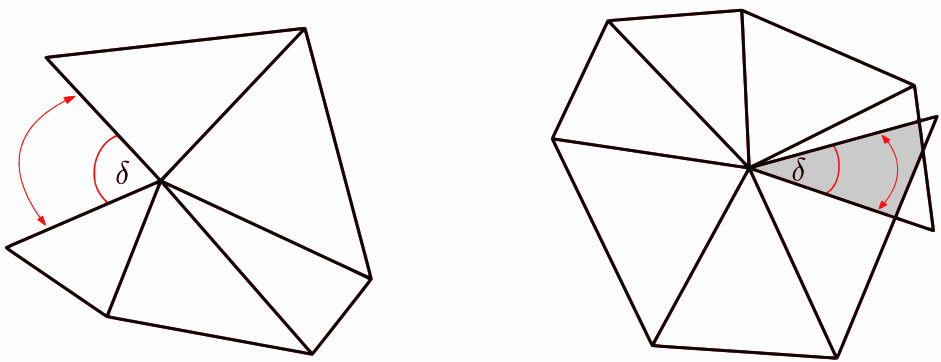}}
\caption{Curvature at a vertex of a $2$-dimensional simplicial manifold: positive on the left ($\delta>0$) and negative on the right ($\delta<0$); the general case, with non-equilateral triangles\label{fig:12}}
\end{figure}

\indent Since (\ref{eq:15}) should be (and -- as it was e.g. explicitly shown  \cite{walk} for $d=2$ -- is indeed) independent of a polytopialisation usually the simplest and most convient choice is made by taking only simplicial manifolds (i.e. built from simplices), called triangulations, into consideration. Moreover, the simplices are assumed to be equilateral (in \textit{Minkowskian} sense). This is exactly what will be done in all the following. Hence the symbol $\mathcal{T}$ above as well as $S_{R}$, denoting the Regge action (see (\ref{eq:19}) here below), and the very name of dynamical triangulations. \\ 
\indent The detailed way of encoding of geometry in a simplicial manifold is as follows (cf. \cite{hist}, \cite{rev}). Actually, it is enough to consider curvature alone. In general, for $d=2$ sectional curvature on a manifold coincides with so-called Gaussian curvature and Ricci scalar is equal $2$ times the latter. The Gaussian curvature can be determined from the parallel transport of a vector around a closed loop. On a simplicial manifold such a loop has to circle a vertex, otherwise no effect will be ever observed (since simplices are flat). A vector will then change its orientation if and only if a sum of the adjacent angles $\neq 2\pi$. A difference is called the deficit angle $\delta$. This is exemplified in Fig~\ref{fig:12}. For $d>2$ one has to sum over all $2$-dimensional submanifolds dual to a certain $(d-2)$-dimensional subsimplex to obtain the nontrivial Ricci scalar, localised at this subsimplex. Hence, curvature is associated with $(d-2)$-subsimplices and expressed by deficit angles $\delta$ around them, which -- by generalising a notion for $d=2$ -- are defined through sums
\begin{align}\label{eq:16}
\sum_{k\in\triangle^{(d-2)}}\Theta_{k}=2\pi-\delta,
\end{align}
where $\Theta_{k}$ are ditopial (a generalisation of dihedral) angles adjacent to a given $(d-2)$-subsimplex $\triangle^{(d-2)}$. Deficit angles, as can be observed, are determined by the manner in which simplices are glued together -- their connectivity. \\ 
\indent From the above discussion, there is the following correspondence between terms of the non-regularised (formula (\ref{eq:11})) and regularised Einstein-Hilbert action
\begin{align}\label{eq:17}
\frac{1}{2}\int_{M}\!\!d^{d}x\ \sqrt{\vert\det g\vert}\,R\ \longmapsto\ \sum_{i\in\mathcal{T}}\mathrm{vol}(\triangle^{(d-2)}_{i})\,\delta_{i}, \\ \label{eq:18}\int_{M}\!\!d^{d}x\ \sqrt{\vert\det g\vert}\ \longmapsto\ \sum_{i\in\mathcal{T}}\mathrm{vol}(\triangle^{(d)}_{i}),
\end{align}
where $\mathrm{vol}$ denotes the volume of a simplex/subsimplex (cf. Section~1.4). Consequently, the Regge action amounts to
\begin{align}\label{eq:19}
S_{R}[\mathcal{T}]=k\sum_{i\in\mathcal{T}}\mathrm{vol}(\triangle^{(d-2)}_{i})\,\delta_{i}-
\lambda\sum_{j\in\mathcal{T}}\mathrm{vol}(\triangle^{(d)}_{j}).
\end{align}
Its precise form for $d=4$ will be given in Section~1.4 (in the Wick-rotated version, yet for the modified regularisation, as will become clear soon).
\section{EDT and their failure}
\begin{figure}[ht]
\centering
\scalebox{0.18}[0.18]{\includegraphics{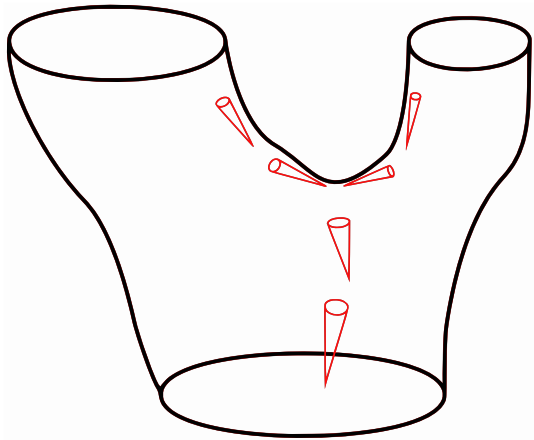}}
\caption{A change of spatial topology by branching of spacetime with the resulting behaviour of light cones shown\label{fig:13}}
\end{figure}

The model whose regularisation scheme was outlined in the previous section is called Euclidean dynamical triangulations (EDT). Unfortunately, there are severe problems (cf. \cite{hist}, \cite{rev}) with moving further in EDT, towards a certain nontrivial and physically relevant continuum limit, which is the ultimate aim of the framework. Everything starts with the absence of a prescription for Wick rotation of the partition function (\ref{eq:15}),
\begin{align}\label{eq:110}
\sum_{\mathcal{T}}\frac{1}{C_{\mathcal{T}}} \exp(i\, S_{R}[\mathcal{T}])\longmapsto\sum_{\mathcal{T}^{\prime}}\frac{1}{C_{\mathcal{T}^{\prime}}} \exp(-\, S_{R}^{E}[\mathcal{T^{\prime}}]),
\end{align}
where $S_{R}^{E}[\mathcal{T}^{\prime}]$ is the Euclidean (Wick-rotated) action. Therefore, the \textit{ad hoc} Wick rotation is applied, in which Lorentzian geometries are replaced with Euclidean ones just by hand. As a result, the summation has to go over simplicial manifolds devoid of the notion of time and causal structure. It is not known how to retrieve the latter afterwards. Almost certainly, it can not be done by straightforward inverse of the Wick rotation, also because of the occurrence of Euclidean geometries anomalous from the perspective of causality. Namely, the problem consists in changing of the topology \cite{top} between $(d-1)$-dimensional submanifolds, especially by events of splitting of such a submanifold into at least two disconnected ones, as it is exemplified in Fig.~\ref{fig:13}. For a Lorentzian manifold this amounts to disintegrating of space during its time evolution, in other words, branching of spacetime into so-called baby universes. (Obviously, the time inverse of the process in the form of joining of the disconnected space components should be identically likely yet it is excluded by the requirement of the fixed spacetime topology.) A consequence is the degeneration of light cones; causality is violated because at a branching point there is more than one possibility for further course of a test particle's world line. Moreover, the above phenomenon turns out to be generic and highly degenerate geometries are dominating in the Wick-rotated (formula (\ref{eq:110})) counterpart of the sum (\ref{eq:15}). \\ 
\indent Hence, we are stuck in the Euclidean theory and this is where the very name of EDT derives from. Bad news has been confirmed by results (see \cite{appr}, \cite{hist}) from numerical simulations. Namely, there is no relevant continuum limit for $d\geq 3$. Rather than that, the space of coupling constants $k$, $\lambda$ for the model of $d=3,4$ (and probably any $d\geq 3$) contains a critical line $\lambda_{\mathrm{crit}}(k)$ (where the continuum limit exists), at which there are two extreme, unphysical phases, separated by a bicritical point. A useful tool to study them is a standard notion of (effective -- in the continuum limit sense) dimensionality, called Hausdorff dimension, which works well also on discretised manifolds. In the phase of small $k$'s, called the crumpled phase, simplicial spacetimes (or rather Euclidean spaces) of a tiny size are produced. Moreover, there exist several vertices to which almost all simplices are connected. Therefore distances between the latter are small and in the continuum Hausdorff dimension of space is going to $\infty$. On the contrary, in the phase of large $k$'s, known as the branched polymer phase, generation of baby universes is the dominating process. Therefore every simplex has only a limited number of neighbours and in the continuum Hausdorff dimension approaches $2$, which enhances interpretation of spaces in this phase as a particular kind of fractals -- namely branched polymers. Neither of the phases yields Hausdorff dimension equal or at least close to the expected $d$.
\section{Imposing causality on DT}
Causal (or Lorentzian) dynamical triangulations (CDT) were invented \cite{caus} as a response to the failure of EDT and constitute their modification \cite{scra}, \cite{sum}. The essential idea behind this approach is to consider from the outset only causally well-behaving simplicial manifolds. More precisely, a triangulation must admit a global proper-time foliation. It is only when it is built from layers of simplices to which a proper time variable can be assigned -- hence space globally evolves in time (spacetime is globally hiperbolic). Consequently, the topology of space is conserved, which means the exclusion of spacetimes branching into baby universes. However, it ought to be stressed that changes of spatial topology are not excluded from consideration \textit{in principle}. Individual geometries in the integral (\ref{eq:12})/sum (\ref{eq:15}) are in general highly nonclassical, as it is always in quantum physics, and possibly could also include fluctuations of spatial topology. Causality which would be lost in such a case might as well be restored for a superposition of quantum spacetimes, corresponding to a classical solution of gravitational theory. On the other hand, we do not even know whether the restriction on topology survives in the continuum limit of the CDT model. Nevertheless, the failure of EDT has shown that in the method of dynamical triangulations variability of spatial topology has to be treated exceptionally (see \cite{rev}) if at all (and its total rejection is worth trying). The same concerns variability of topology of the entire spacetime, discussed at the end of Section~1.1, which leads to the super-exponential growth of the number of geometries in (\ref{eq:15}). \\ 
\begin{figure}[hb]
\centering
\scalebox{0.16}[0.16]{\includegraphics{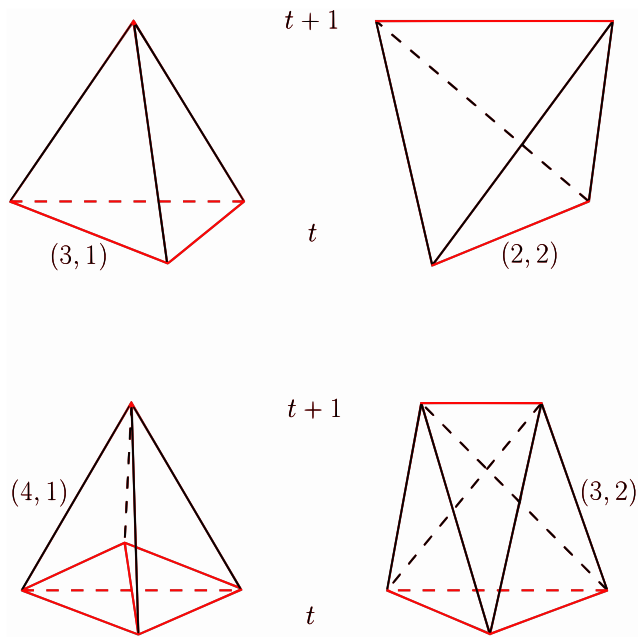}}
\caption{Types of simplicial building blocks in CDT (without the dual ones) for $d=3$ (up) and $d=4$ (bottom)\label{fig:14}}
\end{figure}

\begin{figure}[ht]
\centering
\scalebox{0.94}[0.94]{\includegraphics{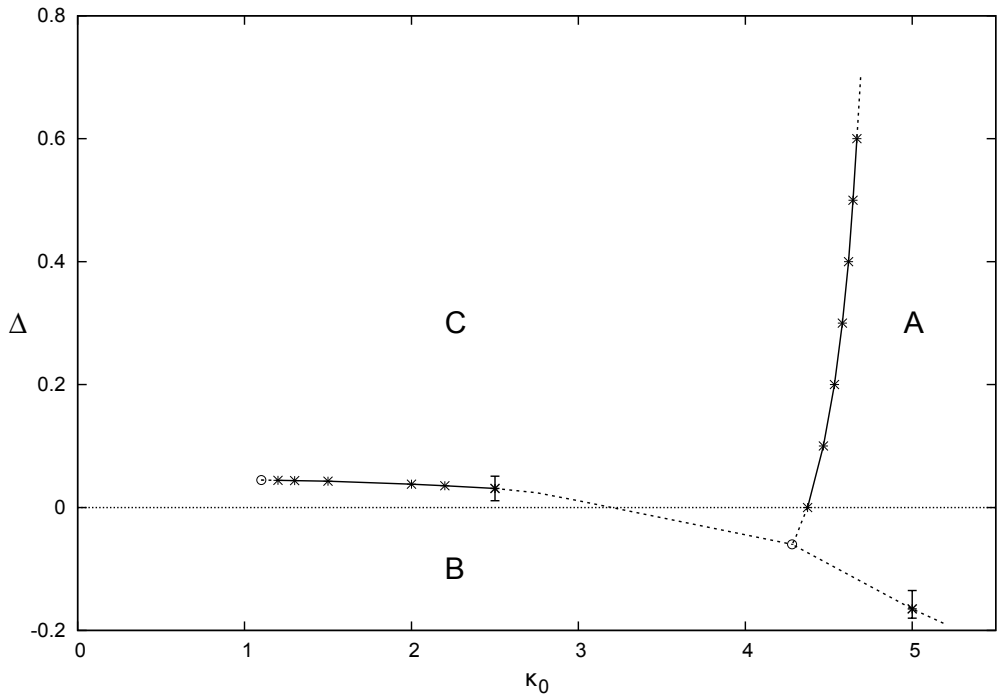}}
\caption{Phase space of CDT for $d=4$ with the measured boundaries between phases A, B, C (from \cite{hora})\label{fig:15}}
\end{figure}

\indent Due to the layered structure of simplicial manifolds in CDT there is a limited number of allowed building blocks from which they are constructed -- simplices placed differently with respect to the time direction. Specifying the statement from the previous paragraph, a proper time variable is assigned to simplicial vertices and $(d-1)$-dimensional surfaces spanned between such at a given time (which obviously are forbidden to intersect). As can be seen, time is discrete. As a result, we have $d$ types of simplicial building blocks, denoted by numbers of their vertices lying at times $t$ and $t+1$ (where $t$ is arbitrary). In the most interesting, lowest dimensions we have the following ones. For $d=2$ there are simplices of types $(2,1)$ ($2$ vertices at a time $t$, joined by a spatial link and one at $t+1$; the other two links are timelike) and $(1,2)$, which is just time inverse of $(2,1)$ (I call them dual to each other). Types for $d=3,4$ are depicted in Fig.~\ref{fig:14}. For $d=3$ there is a pair of dual types of simplices, $(3,1)$, $(1,3)$ (with a spatial triangle) and type $(2,2)$ (with two spatial links only). For $d=4$ there are two dual pairs: $(4,1)$, $(1,4)$ (with a spatial tetrahedron) and $(3,2)$, $(2,3)$ (with a spatial triangle and a spatial link). In all the following I will be denoting collectively the latter as $(\mathit{4,1})$ and $(\mathit{3,2})$, respectively. \\ 
\indent The next step in construction of the CDT model (discussed extensively in \cite{dyna}) is the introduction of anisotropy between time and space. Squared lengths of timelike and spacelike simplicial links are set as, respectively,
\begin{align}\label{eq:111}
l_{\mathrm{timelike}}^{2}=-\alpha\,a^{2},\ l_{\mathrm{spacelike}}^{2}=a^{2},
\end{align}
where $a$ is the cutoff size and $\alpha>0$ the anisotropy parameter. Hence simplices are generally no longer equilateral in the sense of the Wick-rotated model (in EDT we had $\alpha=1$). Now according to the correspondence (\ref{eq:17}), (\ref{eq:18}) one has to calculate volumes of timelike and spacelike simplices and subsimplices as well as values of ditopial angles -- all as functions of $\alpha$. (Different conventions may be used.) Irrespective of what might seem, all Lorentzian volumes are real and $>0$ but angles in general are complex. Yet it is not a problem, as we eventually obtain a \textit{real} regularised action. Its form without substituted volumes and angles (for simplicity) -- an incarnation of (\ref{eq:19}) -- for $d=4$ is
\begin{align}\label{eq:112}
S_{R}[\mathcal{T}]=k\sum_{\triangle^{(\mathit{3,0})}_{a}\subset\mathcal{T}}
\mathrm{vol}(\triangle^{(\mathit{3,0})}_{a})\,\frac{\delta_{a}}{i}+
k\sum_{\triangle^{(\mathit{2,1})}_{b}\subset\mathcal{T}}
\mathrm{vol}(\triangle^{(\mathit{2,1})}_{b})\,\delta_{b}- \nonumber\\ 
\lambda\sum_{\triangle^{(\mathit{4,1})}_{c}\subset\mathcal{T}}
\mathrm{vol}(\triangle^{(\mathit{4,1})}_{c})-
\lambda\sum_{\triangle^{(\mathit{3,2})}_{d}\subset\mathcal{T}}
\mathrm{vol}(\triangle^{(\mathit{3,2})}_{d}),
\end{align}
where by $(\mathit{3,0})$ I denoted spacelike triangles and by $(\mathit{2,1})$ timelike ones. What is essential is that the action (\ref{eq:112}) can be naturally Wick-rotated. The analytic continuation, such that $\alpha\mapsto-\alpha$, goes through the lower half of the complex plane and in the lowest dimensions can be defined: for $d=2$ always, for $d=3$ if $\alpha>\frac{1}{2}$, for $d=4$ if $\alpha>\frac{7}{12}$. The bounds on $\alpha$ originate from the form of ditopial angles. Next, we perform sums in (\ref{eq:112}) and use topological identities (by Dehn and Sommerville), such as the Euler identity, for the numbers of simplices and subsimplices of all types. Finally, for $d=4$ and the assumed spacetime topology $S^{1}\times\Sigma$ (the periodic boundary condition in proper time) where $\Sigma$ is $3$-dimensional and compact we get the Regge action which has the Euclidean form
\begin{align}\label{eq:113}
S_{R}^{E}(\mathcal{T})=-(\kappa_{0}+6\Delta)N_{0}+\kappa_{4}(N_{4}^{(\mathit{4,1})}+
N_{4}^{(\mathit{3,2})})+ \nonumber\\ \Delta(2N_{4}^{(\mathit{4,1})}+N_{4}^{(\mathit{3,2})}),
\end{align}
where $N_{0}$ denotes the number of $0$-simplices (vertices), $N_{4}^{(\mathit{4,1})}$, $N_{4}^{(\mathit{3,2})}$ the numbers of $4$-simplices of the corresponding type and $\kappa_{0}$, $\kappa_{4}$, $\Delta$ are coupling constants and quite complicated functions of $k$, $\lambda$ and $\alpha$. However, their dominating behaviour is determined by single variables -- $k$, $\lambda$ and $\alpha$, respectively, and in particular $\Delta=0\Leftrightarrow\alpha=1$. Consequently, the overall form of the (Wick-rotated here) partition function (\ref{eq:15}) is slightly altered into
\begin{align}\label{eq:114}
\mathcal{Z}(\kappa_{0},\kappa_{4},\Delta):=\sum_{\mathcal{T}}\frac{1}{C_{\mathcal{T}}} \exp(-S_{R}^{E}[\mathcal{T}]),\ C_{\mathcal{T}}\equiv|\mathrm{Aut}(\mathcal{T})|.
\end{align}
\indent Phase structure of the CDT model for $d=3,4$, which can be inferred from numerical simulations (whose technique will be brought up in Chapter~2), differs widely from that of EDT. Actually, there is a disparate continuum limit already for $d=2$ \cite{caus} but I will not discuss this fact here. Three coupling constants instead of two amount to the $3$-dimensional space, containing a critical surface $\kappa_{4}^{\mathrm{crit}}(\kappa_{0},\Delta)$ (where the continuum can be approached) i.e. the phase space of the model. As Fig.~\ref{fig:15} indicates, we observe \cite{reco}, \cite{hora} three phases: A, B and C. They have the following characteristics. Spacetimes in phase A (of large $\kappa$'s) consist of numerous small, disconnected universes, lying along the time direction and as such they are physically irrelevant. This phase is considered a remnant of the branched polymer phase from EDT. Spacetimes in phase B (of small $\Delta$'s) are no more useful, being very short in time and having at each temporal end a vertex of a very large order -- the same as in the crumpled phase from EDT. Therefore phase B can be regarded as a remnant of the latter. Fortunately, there is an additional phase C, in which relevant spacetimes -- extending both in time and space -- are observed. Its existence has been a promising result of the CDT model. It was shown \cite{emer} that a universe from phase C has indeed Hausdorff dimension $d$. Furthermore, recently it turned out \cite{hora} the above phase structure is related to so-called Ho\v{r}ava-Lifshitz gravity.
 \\ 
\indent One remark has to be added. All along, from the very beginning we are dealing with gravitational field in vacuum i.e. without any matter fields. This is the work on pure quantum gravity (or quantum geometry). However, the inclusion of other fields in the gravitational path integral has always been considered and also in CDT is subject to ongoing research.
\chapter{Monte Carlo simulations and the semiclassical solution}
\section{Numerical setup of CDT}
To solve the model of causal dynamical triangulations, as it is formulated in the previous chapter, one has to compute the sum (\ref{eq:114}). Thus far, it was carried out analytically only for $d=2$ \cite{caus} (see also \cite{sum}) and there is little hope for moving towards higher dimensions. However, like every lattice gauge theory CDT (as well as EDT) is well-suited to be investigated \textit{numerically}, provided one can ensure the regularised version of space $\mathrm{Geom}(M)$ is sampled thoroughly in simulations. Hence in this context the problem with the regularisation may be described as twofold: not only has the space of all possible triangulations to be appropriately dense in the $\mathrm{Geom}(M)$ but also numerical simulations enable us to approach every element (in the certain class) of the latter. The first issue is so far assumed, the same as the fact that $\frac{1}{C_{\mathcal{T}}}$ in (\ref{eq:114}) is the measure. To address the second one, Monte Carlo methods are applied \cite{dyna}. More specifically, to approximate (\ref{eq:114}) we generate some number of simplicial manifolds, which are obtained consecutively from the previous ones -- starting from a certain ad hoc constructed, very simple triangulation -- by local transformations called moves. Factor $\frac{1}{C_{\mathcal{T}}}$ is automatically included there. Now, the moves are hoped to be ergodic in the sense of the possibility of reaching every triangulation from any other -- in a certain class -- by the successive application of them. What I mean by the "certain class" will become clear below. Concerning Monte Carlo moves, they consist in replacing of a complex built from some number of simplices with another complex, in general constructed from a different number of building blocks yet respecting the topological and causal structure of a triangulation, in conformity with the assumptions of CDT. For $d=2$ there is a set of $3$ such moves, for $d=3$ there are $5$ of them and for $d=4$ there are $7$ -- their detailed description (omitting $d=2$) is contained in \cite{dyna}. \\ 
\indent From now on the discussion will be restricted to $d=4$, which I actually studied. Regarding the argument from Section~1.4, the topology in the regularised path integral (\ref{eq:114}) is fixed. For convenience one may consider a topology with a periodic boundary condition in proper time, $S^{1}\times\Sigma$ where $\Sigma$ is $3$-dimensional and compact, as it is in the case of the action (\ref{eq:113}). This is exactly what we do in the discussed Monte Carlo simulations, choosing $\Sigma=S^{3}$ (thus far -- as the simplest case). Hence the topology is $S^{1}\times S^{3}$, which is the first constraint characterising the class of triangulations which I investigated, refering to what was said in the previous paragraph. Another possible condition is the fixed number of simplicial temporal layers, in other words, the number of discrete proper times $i$, which in the discussed Monte Carlo simulations is set as $T=80$. The parametrisation of time I will be using here is such that $i\in[0,79]$ and $i=T=80\,\equiv\,i=0$ (the periodic boundary condition). Furthermore, with the aim of keeping the number of $4$-simplices in a triangulation roughly within a certain range, in the case considered here we modify the action (\ref{eq:113}),
\begin{align}\label{eq:21}
S_{R}^{E}(\mathcal{T})\longmapsto S_{R}^{E}(\mathcal{T})+\varepsilon\,(\langle N_{4}^{(4,1)}\rangle-\mathcal{N}_{4})^{2},
\end{align}
where $\langle N_{4}^{(4,1)}\rangle$ is the average total number of $4$-simplices of type $(4,1)$ (see the next section), $\varepsilon$ is a small constant and $\mathcal{N}_{4}=80000$ (some other values were also studied -- as one can check in \cite{emer}, \cite{semi}, \cite{birt}, \cite{nonp} -- yet for the modification of the action different from (\ref{eq:21})). It ought to be noted that in the considered topology the $N_{4}^{(1,4)}$ is exactly the same as the $N_{4}^{(4,1)}$, because only those types of $4$-simplices contain spatial tetrahedra (cf. Fig.~\ref{fig:14}) and hence every $(4,1)$ simplex has its counterpart $(1,4)$, sharing a spatial tetrahedron with it. On the other hand, the $N_{4}^{(\mathit{3,2})}$ is by construction related to the $N_{4}^{(1,4)}$. Thus the number of all $4$-simplices fluctuates in a certain range of values. Last but not least, concerning the discussion of the phase structure of CDT presented in Section~1.4, in order to find ourselves at the critical surface we have to fine-tune $\kappa_{4}$. This is done here using the requirement $\langle N_{4}^{(4,1)}\rangle=\mathcal{N}_{4}$ i.e. the average number of $(4,1)$ (as well as $(1,4)$) simplices is kept fixed. Consequently, $\varepsilon$ characterises fluctuations of $\langle N_{4}^{(4,1)}\rangle$ around $\mathcal{N}_{4}$. \\ 
\indent All what follows concerns numerical results from the Monte Carlo simulations carried out in the physically relevant phase C. I conducted the most extensive analysis for values of the coupling constants $\kappa_{0}=2.2$, $\Delta=0.6$ but all what will be said in this and subsequent chapters is true at least for a certain range of $\kappa_{0}$ and $\Delta$ in phase C. All of the presented graphs will be given for $\varepsilon=0.00004$ yet $\varepsilon=0.00008,0.00002$ alter them only slightly. Furthermore, the results are practically the same also for the variant of (\ref{eq:21}) with $\langle N_{4}\rangle$ (i.e. all simplices) instead of $\langle N_{4}^{(4,1)}\rangle$, with $\varepsilon=0.00008$. The number of statistically independent (which means separated by a sufficient number of moves) Monte Carlo configurations (i.e. triangulations) generated in each of the above cases was about $10^{5}$.
\section{The semiclassical limit}
From now on I will be using the notation $N_{p,q}\equiv N_{4}^{(p,q)}$ (and blank indices will mean an undefined type of simplices). The most fundamental observable in numerical simulations of CDT is the average number of simplices of a given type at a given time
\begin{align}\label{eq:22}
\langle N_{p,q}(t)\rangle=\frac{1}{n}\sum_{k=1}^{n}N_{p,q}^{(k)}(t),
\end{align}
where $n$ is the number of configurations we average over. For the already mentioned total number of $(p,q)$ simplices $\langle N_{p,q}\rangle$ one simply has to sum over all times $t$. The above expression is an estimate of a grand canonical observable for the system fluctuating around $\langle N_{4,1}\rangle=\mathcal{N}_{4}$, provided all configurations are typical i.e. of large probability. Actually, calculating (\ref{eq:22}) one has to take one more thing into account. (I restrict myself to $(4,1)$ simplices at the moment -- which is the same as restricting to $(1,4)$ -- but cf. Section~3.1 and Appendix A.) Namely, looking at an arbitrary temporal distribution $N_{4,1}^{(k)}(i)$ in phase C we observe two regions: one where a universe is well extended in spatial directions and one where it is almost not at all (since $(4,1)$ simplices correspond to spatial tetrahedra). In other words, the extended universe is localised in a certain range of discrete times $i$. Since CDT is a priori symmetric with respect to translation in the time direction, we reparametrise time for every triangulation in such a way that the so-called center of volume of $N_{4,1}(i)$ lies at $i=39$. (However, as it will be seen below, the effective center of volume is between $i=39$ and $i=40$, at $t=39+\frac{1}{2}$.) More precisely, the center of volume $i_{\mathrm{CV}}$ is defined here as $i^{\prime}$ for which
\begin{align}\label{eq:23}
\Bigg\vert\sum_{i=-\frac{T}{2}}^{\frac{T}{2}-1}(i+\tfrac{1}{2})\,N_{4,1}\big(1+
\mathrm{mod}(T+i^{\prime}+i-1,T)\big)\Bigg\vert
\end{align}
reaches a minimum; if there is more than one, $N_{4,1}(i_{\mathrm{CV}})$ must be the greatest among them. (There are alternatives to (\ref{eq:23}) but they all yield the same results up to a difference of $1$.) Now we may calculate the average distribution $\langle N_{4,1}(i)\rangle$. The result is shown in Fig.~\ref{fig:21}, along with the average fluctuations of $N_{4,1}(i)$ (see Section~3.1). One can distinguish three regions in time evolution: the "stalk", where the spatial extension of the universe is very small, consisting of only several spatial tetrahedra (but not less than $5$, which is the minimal number for a simplicial $S^{3}$), the "blob", where the universe is well spatially-extended (with the \textit{double} maximum at $i=39,40$) and the intermediate "tail" (see below). Moreover, the calculated $\langle N_{4,1}\rangle=\mathcal{N}_{4}$ up to a very small deviation. It has yet to be added that $\langle N_{4,1}(i)\rangle$ should be invariant under time inversion. Therefore, it is additionally symmetrised with respect to the effective maximum $t=39+\frac{1}{2}$, which corresponds just to doubling of the number of configurations we average over (and this is justified by the fact $\langle N_{4,1}(i)\rangle$ is very symmetric even without that).
\begin{figure}[hb]
\centering
\scalebox{1}[1]{\input{avsng.tex}}
\caption{The average distribution $\langle N_{4,1}(i)\rangle$ with the semiclassical solution fitted and the average quantum fluctuations (see Section~3.1) denoted by vertical lines (not to be confused with errors, which are invisible here!)\label{fig:21}}
\end{figure}
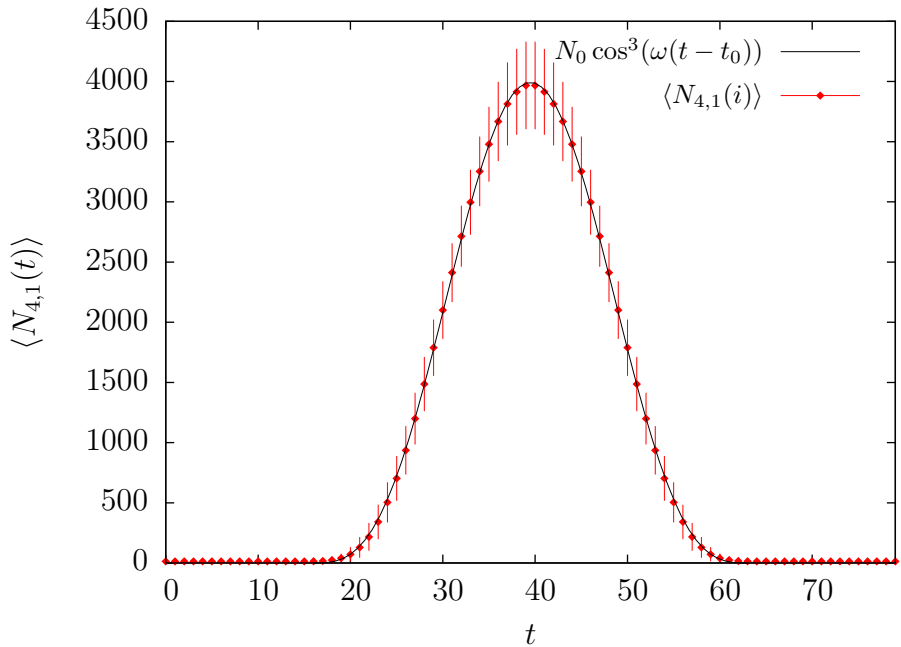 \\  
\indent It turns out \cite{birt} that the blob region of our spacetime can be fitted surprisingly well with a function
\begin{align}\label{eq:24}
N_{0}\cos^{3}(\omega(t-t_{0})),
\end{align}
where $t_{0}=39+\frac{1}{2}$ and $N_{0}$, $\omega$ are certain constants. This is illustrated in Fig.~\ref{fig:21}. As one can see, at least in magnification, the fit starts to deteriorate close to the stalk and this is one of the symptoms of being in the transition region called the tail above. Furthermore, taking the dependence on $\mathcal{N}_{4}$ into consideration, (\ref{eq:24}) has the form with explicit scaling
\begin{align}\label{eq:25}
A_{s}\,\mathcal{N}_{4}^{\frac{3}{4}}\cos^{3}(A_{t}\,\mathcal{N}_{4}^{-\frac{1}{4}}(t-t_{0})),
\end{align}
where $A_{t}$, $A_{s}$ are certain constants. Hence it is a limit of the CDT model, an effective solution. What is crucial in this statement is that eventually the limit may be called semiclassical \cite{semi} -- the semiclassical solution of CDT -- because it is \cite{birt}, \cite{nonp} the (classical) solution of the Wick-rotated mini-superspace model, whose action
\begin{align}\label{eq:26}
S_{\mathrm{cont}}=\frac{1}{24\pi G}\int\!\!dt\ \sqrt{g_{tt}}\bigg(\frac{g^{tt}\dot{V}_{3}(t)^{2}}{V_{3}(t)}+
k_{2}V_{3}(t)^{\frac{1}{3}}-\lambda V_{3}(t)\bigg),
\end{align}
where $V_{3}(t)=2\pi^{2}a(t)^{3}$ (where $a(t)$ denotes the scale factor of a homogeneous, isotropic universe) is the spatial $3$-volume, $k_{2}=9(2\pi^{2})^\frac{2}{3}$ and $\lambda$ is a Lagrange multiplier fixing
\begin{align}\label{eq:27}
\int\!\!dt\ \sqrt{g_{tt}}\,V_{3}(t)=V_{4},
\end{align}
where $V_{4}$ is a certain $4$-volume of spacetime. The solution of (\ref{eq:26}) is a geometrical $4$-sphere, which amounts to the Euclidean de Sitter universe, the maximally symmetric spacetime in the case of a positive cosmological constant. This comes as a great surprise, since individual triangulations -- as it was stressed before -- are highly nonclassical and a priori devoid of any symmetry, except the one with respect to a time translation. What is even more important is that also dynamics of the model i.e. in particular (quantum) fluctuations of $N_{4,1}(i)$ can be described in that framework -- by the effective, semiclassical action (\ref{eq:26}) itself \cite{semi}. This is what I will deal with -- to some extent -- in the remaining chapters.
\chapter{The single slice structure}
\section{Expansion of the semiclassical action}
By construction, simplicial building blocks in temporal layers of a triangulation have to be connected in an appropriate order. Namely, in the simplest case, between a $(4,1)$ simplex at a time $i$ and another one at $i+1$ one has to put consecutively a $(3,2)$, a $(2,3)$ and a $(1,4)$, with successive simplices sharing a face (i.e. a tetrahedron), as it was already said for the case of $(1,4)$ and $(4,1)$ simplices, which share a spatial tetrahedron. (Generally, each of the above mentioned five simplices may occur multiple times there.) Hence the layer has some internal structure. However, in this chapter I will restrict the discussion to spatial hypersurfaces formed by common faces of $(4,1)$ and $(1,4)$ building blocks -- forgetting for a while about the other types of simplices -- in accordance with Section~2.2. This is what I call the single slice structure: the structure in which a temporal layer is contained between two spatial slices (hypersurfaces) at times $i$ and $i+1$, the second of which belongs already to the next layer and there is nothing else, $(\mathit{3,2})$ (alternatively $(3,2)$ and $(2,3)$) simplices are in some sense integrated out. As it will be shown in the next chapter and sketched out in Appendix A, actually there is a nontrivial structure emerging from $(\mathit{3,2})$ simplices, as well as $(3,2)$ and $(2,3)$ considered separately, compatible with that of $(\mathit{4,1})$. \\ 
\indent According to Section~2.2, the average distribution $\langle N_{4,1}(i)\rangle$ is well described in the blob by the semiclassical solution (\ref{eq:25}), resulting from the classical, continuous action (\ref{eq:26}). Since time $i$ is discrete $\langle N_{4,1}(i)\rangle$ actually corresponds to a discrete version of (\ref{eq:25}). Therefore we expect that in the analysed semiclassical limit CDT can be described (at least in the blob) by the discretised mini-superspace action $S_{\mathrm{dis}}$, whose explicit form will be discussed in the next section. More specifically, to take the \textit{quantum} effects -- fluctuactions around the effective, semiclassical geometry -- into account we make the expansion of $S_{\mathrm{dis}}$ around the distribution $\bar{N}_{4,1}(t)\equiv\langle N_{4,1}(i)\rangle$, which should extremise the action, with an increment $\delta N_{4,1}(t)$:
\begin{align}\label{eq:31}
S_{\mathrm{dis}}[\bar{N}_{4,1}(t)+\delta N_{4,1}(t)]=S_{\mathrm{dis}}[\bar{N}_{4,1}(t)]+ \nonumber\\ 
\frac{1}{2!}\sum_{j,i=0}^{T-1}\partial_{N_{4,1}(j)}\partial_{N_{4,1}(i)}
S_{\mathrm{dis}}[\bar{N}_{4,1}(t)]\,\delta N_{4,1}(i)\delta N_{4,1}(j)+
O(\delta N_{4,1}(t)^{3}),
\end{align}
which in particular implies
\begin{align}\label{eq:32}
\forall i\!:\partial_{N_{4,1}(i)}S_{\mathrm{dis}}[\bar{N}_{4,1}(t)]=0,
\end{align}
vanishing of first derivative of the action. \\ 
\indent The second most fundamental observable in our Monte Carlo simulations of CDT is the covariance matrix of the number of simplices of a given type, whose element -- the correlation between given times $t^{\prime}$ and $t^{\prime\prime}$ -- is defined as
\begin{align}\label{eq:33}
C_{p,q}(t^{\prime},t^{\prime\prime})=\frac{1}{n}\sum_{k=1}^{n}
\big(N_{p,q}^{(k)}(t^{\prime})-\langle N_{p,q}^{(k)}(t^{\prime})\rangle\big)
\big(N_{p,q}^{(k)}(t^{\prime\prime})-\langle N_{p,q}^{(k)}(t^{\prime\prime})\rangle\big)
\end{align}
(cf. \ref{eq:22}). Quantities $\sqrt{C_{p,q}(t,t)}$, included in Fig.~\ref{fig:21} for $(p,q)=(4,1)$, are the average fluctuations around the expectation values $\langle N_{p,q}(t)\rangle$. Now, let us again restrict ourselves to $(4,1)$ simplices. Analogically with $\langle N_{4,1}(t)\rangle$, $N_{4,1}^{(k)}$ are centered at $i_{CV}$'s and $C_{4,1}$ is additionally -- but on the justified basis -- symmetrised with respect to $t=39+\frac{1}{2}$, which corresponds to doubling of the number of configurations. In our numerical setup matrix $C_{4,1}$ is invertible and hence the inverse (of the) covariance matrix
\begin{align}\label{eq:34}
P_{4,1}=C_{4,1}^{-1}
\end{align}
can be found. What is essential is that there is the relation
\begin{align}\label{eq:35}
(P_{4,1})_{i,j}=\partial_{N_{4,1}(j)}\partial_{N_{4,1}(i)}S_{\mathrm{dis}}[\bar{N}_{4,1}(t)],
\end{align}
where I introduce the notation $i,j\equiv i+1,j+1$ for matrix elements (since time $\in[0,T-1]$). As a result, from analysis of matrix $P_{4,1}$ one can infere, at least to some extent, the exact form of the action $S_{\mathrm{dis}}$. It turns out the inverse covariance matrix has only simple, tridiagonal structure
\begin{align}\label{eq:36}
P_{4,1}=\left(
\begin{array}{ccccc}
a_{1}\! & b_{1}\! & \ast\!\! & \ast\!\! & b_{T} \\[-5pt] 
b_{1}\! & a_{2}\! & \ddots\!\! & \ast\!\! & \ast \\[-5pt] 
\ast\! & \ddots\! & \ddots\!\! & \ddots\!\! & \ast \\[-5pt] 
\ast\! & \ast\! & \ddots\!\! & a_{T-1}\!\! & b_{T-1} \\[1pt] 
b_{T}\! & \ast\! & \ast\!\! & b_{T-1}\!\! & a_{T}
\end{array}
\right),
\end{align}
with two additional elements at the antidiagonal resulting from the periodic boundary condition in time (neglecting numerical noise, denoted by the asterisks) and $\forall i\!:a_{i}>0$, $\forall i\!:b_{i}<0$. It is also by construction symmetric with respect to both the diagonal and antidiagonal ($a_{i}=a_{T+1-i}$, $b_{i}=b_{T-\mathrm{mod}(i,T)}$).
\section{Form of the discrete action}
The discretised action, with the modification (\ref{eq:21}) taken into account, has the general form
\begin{align}\label{eq:37}
S_{\mathrm{dis}}=k_{1}\sum_{t=0}^{T-1}\big(\tilde{S}_{k}(t)+\tilde{S}_{p}(t)\big)
+\varepsilon\,(\langle N_{4,1}\rangle-\mathcal{N}_{4})^{2},
\end{align}
where $k_{1}\sum_{t}\tilde{S}_{k}(t)\equiv k_{1}\tilde{S}_{k}[N_{4,1}(t)]$, $k_{1}\sum_{t}\tilde{S}_{p}(t)\equiv k_{1}\tilde{S}_{p}[N_{4,1}(t)]$ are respectively its kinetic and potential part, corresponding to the terms $\frac{g^{tt}\dot{V}_{3}(t)^{2}}{V_{3}(t)}$ and $k_{2}V_{3}(t)^{\frac{1}{3}}-\lambda V_{3}(t)$ in the continuous action (\ref{eq:26}) and $k_{1}$ denotes a coupling constant proportional to the inverse of $G$. In all the following I call $\tilde{S}_{k}[N_{4,1}(t)]$, $\tilde{S}_{p}[N_{4,1}(t)]$ the kinetic and potential term, respectively. I assume that the potential term $\tilde{S}_{p}(t)=\tilde{S}_{p}(N_{4,1}(t))$ and has the form
\begin{align}\label{eq:38}
\tilde{S}_{p}(t)=\tilde{k}_{2}N_{4,1}(t)^{\frac{1}{3}}-\tilde{\lambda} N_{4,1}(t),\ V(t)\equiv k_{1}\tilde{S}_{p}(t),
\end{align}
where $\tilde{k}_{2}$, $\tilde{\lambda}$ are coupling constants proportional to $k_{2}$ and $\lambda$, respectively. The kinetic term is more involved. However, from (\ref{eq:35}), (\ref{eq:36}) one immediately infers that $\tilde{S}_{k}(t)=\tilde{S}_{k}(N_{4,1}(t),N_{4,1}(t+1))$. Consequently, the diagonal of $P_{4,1}$ depends on both the kinetic and potential part of the action $S_{\mathrm{dis}}$, whereas its super-/subdiagonal on the kinetic one alone. I assume that the kinetic term
\begin{align}\label{eq:39}
\tilde{S}_{k}(t)=\frac{(N_{4,1}(t+1)-N_{4,1}(t))^{2}}{h(N_{4,1}(t),N_{4,1}(t+1))}=
\frac{(N_{4,1}(t+1)-N_{4,1}(t))^{2}}{h(N_{4,1}(t+1),N_{4,1}(t))}
\end{align}
and satisfies the relation
\begin{align}\label{eq:310}
\partial^{2}_{N_{4,1}(i)}\tilde{S}_{k}[\bar{N}_{4,1}(t)]+
\frac{\bar{N}_{4,1}(i+1)}{\bar{N}_{4,1}(i)}\,\partial_{N_{4,1}(i+1)}\partial_{N_{4,1}(i)}
\tilde{S}_{k}[\bar{N}_{4,1}(t)]+ \nonumber\\ \frac{\bar{N}_{4,1}(i-1)}{\bar{N}_{4,1}(i)}\,\partial_{N_{4,1}(i-1)}\partial_{N_{4,1}(i)}
\tilde{S}_{k}[\bar{N}_{4,1}(t)]=0,
\end{align}
which obviously happens at least for every $h(N_{4,1}(t),N_{4,1}(t+1))$ that is a quotient of combinations of power functions. According to (\ref{eq:35}) I may define the analogue of (\ref{eq:310}):
\begin{align}\label{eq:311}
\mathcal{V}(i)\equiv(P_{4,1})_{i,i}+\frac{\bar{N}_{4,1}(i+1)}{\bar{N}_{4,1}(i)}(P_{4,1})_{i,i+1}+
\frac{\bar{N}_{4,1}(i-1)}{\bar{N}_{4,1}(i)}(P_{4,1})_{i,i-1},
\end{align}
which -- keeping in mind the form of (\ref{eq:38}) and (\ref{eq:39}) -- in the relevant region i.e. at least in the blob, should be equivalent to the non-eliminated $k_{1}\partial^{2}_{N_{4,1}(i)}\tilde{S}_{p}[\bar{N}_{4,1}(t)]=
-\frac{2}{9}k_{1}\tilde{k}_{2}N_{4,1}(t)^{-\frac{5}{3}}$. Hence I can check the potential term (\ref{eq:38}) independently of the explicit form of $\tilde{S}_{k}(t)$ i.e. form of the function $h(N_{4,1}(t),N_{4,1}(t+1))$ (Fig.~\ref{fig:31}). (In figures I use the abbreviated notation $\partial_{i}\equiv\partial_{N_{4,1}(i)}$, $[t]\equiv[\bar{N}_{4,1}(t)]$.) The obtained $\mathcal{V}(i)$ exhibits in the blob the considerable numerical noise. As it will become more evident below, the underlying reason is a small magnitude of the potential term.
\begin{figure}[ht]
\centering
\scalebox{1}[1]{\input{sdpJGS41.tex}}
\caption{$-\frac{2}{9}\tilde{k}N_{4,1}(i)^{-\frac{5}{3}}$ fitted to $\mathcal{V}(i)$, in the relevant region where $\mathcal{V}(i)<0$\label{fig:31}}
\end{figure}
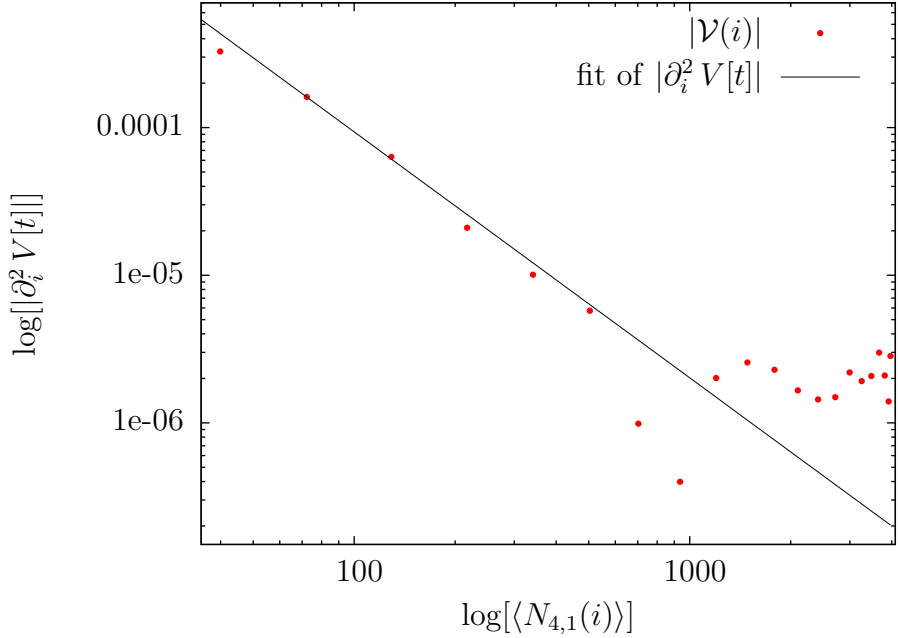
\begin{figure}[ht]
\centering
\scalebox{1}[1]{\input{ftexp41.tex}}
\caption{$\frac{1}{3}\tilde{k}_{2}N_{4,1}(i)^{-\frac{2}{3}}-\tilde{\lambda}$ fitted to $-\partial_{N_{4,1}(i)}\tilde{S}_{k}[\bar{N}_{4,1}(t)]$; for the illustrative purposes the term with $\varepsilon$ is absorbed into $\tilde{\lambda}$\label{fig:32}}
\end{figure}
\begin{figure}[p]
\centering
\scalebox{1}[1]{\input{smdkxi41.tex}}
\caption{$k_{1}\,\partial_{N_{4,1}(i+1)}\partial_{N_{4,1}(i)}\tilde{S}_{k}[\bar{N}_{4,1}(t)]$ fitted to $(P_{4,1})_{i,i+1}-2\varepsilon$\label{fig:33}}
\end{figure}
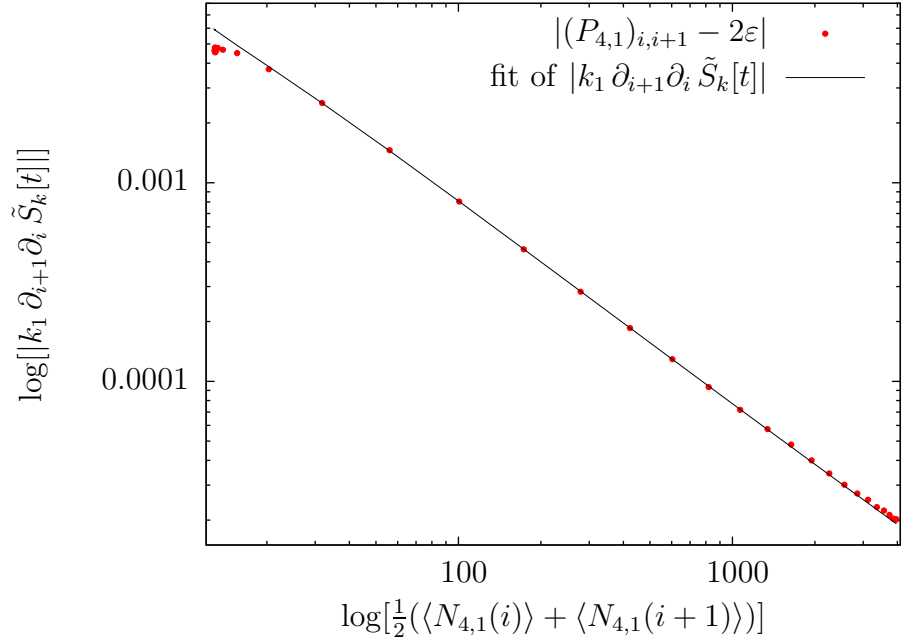
\begin{figure}[p]
\centering
\scalebox{1}[1]{\input{sdpxi41.tex}}
\caption{$-\frac{2}{9}\tilde{k}N_{4,1}(i)^{-\frac{5}{3}}$ fitted to $(P_{4,1})_{i,i}-2\varepsilon-k_{1}\,\partial^{2}_{N_{4,1}(i)}\tilde{S}_{k}[\bar{N}_{4,1}(t)]$, in the relevant region where the latter $<0$\label{fig:34}}
\end{figure}
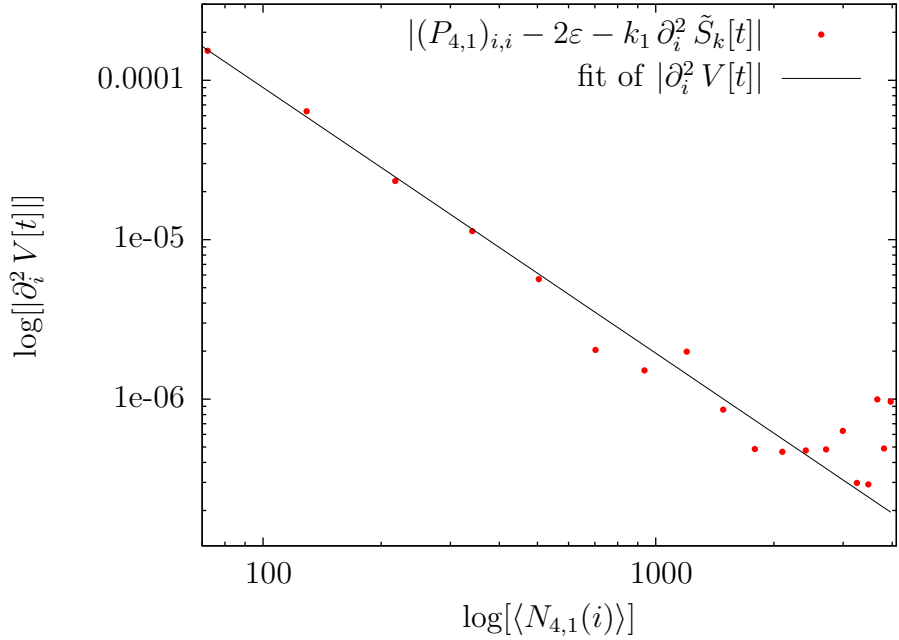 \\ 
\indent The simplest form of the kinetic term (\ref{eq:39}) one can consider is
\begin{align}\label{eq:312}
\tilde{S}_{k}(t)=\frac{2(N_{4,1}(t+1)-N_{4,1}(t))^{2}}{N_{4,1}(t)+N_{4,1}(t+1)}.
\end{align}
Now, regarding the form of the semiclassical action (\ref{eq:37}), the extremum condition (\ref{eq:32}) in the expansion (\ref{eq:31}) yields
\begin{align}\label{eq:313}
\forall i\!:-\partial_{N_{4,1}(i)}\tilde{S}_{k}[\bar{N}_{4,1}(t)]=
\partial_{N_{4,1}(i)}\tilde{S}_{p}[\bar{N}_{4,1}(t)]+\tfrac{2}{k_{1}}
\varepsilon\,(\langle N_{4,1}\rangle-\mathcal{N}_{4})
\end{align}
and I may examine the validity of (\ref{eq:31}), (\ref{eq:32}) by fitting first derivative of the potential term (\ref{eq:38}) to first derivative of the kinetic one (\ref{eq:312}) (Fig.~\ref{fig:32}). This clearly does not work in the stalk. In the blob and perhaps the tail, making the use of (\ref{eq:35}), for the superdiagonal of $P_{4,1}$ it should be
\begin{align}\label{eq:314}
\forall i\!:k_{1}\,\partial_{N_{4,1}(i+1)}\partial_{N_{4,1}(i)}\tilde{S}_{k}[\bar{N}_{4,1}(t)]=
(P_{4,1})_{i,i+1}-2\varepsilon
\end{align}
(a choice of the subdiagonal would obviously imply the same). However, the kinetic term (\ref{eq:312}) starts to fail there. I analysed different functions $h(N_{4,1}(t),N_{4,1}(t+1))$ and something like $\frac{1}{4}\big(\sqrt{N_{4,1}(i)}+\sqrt{N_{4,1}(i+1)}\big)^{2}$ turns out to behave much better. Alternatively, one may consider a modification of (\ref{eq:312}) in the form of a series with coefficients $\{\xi_{j}\}_{j}$,
\begin{align}\label{eq:315}
\tilde{S}_{k}(t)=\frac{(N_{4,1}(t+1)-N_{4,1}(t))^{2}}{N_{4,1}(t)+N_{4,1}(t+1)}
\bigg(1+ \nonumber\\ \xi_{1}\,\Big(\frac{N_{4,1}(t+1)-N_{4,1}(t)}{N_{4,1}(t)+N_{4,1}(t+1)}\Big)^{2}+\ldots\bigg).
\end{align}
Thus far, our numerical precision is sufficient to find only $\xi_{1}$ and hence constraints $\forall j>1\!:\xi_{j}=0$ are assumed. ($\xi_{1}=\frac{1}{4}$ corresponds then to an approximation of $\frac{1}{4}\big(\sqrt{N_{4,1}(i)}+\sqrt{N_{4,1}(i+1)}\big)^{2}$ and $\xi_{1}=0$ is obviously equivalent to (\ref{eq:312}).) I have examined (\ref{eq:314}) with the above kinetic term fitting simultaneously $k_{1}$ and $\xi_{1}$ (Fig.~\ref{fig:33}) and obtained a value of the latter $\xi_{1}=0.327\pm 0.030$, closer to $\frac{1}{4}$ than to $0$. Concerning the relation (\ref{eq:313}), presented in Fig.~\ref{fig:32}, its behaviour is practically unchanged by the modified kinetic term (\ref{eq:315}). Therefore, I do not show the corresponding version of the graph. At last, taking the values of $k_{1}$ and $\xi_{1}$ from Fig.~\ref{fig:33} I can investigate the potential term (\ref{eq:38}) at the diagonal of $P_{4,1}$, since
\begin{align}\label{eq:316}
\forall i\!:k_{1}\,\partial^{2}_{N_{4,1}(i)}\tilde{S}_{p}[\bar{N}_{4,1}(t)]=
(P_{4,1})_{i,i}-2\varepsilon-k_{1}\,\partial^{2}_{N_{4,1}(i)}\tilde{S}_{k}[\bar{N}_{4,1}(t)],
\end{align}
which results from (\ref{eq:35}) (Fig.~\ref{fig:34}). Now, I may compare the values of the constant $\tilde{k}\equiv k_{1}\tilde{k}_{2}$ obtained in Fig.~\ref{fig:31} and Fig.~\ref{fig:34}. They amount respectively to $\tilde{k}=0.908\pm 0.033$ and $\tilde{k}=0.873\pm 0.021$. Furthermore, I may compare the value of $\tilde{k}_{2}$ from Fig.~\ref{fig:32} (or rather its variant with the modified kinetic term) with the one that I can calculate from Fig.~\ref{fig:34}. They amount respectively to $\tilde{k}_{2}=18.9\pm 1.1$ and $\tilde{k}_{2}=22.97\pm 0.81$. The presented uncertainties include only the accuracy of fits and do not contain estimates of the systematic errors.
\chapter{The double slice structure}
\section{Evidence for the double structure}
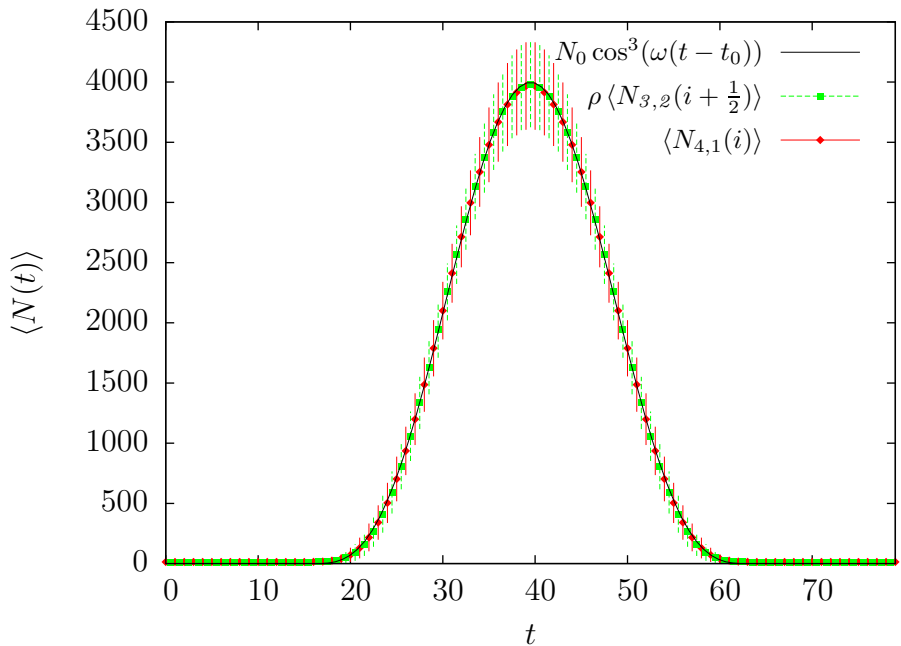
\begin{figure}[hb]
\centering
\scalebox{1}[1]{\input{avdbl.tex}}
\caption{The joint average distribution of $\langle N_{4,1}(i)\rangle$, $\rho\,\langle N_{\mathit{3,2}}(i+\frac{1}{2})\rangle$ with the semiclassical solution fitted and the average quantum fluctuations denoted by vertical lines (not to be confused with errors, which are invisible here!)\label{fig:41}}
\end{figure}

According to Section~3.1, $(\mathit{3,2})$ simplices bond together the adjacent temporal layers of $(\mathit{4,1})$ building blocks. Furthermore, refering to Fig.~\ref{fig:14}, one can observe that $(\mathit{4,1})$ simplices contain -- apart from a spatial tetrahedron -- only $(\mathit{3,1})$ timelike tetrahedra, while $(\mathit{3,2})$ simplices possess both $(\mathit{3,1})$ (two) and $(\mathit{2,2})$ (three) tetrahedra. Hence every $(\mathit{3,2})$ simplex has to share a face with three others. Numerical simulations confirm that this type of building blocks forms closed hypersurfaces in the sense of the dual lattice and thus we may speak about layers of $(\mathit{3,2})$ simplices. Therefore I extend my description to what I call the double slice structure -- characterised by the inclusion of remaining $(\mathit{3,2})$ simplicial building blocks, which implies additional spatial slices between those of the $(\mathit{4,1})$ i.e. an \textit{effective} second slice in every temporal layer, being a spatial section of the $(\mathit{3,2})$. Nevertheless, either $(3,2)$ or $(2,3)$ building blocks are still in some sense integrated out, as both types are considered collectively. As it was already mentioned, Appendix~A makes the outline of their separate treatment. \\ 
\indent I begin with a computation of the average temporal distribution of the number of $(\mathit{3,2})$ simplices, using formula (\ref{eq:22}). In agreement with what one might expect, an arbitrary distribution $N_{\mathit{3,2}}^{(k)}(i)$ turns out to have two regions: the spatialy extended and spatialy unextended. Therefore, according to the case of $(4,1)$ simplices, it requires a translation in time of the "center of volume". However, since I investigate the \textit{double structure}, distributions $N_{4,1}^{(k)}(i)$ are to be reparametrised also and obviously with respect to the same centers of volume. I have found that $i_{\mathrm{CV}}$'s (defined in Section~2.2) for $(4,1)$ and $(\mathit{3,2})$ building blocks are identical or at most different by $1$ and hence decided to consider $i_{\mathrm{CV}}$'s found for $(4,1)$ simplices alone. I obtain the average distribution $\langle N_{\mathit{3,2}}(i)\rangle$ having the same three regions as $\langle N_{4,1}(i)\rangle$: the stalk (with $12$ simplices at least, which results from the minimal number of $5$ $(4,1)$ building blocks), the blob (with the \textit{single} maximum at $i=39$) and the tail. Hence I assume $\langle N_{\mathit{3,2}}(i)\rangle\equiv\langle N_{\mathit{3,2}}(i+\frac{1}{2})\rangle$ and fulfilling the condition of invariance under time inversion I symmetrise the distribution with respect to $t=39+\frac{1}{2}$, which corresponds to doubling of the number of configurations. Now, it turns out in the blob $\langle N_{\mathit{3,2}}(i+\frac{1}{2})\rangle$ is proportional to $\langle N_{4,1}(i)\rangle$ in the interpolated sense i.e. I can find a scaling factor $\rho$ such that $\rho\,\langle N_{\mathit{3,2}}(i+\frac{1}{2})\rangle=\langle N_{4,1}(i+\frac{1}{2})\rangle$, where $\langle N_{4,1}(i+\frac{1}{2})\rangle$ denotes the interpolating distribution i.e. $\langle N_{4,1}(i)\rangle$ symmetrised with respect to $i=39$. The joint distribution of $\langle N_{4,1}(i)\rangle$ and $\rho\,\langle N_{\mathit{3,2}}(i+\frac{1}{2})\rangle$ is shown in Fig.~\ref{fig:41}, along with the average fluctuations, also appropriately rescaled in the case of $(\mathit{3,2})$ simplices. Furthermore, the smoothed distribution can automatically be fitted with the semiclassical solution (\ref{eq:25}) and this is done indeed in Fig.~\ref{fig:41}, up to even smaller deviation than in Fig.~\ref{fig:21}. The average total number of $(\mathit{3,2})$ simplices $\langle N_{\mathit{3,2}}\rangle\approx 207000$. Consequently, the average total number of all simplices (both $(\mathit{4,1})$ and $(\mathit{3,2})$) $\langle N\rangle\approx 367000$ (cf. Section~2.1). \\ 

\indent Using formula (\ref{eq:33}) one may also calculate the covariance matrix of $(\mathit{3,2})$ simplices number $C_{\mathit{3,2}}$ and in particular obtain the average fluctuations around $\langle N_{\mathit{3,2}}(i+\frac{1}{2})\rangle$, which are included above in Fig.~\ref{fig:41}. Further, I find the inverse covariance matrix $P_{\mathit{3,2}}=C_{\mathit{3,2}}^{-1}$ and it turns out to have the same simple structure as $P_{4,1}$ (cf. (\ref{eq:36})), enhancing the notion of the double slice structure. However, the analysis such as in Section~3.2 will not be carried out here. Rather than that, in the next section I will focus on the joint, double structure of $(4,1)$ and $(\mathit{3,2})$ simplices.
\section{The double discrete action}
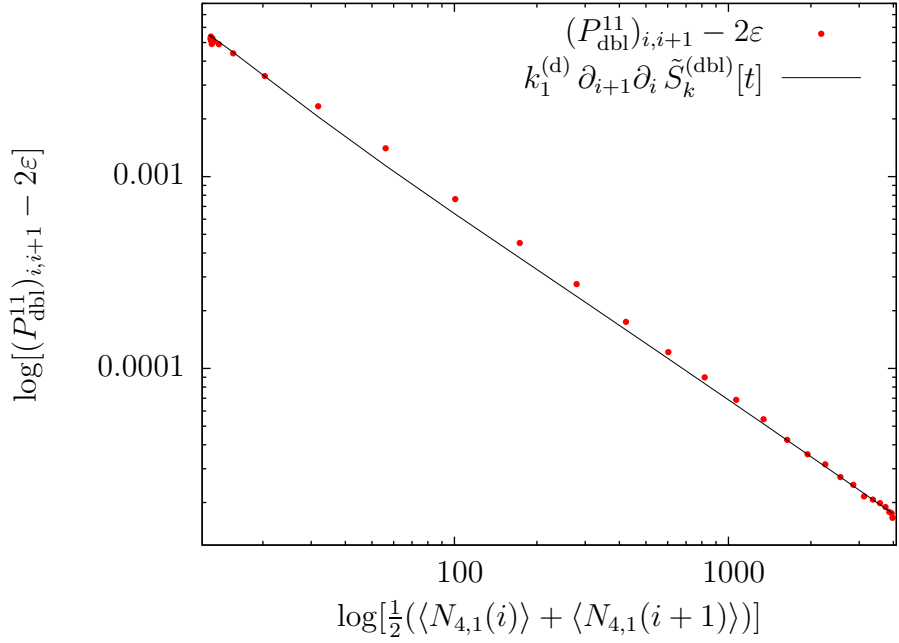
\begin{figure}[p]
\centering
\scalebox{1}[1]{\input{smdk41db.tex}}
\caption{$k^{(\mathrm{d})}_{1}\,\partial_{N_{4,1}(i+1)}\partial_{N_{4,1}(i)}
\tilde{S}^{(\mathrm{dbl})}_{k}[\bar{N}(t)]$ compared to the superdiagonal of $P_{\mathrm{dbl}}^{11}-2\varepsilon$\label{fig:44}}
\end{figure}

Since the semiclassical solution (\ref{eq:25}) works well in the blob also in the case of the double slice structure I expect that the expansion (\ref{eq:31}) (as well as the condition (\ref{eq:32})) remains valid for the extended discretised action $S^{(\mathrm{dbl})}_{\mathrm{dis}}$, which includes the dynamics of $(\mathit{3,2})$ simplices. More specifically,
\begin{align}\label{eq:41}
S^{(\mathrm{dbl})}_{\mathrm{dis}}[\bar{N}(t)+\delta N(t)]=S^{(\mathrm{dbl})}_{\mathrm{dis}}[\bar{N}(t)]+ \nonumber\\ 
\frac{1}{2!}\sum_{(r,s)}\sum_{(p,q)}\sum_{j,i=0}^{T-1}\partial_{N_{r,s}(j)}\partial_{N_{p,q}(i)}
S^{(\mathrm{dbl})}_{\mathrm{dis}}[\bar{N}(t)]\,\delta N_{p,q}(i)\delta N_{r,s}(j)+O(\delta N(t)^{3})
\end{align}
and
\begin{align}\label{eq:42}
\forall i\!:\partial_{N_{p,q}(i)}S^{(\mathrm{dbl})}_{\mathrm{dis}}[\bar{N}(t)]=0,
\end{align}
where $\bar{N}(t)=(\bar{N}_{4,1}(t),\bar{N}_{\mathit{3,2}}(t+\frac{1}{2}))$, $\delta N(t)=(\delta N_{4,1}(t),\delta N_{\mathit{3,2}}(t+\frac{1}{2}))$ and $(p,q)$, $(r,s)$ go over the set $\{(4,1),(\mathit{3,2})\}$ i.e. in particular there are four combinations of derivatives $\forall i,j$ in the second term of the expansion. \\ 
\indent The corresponding "double" covariance matrix is defined as the joint covariance matrix of both types of simplices i.e. (\ref{eq:33}) is extended to
\begin{align}\label{eq:43}
C_{\mathrm{dbl}}(t^{\prime},t^{\prime\prime})=\frac{1}{n}\sum_{k=1}^{n}
\big(N_{p,q}^{(k)}(t^{\prime})-\langle N_{p,q}^{(k)}(t^{\prime})\rangle\big)\big(N_{r,s}^{(k)}(t^{\prime\prime})-\langle N_{r,s}^{(k)}(t^{\prime\prime})\rangle\big).
\end{align}
Hence $C_{\mathrm{dbl}}$ may be arranged into four blocks, two of which are the ("single") covariance matrices $C_{4,1}$, $C_{\mathit{3,2}}$ and two mix both types of simplices and which will be denoted by superscripts $11$, $22$, $12$ and $21$, respectively. (Obviously $C_{\mathrm{dbl}}^{12}=(C_{\mathrm{dbl}}^{21})^{\mathrm{T}}$.) Each of them is appropriately symmetrised with respect to $t=39+\frac{1}{2}$. Matrix $C_{\mathrm{dbl}}$ is invertible and the double inverse covariance matrix $P_{\mathrm{dbl}}=C_{\mathrm{dbl}}^{-1}$. Now, (\ref{eq:35}) generalises to
\begin{align}\label{eq:44}
(P_{\mathrm{dbl}}^{qs})_{i,j}=\partial_{N_{r,s}(j)}\partial_{N_{p,q}(i)}
S^{(\mathrm{dbl})}_{\mathrm{dis}}[\bar{N}(t)],
\end{align}
where blocks of $P_{\mathrm{dbl}}$, denoted by superscripts, correspond to those of $C_{\mathrm{dbl}}$. Like $P_{4,1}$ (cf. (\ref{eq:36})), the double inverse covariance matrix is very simple
\begin{align}\label{eq:45}
P_{\mathrm{dbl}}=\left(
\begin{array}{cccccccccc}
c_{1}\! & d_{1}\!\! & \ast\!\! & \ast\!\! & d_{T} & f_{1}\!\! & \ast\! & \ast\!\! & \ast\!\! & f_{T} \\[-5pt] 
d_{1}\! & c_{2}\!\! & \ddots\!\! & \ast\!\! & \ast & f_{T-1}\!\! & f_{2}\! & \ast\!\! & \ast\!\! & \ast \\[-5pt] 
\ast\! & \ddots\!\! & \ddots\!\! & \ddots\!\! & \ast & \ast\!\! & \ddots\! & \ddots\!\! & \ast\!\! & \ast \\[-5pt] 
\ast\! & \ast\!\! & \ddots\!\! & c_{T-1}\!\! & d_{T-1} & \ast\!\! & \ast\! & \ddots\!\! & f_{T-1}\!\! & \ast \\[1pt] 
d_{T}\! & \ast\!\! & \ast\!\! & d_{T-1}\!\! & c_{T} & \ast\!\! & \ast\! & \ast\!\! & f_{1}\!\! & f_{T} \\[10pt] 
f_{1}\! & f_{T-1}\!\! & \ast\!\! & \ast\!\! & \ast & e_{1}\!\! & \ast\! & \ast\!\! & \ast\!\! & \ast \\[-5pt] 
\ast\! & f_{2}\!\! & \ddots\!\! & \ast\!\! & \ast & \ast\!\! & e_{2}\! & \ast\!\! & \ast\!\! & \ast \\[-5pt] 
\ast\! & \ast\!\! & \ddots\!\! & \ddots\!\! & \ast & \ast\!\! & \ast\! & \ddots\!\! & \ast\!\! & \ast \\[1pt] 
\ast\! & \ast\!\! & \ast\!\! & f_{T-1}\!\! & f_{1} & \ast\!\! & \ast\! & \ast\!\! & e_{T-1}\!\! & \ast \\[1pt] 
f_{T}\! & \ast\!\! & \ast\!\! & \ast\!\! & f_{T} & \ast\!\! & \ast\! & \ast\!\! & \ast\!\! & e_{T}
\end{array}
\right),
\end{align}
with blocks having band structure (neglecting numerical noise, denoted by the asterisks) -- except additional elements at their antidiagonals, resulting from the periodic boundary condition in time -- and $\forall i\!:c/d/e_{i}>0$, $\forall i\!:f_{i}<0$. Blocks $P_{\mathrm{dbl}}^{11}$, $P_{\mathrm{dbl}}^{22}$ are also by construction symmetric with respect to both the diagonal and antidiagonal and the entire $P_{\mathrm{dbl}}$ is symmetric with respect to the diagonal ($c/e_{i}=c/e_{T+1-i}$, $d_{i}=d_{T-\mathrm{mod}(i,T)}$ but $f_{i}\approx f_{T-\mathrm{mod}(i,T)}$ merely).
\begin{figure}[ht]
\centering
\scalebox{1}[1]{\input{ftexp41db.tex}}
\caption{$-\frac{1}{3}\tilde{k}^{11}_{2}N_{4,1}(i)^{-\frac{2}{3}}+\tilde{\lambda}^{11}$ fitted to $-\partial_{N_{4,1}(i)}\tilde{S}^{(\mathrm{dbl})}_{k}[\bar{N}(t)]$; for the illustrative purposes the term with $\varepsilon$ is absorbed into $\tilde{\lambda}^{11}$\label{fig:42}}
\vspace{-0.07cm}
\end{figure} \\ 
\indent Generalising (\ref{eq:37}), the double discrete action has the general form
\begin{align}\label{eq:46}
S^{(\mathrm{dbl})}_{\mathrm{dis}}=k^{(\mathrm{d})}_{1}\sum_{t=0}^{T-1}
\big(\tilde{S}^{(\mathrm{dbl})}_{k}(t)+\tilde{S}^{(\mathrm{dbl})}_{p}(t)\big)+
\varepsilon\,(\langle N_{4,1}\rangle-\mathcal{N}_{4})^{2}.
\end{align}
In accordance with (\ref{eq:44}) and the structure of $P_{\mathrm{dbl}}$, I consider the following (probably the simplest possible) kinetic term
\begin{align}\label{eq:47}
\tilde{S}^{(\mathrm{dbl})}_{k}(t)=
-a\,\frac{2(N_{4,1}(t+1)-N_{4,1}(t))^{2}}{N_{4,1}(t)+N_{4,1}(t+1)}+ \nonumber\\ 
\frac{2(\rho\,N_{\mathit{3,2}}(t+\frac{1}{2})-N_{4,1}(t))^{2}}{N_{4,1}(t)+
\sigma\,N_{\mathit{3,2}}(t+\frac{1}{2})}+
\frac{2(N_{4,1}(t)-\rho\,N_{\mathit{3,2}}(t-\frac{1}{2}))^{2}}{\sigma\, N_{\mathit{3,2}}(t-\frac{1}{2})+N_{4,1}(t)}
\end{align}
and the potential one
\begin{align}\label{eq:48}
\tilde{S}^{(\mathrm{dbl})}_{p}(t)=-\tilde{k}^{11}_{2}N_{4,1}(t)^{\frac{1}{3}}
+\tilde{\lambda}^{11}N_{4,1}(t)+  \nonumber\\ 
\tilde{k}^{22}_{2}N_{\mathit{3,2}}(t+\tfrac{1}{2})^{\frac{1}{3}}
-\tilde{\lambda}^{22}N_{\mathit{3,2}}(t+\tfrac{1}{2}),
\end{align}
where $a$, $\sigma$, $\tilde{k}^{11}_{2}$, $\tilde{\lambda}^{11}$, $\tilde{k}^{22}_{2}$, $\tilde{\lambda}^{22}$ (as well as $k^{(\mathrm{d})}$) are the "double" coupling constants and a value of $\rho$ is taken from the previous section. Consequently, all elements of $P_{\mathrm{dbl}}$ depend on the kinetic part of the action $S^{(\mathrm{dbl})}_{\mathrm{dis}}$ while the potential part influences merely the diagonal (of the blocks $P_{\mathrm{dbl}}^{11}$, $P_{\mathrm{dbl}}^{22}$).
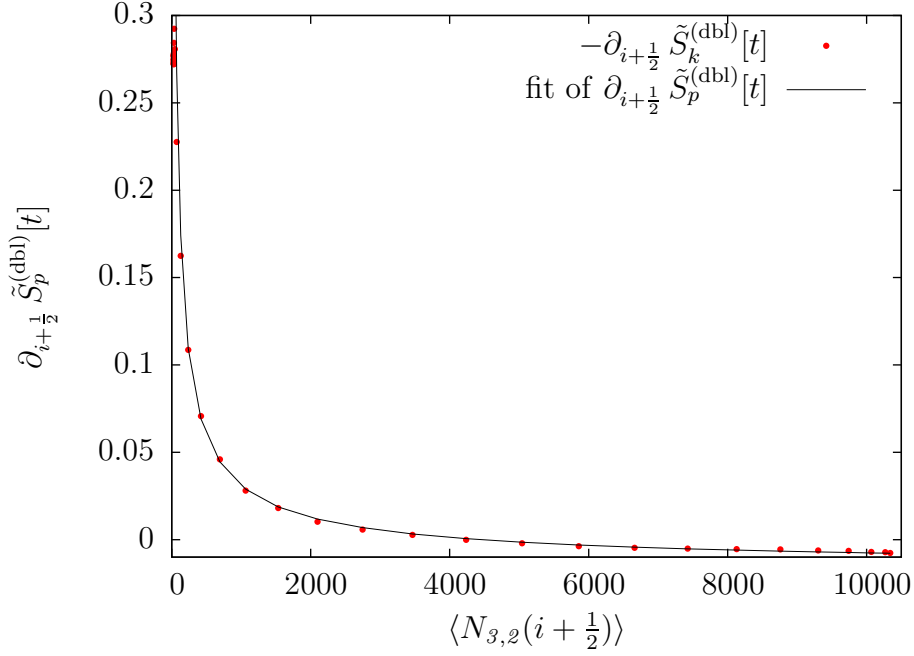
\begin{figure}[ht]
\centering
\scalebox{1}[1]{\input{ftexp32db.tex}}
\caption{$\frac{1}{3}\tilde{k}^{22}_{2}N_{\mathit{3,2}}(i+\frac{1}{2})^{-\frac{2}{3}}-
\tilde{\lambda}^{22}$ fitted to $-\partial_{N_{\mathit{3,2}}(i+\frac{1}{2})}\tilde{S}^{(\mathrm{dbl})}_{k}[\bar{N}(t)]$\label{fig:43}}
\end{figure} \\ 
\indent The extremum condition (\ref{eq:42}) in the expansion (\ref{eq:41}) yields
\begin{align}\label{eq:49}
\forall i\!:-\partial_{N_{p,q}(i)}\tilde{S}^{(\mathrm{dbl})}_{k}[\bar{N}(t)]=
\partial_{N_{p,q}(i)}\tilde{S}^{(\mathrm{dbl})}_{p}[\bar{N}(t)]+\tfrac{2}{k^{(\mathrm{d})}_{1}}
\varepsilon\,(\langle N_{4,1}\rangle-\mathcal{N}_{4})\delta_{4,1}^{p,q}.
\end{align}
Hence, similarly as in the case of Fig.~\ref{fig:32}, I may fit first derivative of the potential term (\ref{eq:48}) to first derivative of the kinetic one (\ref{eq:47}) -- for $(p,q)=(4,1)$ (Fig.~\ref{fig:42}) and $(p,q)=(\mathit{3,2})$ (Fig.~\ref{fig:43}). (The abbreviated notation in figures is the same as in the previous chapter.) It is worth to be noted that the agreement between them (and hence validity of (\ref{eq:42})) is good in the whole relevant region of the universe, almost right up to the stalk and hence much better than it was for the single slice structure. As a result, it is well justified to examine, making the use of (\ref{eq:44}) (analogically with Section~3.2), the substitution of the assumed kinetic and potential term into the second term of the expansion (\ref{eq:41}). In particular, the suitability of the kinetic term and values of its coupling constants can be directly inferred from
\begin{align}\label{eq:410}
k^{(\mathrm{d})}_{1}\,\partial_{N_{4,1}(i+1)}\partial_{N_{4,1}(i)}
\tilde{S}^{(\mathrm{dbl})}_{k}[\bar{N}(t)]=
(P_{\mathrm{dbl}}^{11})_{i,i+1}-2\varepsilon, \nonumber\\ 
k^{(\mathrm{d})}_{1}\,\partial_{N_{\mathit{3,2}}(i+\frac{1}{2})}\partial_{N_{4,1}(i)}
\tilde{S}^{(\mathrm{dbl})}_{k}[\bar{N}(t)]=
(P_{\mathrm{dbl}}^{12})_{i,i}.
\end{align}
Then, the potential term can be obtained through
\begin{align}\label{eq:411}
k^{(\mathrm{d})}_{1}\,\partial^{2}_{N_{4,1}(i)}
\tilde{S}^{(\mathrm{dbl})}_{p}[\bar{N}(t)]=(P_{\mathrm{dbl}}^{11})_{i,i}-2\varepsilon-
k^{(\mathrm{d})}_{1}\,\partial^{2}_{N_{4,1}(i)}\tilde{S}^{(\mathrm{dbl})}_{k}[\bar{N}(t)], \nonumber\\ 
k^{(\mathrm{d})}_{1}\,\partial^{2}_{N_{\mathit{3,2}}(i+\frac{1}{2})}
\tilde{S}^{(\mathrm{dbl})}_{p}[\bar{N}(t)]=
(P_{\mathrm{dbl}}^{22})_{i,i}-k^{(\mathrm{d})}_{1}\,
\partial^{2}_{N_{\mathit{3,2}}(i+\frac{1}{2})}\tilde{S}^{(\mathrm{dbl})}_{k}[\bar{N}(t)].
\end{align}
However, fitting LHS's of (\ref{eq:410}) to their RHS's does not yield completely reliable results. Therefore I use another method and determine coupling constants one by one. I will not describe that in detail but the scheme is as follows. At the beginning I find $C_{1}\equiv k^{(\mathrm{d})}_{1}(\rho+\sigma)^{2}$ from $(P_{\mathrm{dbl}}^{12})_{i,i}-(P_{\mathrm{dbl}}^{12})_{i+1,i}$. Next -- knowing the preceding -- $\sigma$ from the diagonal $(P_{\mathrm{dbl}}^{12})_{i,i}$. At last, I find $C_{2}\equiv k^{(\mathrm{d})}_{1}a$ from the superdiagonal $(P_{\mathrm{dbl}}^{11})_{i,i+1}$. Hence I get $\sigma$ while $k^{(\mathrm{d})}_{1}$, $a$ can be trivially calculated. Afterwards, the obtained LHS's of (\ref{eq:410}) may be compared to RHS's (Fig.~\ref{fig:44}, Fig.~\ref{fig:45}), with which they should agree. (The quite poor result for Fig.~\ref{fig:44} may indicate the modified $\tilde{S}^{(\mathrm{dbl})}_{k}$.) Furthermore, LHS's of (\ref{eq:411}) with $\tilde{S}^{(\mathrm{dbl})}_{p}$, $\tilde{S}^{(\mathrm{dbl})}_{k}$ exchanged may be compared to RHS's yet with second derivatives of $\tilde{S}^{(\mathrm{dbl})}_{p}[\bar{N}(t)]$ neglected (Fig.~\ref{fig:46}, Fig.~\ref{fig:47}). However, keeping in mind small magnitude of the latter, such a comparison as well gives an insight into the suitability of the kinetic term. At last, one may investigate the potential term (\ref{eq:48}) using (\ref{eq:411}) and obtaining the concurrent results but, for conciseness, I will not show this here.
\begin{figure}[p]
\centering
\scalebox{1}[1]{\input{smdk4132db.tex}}
\caption{$k^{(\mathrm{d})}_{1}\,\partial_{N_{\mathit{3,2}}(i+\frac{1}{2})}\partial_{N_{4,1}(i)}
\tilde{S}^{(\mathrm{dbl})}_{k}[\bar{N}(t)]$ compared to the diagonal of $P_{\mathrm{dbl}}^{12}$\label{fig:45}}
\end{figure}
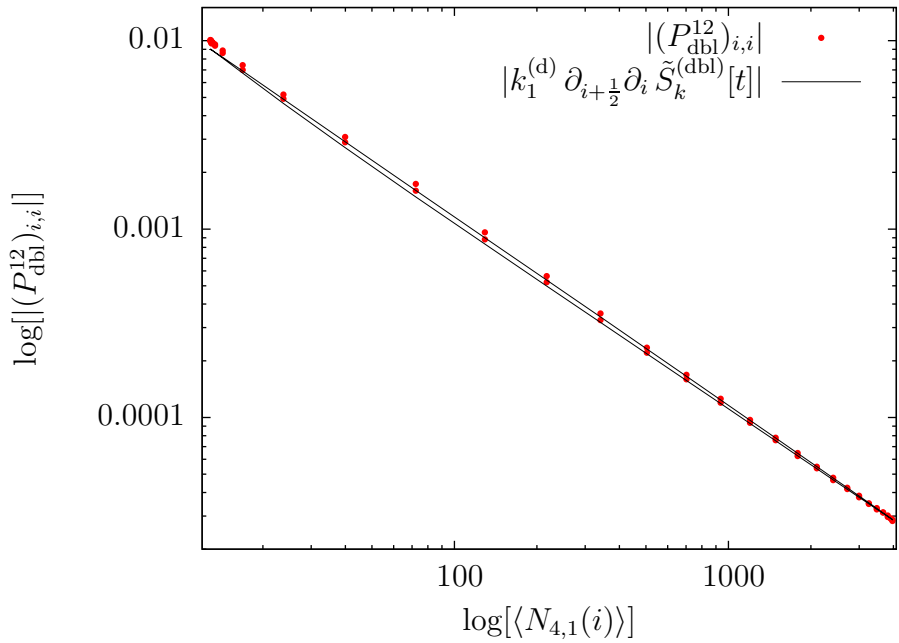
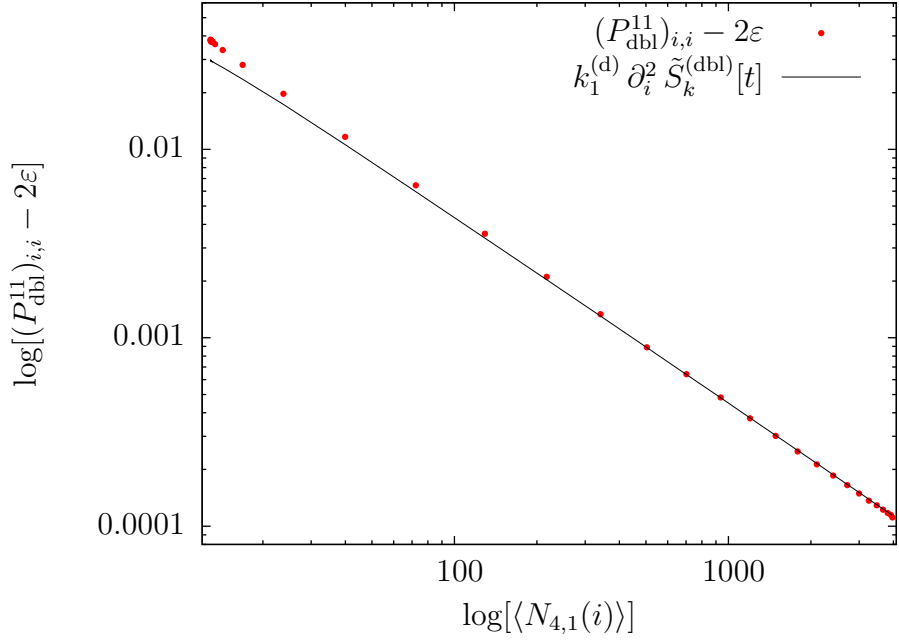
\begin{figure}[p]
\centering
\scalebox{1}[1]{\input{shdk41db.tex}}
\caption{$k^{(\mathrm{d})}_{1}\,\partial^{2}_{N_{4,1}(i)}
\tilde{S}^{(\mathrm{dbl})}_{k}[\bar{N}(t)]$ compared to the diagonal of $P_{\mathrm{dbl}}^{11}-2\varepsilon$ ($\tilde{S}^{(\mathrm{dbl})}_{p}[\bar{N}(t)]$ neglected)\label{fig:46}}
\vspace{-0.37cm}
\end{figure}
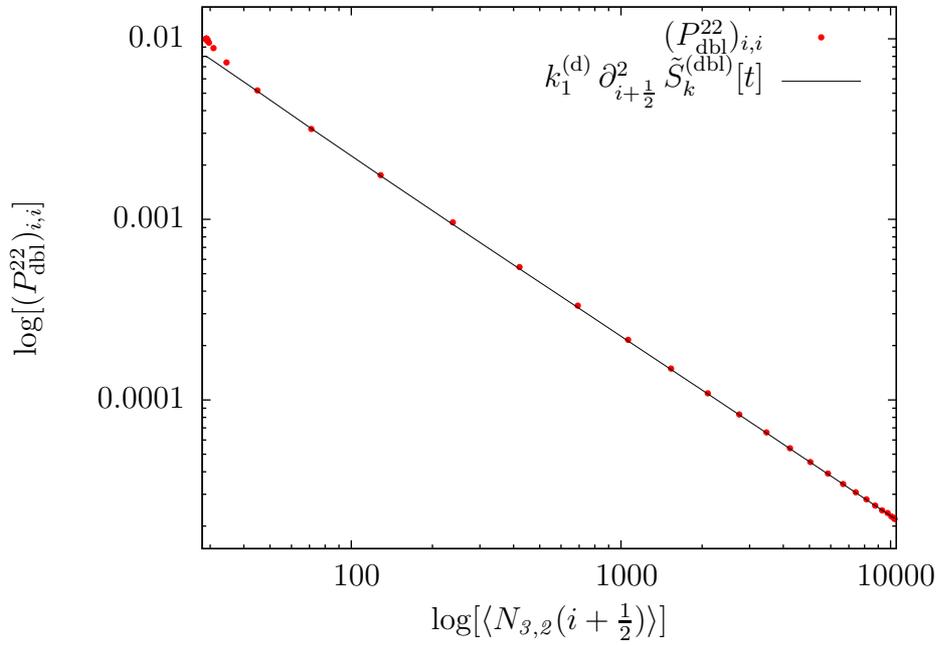
\begin{figure}[p]
\centering
\scalebox{1}[1]{\input{shdk32db.tex}}
\caption{$k^{(\mathrm{d})}_{1}\,\partial^{2}_{N_{\mathit{3,2}}(i+\frac{1}{2})}
\tilde{S}^{(\mathrm{dbl})}_{k}[\bar{N}(t)]$ compared to the diagonal of $P_{\mathrm{dbl}}^{22}$ ($\tilde{S}^{(\mathrm{dbl})}_{p}[\bar{N}(t)]$ neglected)\label{fig:47}}
\vspace{-0.37cm}
\end{figure}
\newpage
\section{Reduction to the single slice structure}
Having two different descriptions of the CDT universe, the latter of which is intended to be an extension of the former, we would like to be able to go in the opposite direction and to reduce the broader framework to the narrower one. In other words, after "integrating out" $(\mathit{3,2})$ simplices we should obtain the single slice structure of $(4,1)$ simplices alone. For the average distribution of the number of simplices this is trivial -- one just removes the second component from the joint distribution of $\langle N_{4,1}(i)\rangle$ and $\rho\,\langle N_{\mathit{3,2}}(i+\frac{1}{2})\rangle$. The situation is the same in the case of the covariance matrix, where simply $C_{4,1}=C_{\mathrm{dbl}}^{11}$, as it was described in the previous section. However, for the inverse covariance matrix such a reduction is inevitably not so trivial and its existence is by no means obvious. Fortunately, it turns out that for an invertible rectangular matrix composed of four square blocks satisfying certain basic conditions there are useful algebraic relations between those blocks and blocks of the inverse matrix, which we may apply to $C_{\mathrm{dbl}}$ and $P_{\mathrm{dbl}}$. The so-called Boltz-Banachiewicz inversion formula has version I,
\begin{align}\label{eq:412}
\left(
\begin{array}{cc} 
A & B \\
C & D
\end{array}
\right)^{-1}= \nonumber\\ \left(
\begin{array}{@{}cc@{}}
A^{-1}+A^{-1}B(D-CA^{-1}B)^{-1}CA^{-1} & -A^{-1}B(D-CA^{-1}B)^{-1} \\
-(D-CA^{-1}B)^{-1}CA^{-1} & (D-CA^{-1}B)^{-1}
\end{array}
\right),
\end{align}
and version II,
\begin{align}\label{eq:413}
\left(
\begin{array}{cc}
A & B \\
C & D
\end{array}
\right)^{-1}= \nonumber\\ \left(
\begin{array}{@{}cc@{}}
(A-BD^{-1}C)^{-1} & -(A-BD^{-1}C)^{-1}BD^{-1} \\
-D^{-1}C(A-BD^{-1}C)^{-1} & D^{-1}+D^{-1}C(A-BD^{-1}C)^{-1}BD^{-1}
\end{array}
\right),
\end{align}
comparison of which yields additionally four algebraic equations. However, I will restrict myself to the statement that transforming (\ref{eq:413}) one can retrieve $P_{4,1}$ from $P_{\mathrm{dbl}}$:
\begin{align}\label{eq:414}
P_{4,1}=\big((P_{\mathrm{dbl}}^{11})^{-1}+
C_{\mathrm{dbl}}^{12}\,P_{\mathit{3,2}}\,(C_{\mathrm{dbl}}^{12})^{\mathrm{T}}\big)^{-1},
\end{align}
where the transposition results from the symmetry of $C_{\mathrm{dbl}}$. \\ 
\indent The next and crucial step is a reduction of the double discrete action $S^{(\mathrm{dbl})}_{\mathrm{dis}}$ to the "single" one $S_{\mathrm{dis}}$. In the double slice structure I consider the kinetic term (\ref{eq:47}) and the potential one (\ref{eq:48}), the only ones which I have discussed in the previous section. Looking at their form one may suspect that in the single slice structure they correspond to the (simplest possible) kinetic term (\ref{eq:312}) and the potential one (\ref{eq:38}). In order to make the desired transition to $S_{\mathrm{dis}}$ I apply a Gaussian approximation to (\ref{eq:47}) i.e. I literally integrate out $(\mathit{3,2})$ simplices, although in a specific way,
\begin{align}\label{eq:415}
\int\!\!dz\ \exp\!\big(\!-(y-x)^{2}H^{-1}-(z-x)^{2}I^{-1}-(y-z)^{2}I^{-1}\big)= \nonumber\\ 
const\exp\!\big(\!-(y-x)^{2}\big(\tfrac{1}{2}H^{-1}+I^{-1}\big)\big)
\end{align}
where $x=N_{4,1}(t)$, $y=N_{4,1}(t+1)$, $z=\rho\,N_{\mathit{3,2}}(t+\frac{1}{2})$ and the denominators $H=-(2k^{(\mathrm{d})}_{1})^{-1}(N_{4,1}(t)+N_{4,1}(t+1))$, $I=(2k^{(\mathrm{d})}_{1})^{-1}(N_{4,1}(t)+\sigma N_{\mathit{3,2}}(t+\frac{1}{2}))\approx (2k^{(\mathrm{d})}_{1})^{-1}\frac{1}{2}(1+\frac{\sigma}{\rho})(N_{4,1}(t)+N_{4,1}(t+1))$ (i.e. the approximation is made before the integration). For the potential term (\ref{eq:48}) I assume $\rho\,N_{\mathit{3,2}}(t+\frac{1}{2})\approx N_{4,1}(t)$ and take the rescaling of $k^{(\mathrm{d})}_{1}$ implied by (\ref{eq:415}) into account. As a result, the "single" coupling constants are expressed through the double ones,
\begin{align}\label{eq:416}
k_{1}=k^{(\mathrm{d})}_{1}\Big(-a+\frac{\rho}{\rho+\sigma}\Big), \\ \label{eq:417}\tilde{k}_{2}=\big(-\tilde{k}^{11}_{2}+\rho^{-\frac{1}{3}}\,\tilde{k}^{22}_{2}\big)
\frac{\rho+\sigma}{\rho-a(\rho+\sigma)}, \\ 
\label{eq:418}\tilde{\lambda}=\big(-\tilde{\lambda}^{11}+\rho^{-1}\,\tilde{\lambda}^{22}\big)
\frac{\rho+\sigma}{\rho-a(\rho+\sigma)}.
\end{align}
The actual investigation produce the following results. Values of the single coupling constants obtained directly from the single slice structure (embodied in Fig.~\ref{fig:32} but not Fig.~\ref{fig:33}, Fig.~\ref{fig:34}, where the kinetic term is (\ref{eq:315})) are
$k_{1}=0.04412\pm 0.00069$, $\tilde{k}_{2}=17.55\pm 0.87$ and $\tilde{\lambda}=0.0543\pm 0.0075$ while values from the double slice structure, calculated using (\ref{eq:416}), (\ref{eq:417}), (\ref{eq:418}), amount to $k_{1}=0.0391\pm 0.0026$, $\tilde{k}_{2}=14.28\pm 0.84$ and $\tilde{\lambda}=0.0530\pm 0.0054$. The presented uncertainities include only the accuracy of fits and do not contain estimates of the systematic errors.
\chapter*{Summary}
\addcontentsline{toc}{chapter}{Summary}
I have verified the existence of the semiclassical limit of CDT in $3+1$ dimensions for spatial hypersurfaces formed by $(\mathit{4,1})$ simplicial building blocks of spacetime. The blob region of the universe generated by the triangulated path integral turns out to behave effectively like a discrete mini-superspace model. This manifests not only in the average distribution of the number of $(4,1)$ simplices but also in the (quantum) dynamics of the system, governed by the discretised mini-superspace action, which is extremised by the average distribution. The latter can be investigated using the inverse covariance matrix of the number of $(4,1)$ simplices. Analysis of its exceptionally simple structure indicates an intuitive form of the potential part of the action and more complicated of the kinetic one. \\ 
\indent Furthermore, I have shown that there is an extension of the above semiclassical description -- which I called the single slice structure -- to the double slice structure, including spatial slices of both $(4,1)$ and $(\mathit{3,2})$ simplices. After the rescaling of the number of $(\mathit{3,2})$ simplices, the joint average distribution of the number of both types of simplices in the blob again corresponds to the discrete mini-superspace solution. Next, the "double" discretised semiclassical action, extremised by the average distributions, can be examined using the inverse joint covariance matrix, which consists of four simple blocks. The only discussed form of the action turns out to work reasonably well. Last but not least, the double slice structure can be reduced to the single slice structure. \\ 
\indent The next step to make is splitting layers of $(\mathit{3,2})$ building blocks into those of $(3,2)$ and $(2,3)$ simplices, which amounts to the triple slice structure. I argue for that in Appendix~A yet this is a matter of further research, the same as e.g. consideration of other forms of the double action. 
\appendix
\chapter{The triple slice structure}
\begin{figure}[ht]
\centering
\scalebox{1}[1]{\input{avtrp.tex}}
\caption{The joint average distribution of $\langle N_{4,1}(i)\rangle$, $2\rho\,\langle N_{3,2}(i+\frac{1}{3})\rangle$, $2\rho\,\langle N_{2,3}(i+\frac{2}{3})\rangle$ with the semiclassical solution fitted and the average quantum fluctuations denoted by vertical lines (not to be confused with errors!)\label{fig:A1}}
\end{figure}
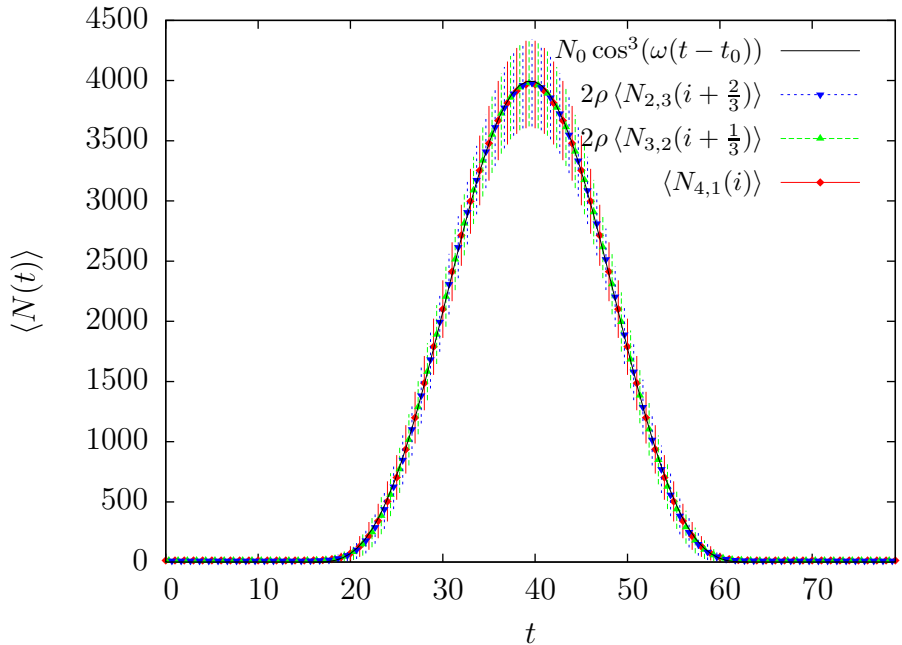

Below, I present some evidence that the description of the CDT universe from Chapters 3, 4 can be extended even more by replacing layers of $(\mathit{3,2})$ building blocks with the separate treatment of $(3,2)$ and $(2,3)$ simplices. Such a framework would be the ultimate one because of the eventual inclusion of all types of simplices -- since $(4,1)$ building blocks yield the single slice structure together with the $(1,4)$ and $N_{4,1}(i)=N_{1,4}(i)$, which means $(1,4)$ simplices contain no additional information. I call it the triple slice structure, expecting $(3,2)$ and $(2,3)$ simplices to form their own effective spatial slices by splitting those of the $(\mathit{3,2})$, in such a way we will obtain three spatial slices in every temporal layer, beginning with spatial tetrahedra which belong to the $(4,1)$. Such a split will be in agreement with construction of a causal triangulation i.e. the order of gluing of simplicial building blocks presented at the beginning of Section~3.1. \\ 
\indent In the first place, let us consider the average temporal distributions of the number of $(3,2)$ and $(2,3)$ simplices, using formula (\ref{eq:22}). Since by definition $N_{\mathit{3,2}}(i)=N_{3,2}(i)+N_{2,3}(i)$ individual distributions $N_{3,2}^{(k)}(i)$, $N_{2,3}^{(k)}(i)$ are reparametrised in the same manner as it was in the case of $(\mathit{3,2})$ building blocks (see Section~4.1). This is justified by the fact that $i_{\mathrm{CV}}$'s for $(3,2)$ and $(2,3)$ simplices differ from those for the $(4,1)$ to the same degree as the $(\mathit{3,2})$. Unsurprisingly, I obtain the average distributions $\langle N_{3,2}(i)\rangle$, $\langle N_{2,3}(i)\rangle$ having the already known three regions: the stalk (both with $6$ simplices at least), the blob (with single maxima at $i=39$ and $i=40$, respectively) and the tail. Now, looking at their profiles, I assume $\langle N_{3,2}(i)\rangle\equiv\langle N_{3,2}(i+\frac{1}{3})\rangle$ and $\langle N_{2,3}(i)\rangle\equiv\langle N_{2,3}(i+\frac{2}{3})\rangle$. Moreover, fulfilling the condition of invariance under time inversion (cf. Sections~2.2, 4.1), I carry out the appropriate procedures of symmetrisation of the distributions with respect to $t=39+\frac{1}{2}$, which actually do not make them strictly symmetric for the obvious reason that sets $\{i+\frac{1}{3}\}_{i}$ and $\{i+\frac{2}{3}\}_{i}$ are \textit{asymmetric} with respect to $39+\frac{1}{2}$. Nevertheless, $\langle N_{3,2}(i+\frac{1}{3})\rangle$, $\langle N_{2,3}(i+\frac{2}{3})\rangle$ are altered only slightly. The invariance under time inversion also implies $\langle N_{3,2}(i+\frac{1}{3})\rangle\approx\langle N_{2,3}(i+\frac{2}{3})\rangle\approx\frac{1}{2}\langle N_{\mathit{3,2}}(i+\frac{1}{2})\rangle$ and it is these almost-equalities that explain the obtained shape of the distributions. Consequently, in the blob $\langle N_{3,2}(i+\frac{1}{3})\rangle$, $\langle N_{2,3}(i+\frac{2}{3})\rangle$ should be proportional to $\langle N_{4,1}(i)\rangle$ in the interpolated sense (see Section~4.1), with the scaling factor $2\rho$. They turn out to be indeed. The joint distribution of $2\rho\,\langle N_{3,2}(i+\frac{1}{3})\rangle$, $2\rho\,\langle N_{2,3}(i+\frac{2}{3})\rangle$ and $\langle N_{4,1}(i)\rangle$ is shown in Fig.~\ref{fig:A1}, along with the average fluctuations, rescaled in the case of $(3,2)$ and $(2,3)$ simplices. Furthermore, it is fitted with the semiclassical solution (\ref{eq:25}) up to yet smaller deviation than in Fig.~\ref{fig:41}. The average total number of simplices $\langle N_{3,2}\rangle\approx\langle N_{2,3}\rangle\approx 103500$. \\ 
\indent Secondly, very briefly, using formula (\ref{eq:33}) I calculate the covariance matrices $C_{3,2}$, $C_{2,3}$ and their inverses $P_{3,2}$, $P_{2,3}$, which obviously turn out to have the same properties as $C_{\mathit{3,2}}$ and $P_{\mathit{3,2}}$, up to a smaller symmetry. Thirdly, also very briefly, naturally extending the definition (\ref{eq:43}) I calculate the "triple" covariance matrix $C_{\mathrm{trl}}$ and its inverse $P_{\mathrm{trl}}$. The triple inverse covariance matrix $P_{\mathrm{trl}}$ has simple form analogical with $P_{\mathrm{dbl}}$ (cf. (\ref{eq:45})) i.e. it consists of nine blocks having band structure. Thus we may expect that it will not be difficult to find the corresponding triple discrete action.

\end{document}

%% file: avsng.tex
\begingroup
  \makeatletter
  \providecommand\color[2][]{%
    \GenericError{(gnuplot) \space\space\space\@spaces}{%
      Package color not loaded in conjunction with
      terminal option `colourtext'%
    }{See the gnuplot documentation for explanation.%
    }{Either use 'blacktext' in gnuplot or load the package
      color.sty in LaTeX.}%
    \renewcommand\color[2][]{}%
  }%
  \providecommand\includegraphics[2][]{%
    \GenericError{(gnuplot) \space\space\space\@spaces}{%
      Package graphicx or graphics not loaded%
    }{See the gnuplot documentation for explanation.%
    }{The gnuplot epslatex terminal needs graphicx.sty or graphics.sty.}%
    \renewcommand\includegraphics[2][]{}%
  }%
  \providecommand\rotatebox[2]{#2}%
  \@ifundefined{ifGPcolor}{%
    \newif\ifGPcolor
    \GPcolortrue
  }{}%
  \@ifundefined{ifGPblacktext}{%
    \newif\ifGPblacktext
    \GPblacktexttrue
  }{}%
  \let\gplgaddtomacro\g@addto@macro
  \gdef\gplbacktext{}%
  \gdef\gplfronttext{}%
  \makeatother
  \ifGPblacktext
    \def\colorrgb#1{}%
    \def\colorgray#1{}%
  \else
    \ifGPcolor
      \def\colorrgb#1{\color[rgb]{#1}}%
      \def\colorgray#1{\color[gray]{#1}}%
      \expandafter\def\csname LTw\endcsname{\color{white}}%
      \expandafter\def\csname LTb\endcsname{\color{black}}%
      \expandafter\def\csname LTa\endcsname{\color{black}}%
      \expandafter\def\csname LT0\endcsname{\color[rgb]{1,0,0}}%
      \expandafter\def\csname LT1\endcsname{\color[rgb]{0,1,0}}%
      \expandafter\def\csname LT2\endcsname{\color[rgb]{0,0,1}}%
      \expandafter\def\csname LT3\endcsname{\color[rgb]{1,0,1}}%
      \expandafter\def\csname LT4\endcsname{\color[rgb]{0,1,1}}%
      \expandafter\def\csname LT5\endcsname{\color[rgb]{1,1,0}}%
      \expandafter\def\csname LT6\endcsname{\color[rgb]{0,0,0}}%
      \expandafter\def\csname LT7\endcsname{\color[rgb]{1,0.3,0}}%
      \expandafter\def\csname LT8\endcsname{\color[rgb]{0.5,0.5,0.5}}%
    \else
      \def\colorrgb#1{\color{black}}%
      \def\colorgray#1{\color[gray]{#1}}%
      \expandafter\def\csname LTw\endcsname{\color{white}}%
      \expandafter\def\csname LTb\endcsname{\color{black}}%
      \expandafter\def\csname LTa\endcsname{\color{black}}%
      \expandafter\def\csname LT0\endcsname{\color{black}}%
      \expandafter\def\csname LT1\endcsname{\color{black}}%
      \expandafter\def\csname LT2\endcsname{\color{black}}%
      \expandafter\def\csname LT3\endcsname{\color{black}}%
      \expandafter\def\csname LT4\endcsname{\color{black}}%
      \expandafter\def\csname LT5\endcsname{\color{black}}%
      \expandafter\def\csname LT6\endcsname{\color{black}}%
      \expandafter\def\csname LT7\endcsname{\color{black}}%
      \expandafter\def\csname LT8\endcsname{\color{black}}%
    \fi
  \fi
  \setlength{\unitlength}{0.0500bp}%
  \begin{picture}(7200.00,5040.00)%
    \gplgaddtomacro\gplbacktext{%
      \csname LTb\endcsname%
      \put(1386,704){\makebox(0,0)[r]{\strut{} 0}}%
      \put(1386,1156){\makebox(0,0)[r]{\strut{} 500}}%
      \put(1386,1609){\makebox(0,0)[r]{\strut{} 1000}}%
      \put(1386,2061){\makebox(0,0)[r]{\strut{} 1500}}%
      \put(1386,2514){\makebox(0,0)[r]{\strut{} 2000}}%
      \put(1386,2966){\makebox(0,0)[r]{\strut{} 2500}}%
      \put(1386,3419){\makebox(0,0)[r]{\strut{} 3000}}%
      \put(1386,3871){\makebox(0,0)[r]{\strut{} 3500}}%
      \put(1386,4324){\makebox(0,0)[r]{\strut{} 4000}}%
      \put(1386,4776){\makebox(0,0)[r]{\strut{} 4500}}%
      \put(1518,484){\makebox(0,0){\strut{} 0}}%
      \put(2207,484){\makebox(0,0){\strut{} 10}}%
      \put(2895,484){\makebox(0,0){\strut{} 20}}%
      \put(3584,484){\makebox(0,0){\strut{} 30}}%
      \put(4272,484){\makebox(0,0){\strut{} 40}}%
      \put(4961,484){\makebox(0,0){\strut{} 50}}%
      \put(5650,484){\makebox(0,0){\strut{} 60}}%
      \put(6338,484){\makebox(0,0){\strut{} 70}}%
      \put(484,2740){\rotatebox{90}{\makebox(0,0){\strut{}$\langle N_{4,1}(t)\rangle$}}}%
      \put(4238,154){\makebox(0,0){\strut{}$t$}}%
    }%
    \gplgaddtomacro\gplfronttext{%
      \csname LTb\endcsname%
      \put(5971,4218){\makebox(0,0)[r]{\strut{}\footnotesize$\langle N_{4,1}(i)\rangle$}}%
      \csname LTb\endcsname%
      \put(5971,4548){\makebox(0,0)[r]{\strut{}\footnotesize$N_{0}\cos^{3}(\omega(t-t_{0}))$}}%
    }%
    \gplbacktext
    \put(0,0){\includegraphics{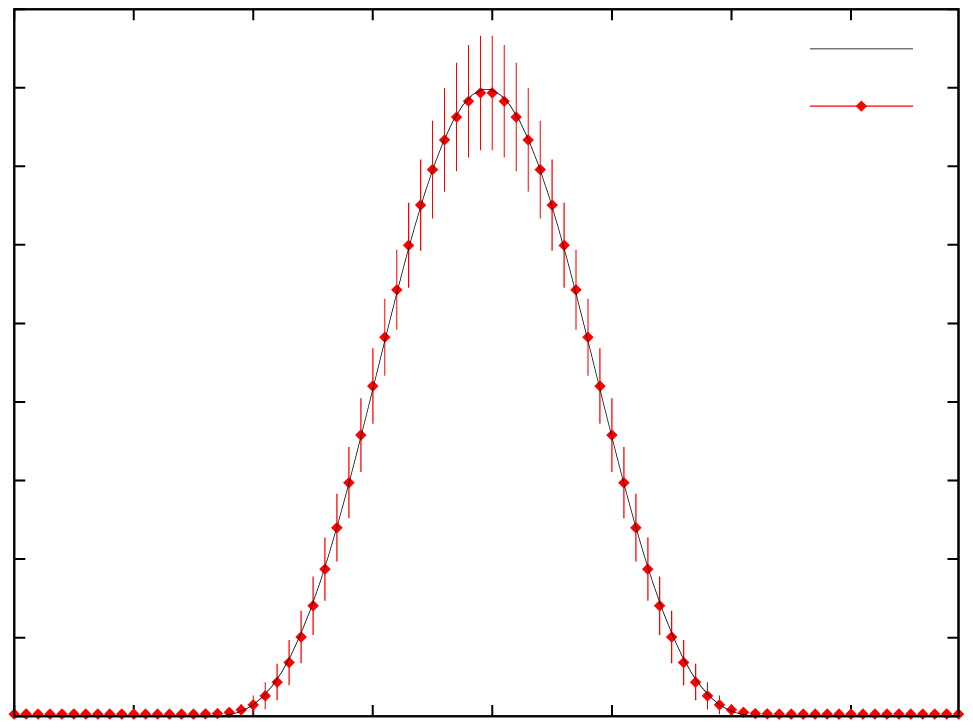}}%
    \gplfronttext
  \end{picture}%
\endgroup

%% file: sdpJGS41.tex
\begingroup
  \makeatletter
  \providecommand\color[2][]{%
    \GenericError{(gnuplot) \space\space\space\@spaces}{%
      Package color not loaded in conjunction with
      terminal option `colourtext'%
    }{See the gnuplot documentation for explanation.%
    }{Either use 'blacktext' in gnuplot or load the package
      color.sty in LaTeX.}%
    \renewcommand\color[2][]{}%
  }%
  \providecommand\includegraphics[2][]{%
    \GenericError{(gnuplot) \space\space\space\@spaces}{%
      Package graphicx or graphics not loaded%
    }{See the gnuplot documentation for explanation.%
    }{The gnuplot epslatex terminal needs graphicx.sty or graphics.sty.}%
    \renewcommand\includegraphics[2][]{}%
  }%
  \providecommand\rotatebox[2]{#2}%
  \@ifundefined{ifGPcolor}{%
    \newif\ifGPcolor
    \GPcolortrue
  }{}%
  \@ifundefined{ifGPblacktext}{%
    \newif\ifGPblacktext
    \GPblacktexttrue
  }{}%
  \let\gplgaddtomacro\g@addto@macro
  \gdef\gplbacktext{}%
  \gdef\gplfronttext{}%
  \makeatother
  \ifGPblacktext
    \def\colorrgb#1{}%
    \def\colorgray#1{}%
  \else
    \ifGPcolor
      \def\colorrgb#1{\color[rgb]{#1}}%
      \def\colorgray#1{\color[gray]{#1}}%
      \expandafter\def\csname LTw\endcsname{\color{white}}%
      \expandafter\def\csname LTb\endcsname{\color{black}}%
      \expandafter\def\csname LTa\endcsname{\color{black}}%
      \expandafter\def\csname LT0\endcsname{\color[rgb]{1,0,0}}%
      \expandafter\def\csname LT1\endcsname{\color[rgb]{0,1,0}}%
      \expandafter\def\csname LT2\endcsname{\color[rgb]{0,0,1}}%
      \expandafter\def\csname LT3\endcsname{\color[rgb]{1,0,1}}%
      \expandafter\def\csname LT4\endcsname{\color[rgb]{0,1,1}}%
      \expandafter\def\csname LT5\endcsname{\color[rgb]{1,1,0}}%
      \expandafter\def\csname LT6\endcsname{\color[rgb]{0,0,0}}%
      \expandafter\def\csname LT7\endcsname{\color[rgb]{1,0.3,0}}%
      \expandafter\def\csname LT8\endcsname{\color[rgb]{0.5,0.5,0.5}}%
    \else
      \def\colorrgb#1{\color{black}}%
      \def\colorgray#1{\color[gray]{#1}}%
      \expandafter\def\csname LTw\endcsname{\color{white}}%
      \expandafter\def\csname LTb\endcsname{\color{black}}%
      \expandafter\def\csname LTa\endcsname{\color{black}}%
      \expandafter\def\csname LT0\endcsname{\color{black}}%
      \expandafter\def\csname LT1\endcsname{\color{black}}%
      \expandafter\def\csname LT2\endcsname{\color{black}}%
      \expandafter\def\csname LT3\endcsname{\color{black}}%
      \expandafter\def\csname LT4\endcsname{\color{black}}%
      \expandafter\def\csname LT5\endcsname{\color{black}}%
      \expandafter\def\csname LT6\endcsname{\color{black}}%
      \expandafter\def\csname LT7\endcsname{\color{black}}%
      \expandafter\def\csname LT8\endcsname{\color{black}}%
    \fi
  \fi
  \setlength{\unitlength}{0.0500bp}%
  \begin{picture}(7200.00,5040.00)%
    \gplgaddtomacro\gplbacktext{%
      \csname LTb\endcsname%
      \put(1650,1618){\makebox(0,0)[r]{\strut{} 1e-06}}%
      \put(1650,2728){\makebox(0,0)[r]{\strut{} 1e-05}}%
      \put(1650,3838){\makebox(0,0)[r]{\strut{} 0.0001}}%
      \put(2923,484){\makebox(0,0){\strut{} 100}}%
      \put(5425,484){\makebox(0,0){\strut{} 1000}}%
      \put(484,2740){\rotatebox{90}{\makebox(0,0){\strut{}$\log[\vert\partial^{2}_{i}\,V[t]\vert]$}}}%
      \put(4370,154){\makebox(0,0){\strut{}$\log[\langle N_{4,1}(i)\rangle]$}}%
    }%
    \gplgaddtomacro\gplfronttext{%
      \csname LTb\endcsname%
      \put(5971,4548){\makebox(0,0)[r]{\strut{}$\vert\mathcal{V}(i)\vert$}}%
      \csname LTb\endcsname%
      \put(5971,4218){\makebox(0,0)[r]{\strut{}$\mathrm{fit\ of}\ \vert\partial^{2}_{i}\,V[t]\vert$}}%
    }%
    \gplbacktext
    \put(0,0){\includegraphics{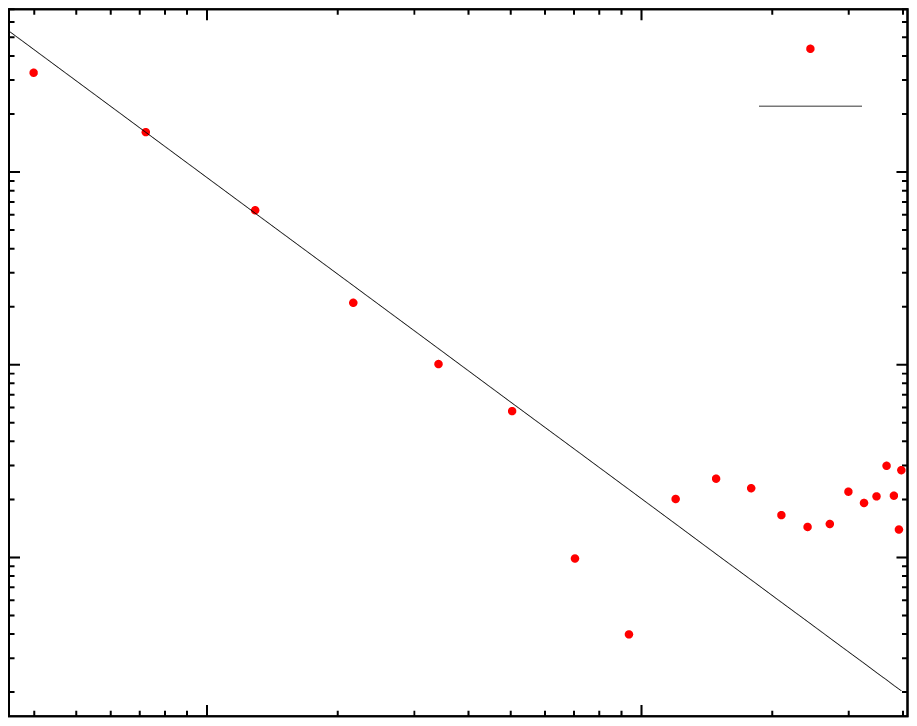}}%
    \gplfronttext
  \end{picture}%
\endgroup

%% file: ftexp41.tex
\begingroup
  \makeatletter
  \providecommand\color[2][]{%
    \GenericError{(gnuplot) \space\space\space\@spaces}{%
      Package color not loaded in conjunction with
      terminal option `colourtext'%
    }{See the gnuplot documentation for explanation.%
    }{Either use 'blacktext' in gnuplot or load the package
      color.sty in LaTeX.}%
    \renewcommand\color[2][]{}%
  }%
  \providecommand\includegraphics[2][]{%
    \GenericError{(gnuplot) \space\space\space\@spaces}{%
      Package graphicx or graphics not loaded%
    }{See the gnuplot documentation for explanation.%
    }{The gnuplot epslatex terminal needs graphicx.sty or graphics.sty.}%
    \renewcommand\includegraphics[2][]{}%
  }%
  \providecommand\rotatebox[2]{#2}%
  \@ifundefined{ifGPcolor}{%
    \newif\ifGPcolor
    \GPcolortrue
  }{}%
  \@ifundefined{ifGPblacktext}{%
    \newif\ifGPblacktext
    \GPblacktexttrue
  }{}%
  \let\gplgaddtomacro\g@addto@macro
  \gdef\gplbacktext{}%
  \gdef\gplfronttext{}%
  \makeatother
  \ifGPblacktext
    \def\colorrgb#1{}%
    \def\colorgray#1{}%
  \else
    \ifGPcolor
      \def\colorrgb#1{\color[rgb]{#1}}%
      \def\colorgray#1{\color[gray]{#1}}%
      \expandafter\def\csname LTw\endcsname{\color{white}}%
      \expandafter\def\csname LTb\endcsname{\color{black}}%
      \expandafter\def\csname LTa\endcsname{\color{black}}%
      \expandafter\def\csname LT0\endcsname{\color[rgb]{1,0,0}}%
      \expandafter\def\csname LT1\endcsname{\color[rgb]{0,1,0}}%
      \expandafter\def\csname LT2\endcsname{\color[rgb]{0,0,1}}%
      \expandafter\def\csname LT3\endcsname{\color[rgb]{1,0,1}}%
      \expandafter\def\csname LT4\endcsname{\color[rgb]{0,1,1}}%
      \expandafter\def\csname LT5\endcsname{\color[rgb]{1,1,0}}%
      \expandafter\def\csname LT6\endcsname{\color[rgb]{0,0,0}}%
      \expandafter\def\csname LT7\endcsname{\color[rgb]{1,0.3,0}}%
      \expandafter\def\csname LT8\endcsname{\color[rgb]{0.5,0.5,0.5}}%
    \else
      \def\colorrgb#1{\color{black}}%
      \def\colorgray#1{\color[gray]{#1}}%
      \expandafter\def\csname LTw\endcsname{\color{white}}%
      \expandafter\def\csname LTb\endcsname{\color{black}}%
      \expandafter\def\csname LTa\endcsname{\color{black}}%
      \expandafter\def\csname LT0\endcsname{\color{black}}%
      \expandafter\def\csname LT1\endcsname{\color{black}}%
      \expandafter\def\csname LT2\endcsname{\color{black}}%
      \expandafter\def\csname LT3\endcsname{\color{black}}%
      \expandafter\def\csname LT4\endcsname{\color{black}}%
      \expandafter\def\csname LT5\endcsname{\color{black}}%
      \expandafter\def\csname LT6\endcsname{\color{black}}%
      \expandafter\def\csname LT7\endcsname{\color{black}}%
      \expandafter\def\csname LT8\endcsname{\color{black}}%
    \fi
  \fi
  \setlength{\unitlength}{0.0500bp}%
  \begin{picture}(7200.00,5040.00)%
    \gplgaddtomacro\gplbacktext{%
      \csname LTb\endcsname%
      \put(1254,985){\makebox(0,0)[r]{\strut{} 0}}%
      \put(1254,1687){\makebox(0,0)[r]{\strut{} 0.1}}%
      \put(1254,2389){\makebox(0,0)[r]{\strut{} 0.2}}%
      \put(1254,3091){\makebox(0,0)[r]{\strut{} 0.3}}%
      \put(1254,3793){\makebox(0,0)[r]{\strut{} 0.4}}%
      \put(1254,4495){\makebox(0,0)[r]{\strut{} 0.5}}%
      \put(1386,484){\makebox(0,0){\strut{} 0}}%
      \put(2083,484){\makebox(0,0){\strut{} 500}}%
      \put(2779,484){\makebox(0,0){\strut{} 1000}}%
      \put(3476,484){\makebox(0,0){\strut{} 1500}}%
      \put(4172,484){\makebox(0,0){\strut{} 2000}}%
      \put(4869,484){\makebox(0,0){\strut{} 2500}}%
      \put(5565,484){\makebox(0,0){\strut{} 3000}}%
      \put(6262,484){\makebox(0,0){\strut{} 3500}}%
      \put(6958,484){\makebox(0,0){\strut{} 4000}}%
      \put(484,2740){\rotatebox{90}{\makebox(0,0){\strut{}$\partial_{i}\,\tilde{S}_{p}[t]$}}}%
      \put(4172,154){\makebox(0,0){\strut{}$\langle N_{4,1}(i)\rangle$}}%
    }%
    \gplgaddtomacro\gplfronttext{%
      \csname LTb\endcsname%
      \put(5971,4548){\makebox(0,0)[r]{\strut{}$-\partial_{i}\,\tilde{S}_{k}[t]$}}%
      \csname LTb\endcsname%
      \put(5971,4218){\makebox(0,0)[r]{\strut{}$\mathrm{fit\ of}\ \partial_{i}\,\tilde{S}_{p}[t]$}}%
    }%
    \gplbacktext
    \put(0,0){\includegraphics{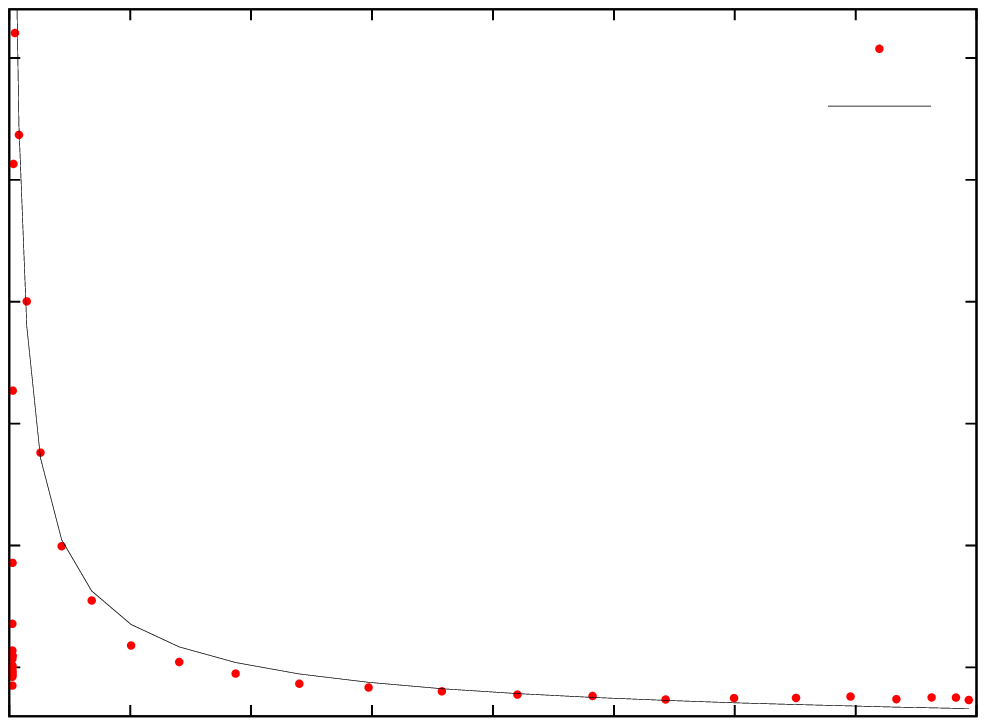}}%
    \gplfronttext
  \end{picture}%
\endgroup

%% file: smdkxi41.tex
\begingroup
  \makeatletter
  \providecommand\color[2][]{%
    \GenericError{(gnuplot) \space\space\space\@spaces}{%
      Package color not loaded in conjunction with
      terminal option `colourtext'%
    }{See the gnuplot documentation for explanation.%
    }{Either use 'blacktext' in gnuplot or load the package
      color.sty in LaTeX.}%
    \renewcommand\color[2][]{}%
  }%
  \providecommand\includegraphics[2][]{%
    \GenericError{(gnuplot) \space\space\space\@spaces}{%
      Package graphicx or graphics not loaded%
    }{See the gnuplot documentation for explanation.%
    }{The gnuplot epslatex terminal needs graphicx.sty or graphics.sty.}%
    \renewcommand\includegraphics[2][]{}%
  }%
  \providecommand\rotatebox[2]{#2}%
  \@ifundefined{ifGPcolor}{%
    \newif\ifGPcolor
    \GPcolortrue
  }{}%
  \@ifundefined{ifGPblacktext}{%
    \newif\ifGPblacktext
    \GPblacktexttrue
  }{}%
  \let\gplgaddtomacro\g@addto@macro
  \gdef\gplbacktext{}%
  \gdef\gplfronttext{}%
  \makeatother
  \ifGPblacktext
    \def\colorrgb#1{}%
    \def\colorgray#1{}%
  \else
    \ifGPcolor
      \def\colorrgb#1{\color[rgb]{#1}}%
      \def\colorgray#1{\color[gray]{#1}}%
      \expandafter\def\csname LTw\endcsname{\color{white}}%
      \expandafter\def\csname LTb\endcsname{\color{black}}%
      \expandafter\def\csname LTa\endcsname{\color{black}}%
      \expandafter\def\csname LT0\endcsname{\color[rgb]{1,0,0}}%
      \expandafter\def\csname LT1\endcsname{\color[rgb]{0,1,0}}%
      \expandafter\def\csname LT2\endcsname{\color[rgb]{0,0,1}}%
      \expandafter\def\csname LT3\endcsname{\color[rgb]{1,0,1}}%
      \expandafter\def\csname LT4\endcsname{\color[rgb]{0,1,1}}%
      \expandafter\def\csname LT5\endcsname{\color[rgb]{1,1,0}}%
      \expandafter\def\csname LT6\endcsname{\color[rgb]{0,0,0}}%
      \expandafter\def\csname LT7\endcsname{\color[rgb]{1,0.3,0}}%
      \expandafter\def\csname LT8\endcsname{\color[rgb]{0.5,0.5,0.5}}%
    \else
      \def\colorrgb#1{\color{black}}%
      \def\colorgray#1{\color[gray]{#1}}%
      \expandafter\def\csname LTw\endcsname{\color{white}}%
      \expandafter\def\csname LTb\endcsname{\color{black}}%
      \expandafter\def\csname LTa\endcsname{\color{black}}%
      \expandafter\def\csname LT0\endcsname{\color{black}}%
      \expandafter\def\csname LT1\endcsname{\color{black}}%
      \expandafter\def\csname LT2\endcsname{\color{black}}%
      \expandafter\def\csname LT3\endcsname{\color{black}}%
      \expandafter\def\csname LT4\endcsname{\color{black}}%
      \expandafter\def\csname LT5\endcsname{\color{black}}%
      \expandafter\def\csname LT6\endcsname{\color{black}}%
      \expandafter\def\csname LT7\endcsname{\color{black}}%
      \expandafter\def\csname LT8\endcsname{\color{black}}%
    \fi
  \fi
  \setlength{\unitlength}{0.0500bp}%
  \begin{picture}(7200.00,5040.00)%
    \gplgaddtomacro\gplbacktext{%
      \csname LTb\endcsname%
      \put(1650,1934){\makebox(0,0)[r]{\strut{} 0.0001}}%
      \put(1650,3427){\makebox(0,0)[r]{\strut{} 0.001}}%
      \put(3663,484){\makebox(0,0){\strut{} 100}}%
      \put(5706,484){\makebox(0,0){\strut{} 1000}}%
      \put(484,2740){\rotatebox{90}{\makebox(0,0){\strut{}$\log[\vert k_{1}\,\partial_{i+1}\partial_{i}\,\tilde{S}_{k}[t]\vert]$}}}%
      \put(4370,154){\makebox(0,0){\strut{}$\log[\frac{1}{2}(\langle N_{4,1}(i)\rangle+\langle N_{4,1}(i+1)\rangle)]$}}%
    }%
    \gplgaddtomacro\gplfronttext{%
      \csname LTb\endcsname%
      \put(5971,4548){\makebox(0,0)[r]{\strut{}$\vert(P_{4,1})_{i,i+1}-2\varepsilon\vert$}}%
      \csname LTb\endcsname%
      \put(5971,4218){\makebox(0,0)[r]{\strut{}$\mathrm{fit\ of}\ \vert k_{1}\,\partial_{i+1}\partial_{i}\,\tilde{S}_{k}[t]\vert$}}%
    }%
    \gplbacktext
    \put(0,0){\includegraphics{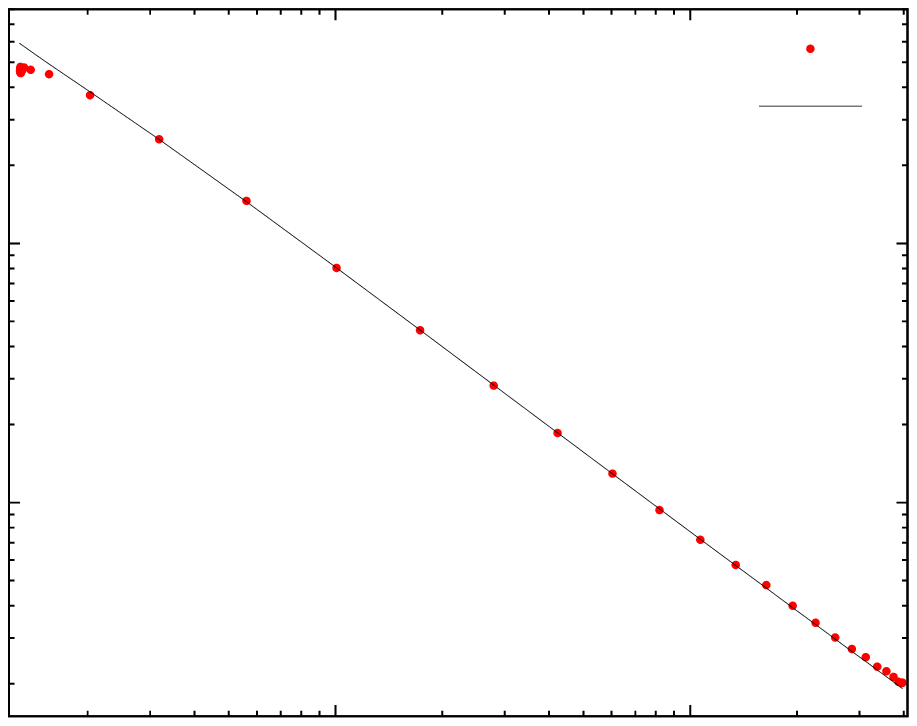}}%
    \gplfronttext
  \end{picture}%
\endgroup

%% file: sdpxi41.tex
\begingroup
  \makeatletter
  \providecommand\color[2][]{%
    \GenericError{(gnuplot) \space\space\space\@spaces}{%
      Package color not loaded in conjunction with
      terminal option `colourtext'%
    }{See the gnuplot documentation for explanation.%
    }{Either use 'blacktext' in gnuplot or load the package
      color.sty in LaTeX.}%
    \renewcommand\color[2][]{}%
  }%
  \providecommand\includegraphics[2][]{%
    \GenericError{(gnuplot) \space\space\space\@spaces}{%
      Package graphicx or graphics not loaded%
    }{See the gnuplot documentation for explanation.%
    }{The gnuplot epslatex terminal needs graphicx.sty or graphics.sty.}%
    \renewcommand\includegraphics[2][]{}%
  }%
  \providecommand\rotatebox[2]{#2}%
  \@ifundefined{ifGPcolor}{%
    \newif\ifGPcolor
    \GPcolortrue
  }{}%
  \@ifundefined{ifGPblacktext}{%
    \newif\ifGPblacktext
    \GPblacktexttrue
  }{}%
  \let\gplgaddtomacro\g@addto@macro
  \gdef\gplbacktext{}%
  \gdef\gplfronttext{}%
  \makeatother
  \ifGPblacktext
    \def\colorrgb#1{}%
    \def\colorgray#1{}%
  \else
    \ifGPcolor
      \def\colorrgb#1{\color[rgb]{#1}}%
      \def\colorgray#1{\color[gray]{#1}}%
      \expandafter\def\csname LTw\endcsname{\color{white}}%
      \expandafter\def\csname LTb\endcsname{\color{black}}%
      \expandafter\def\csname LTa\endcsname{\color{black}}%
      \expandafter\def\csname LT0\endcsname{\color[rgb]{1,0,0}}%
      \expandafter\def\csname LT1\endcsname{\color[rgb]{0,1,0}}%
      \expandafter\def\csname LT2\endcsname{\color[rgb]{0,0,1}}%
      \expandafter\def\csname LT3\endcsname{\color[rgb]{1,0,1}}%
      \expandafter\def\csname LT4\endcsname{\color[rgb]{0,1,1}}%
      \expandafter\def\csname LT5\endcsname{\color[rgb]{1,1,0}}%
      \expandafter\def\csname LT6\endcsname{\color[rgb]{0,0,0}}%
      \expandafter\def\csname LT7\endcsname{\color[rgb]{1,0.3,0}}%
      \expandafter\def\csname LT8\endcsname{\color[rgb]{0.5,0.5,0.5}}%
    \else
      \def\colorrgb#1{\color{black}}%
      \def\colorgray#1{\color[gray]{#1}}%
      \expandafter\def\csname LTw\endcsname{\color{white}}%
      \expandafter\def\csname LTb\endcsname{\color{black}}%
      \expandafter\def\csname LTa\endcsname{\color{black}}%
      \expandafter\def\csname LT0\endcsname{\color{black}}%
      \expandafter\def\csname LT1\endcsname{\color{black}}%
      \expandafter\def\csname LT2\endcsname{\color{black}}%
      \expandafter\def\csname LT3\endcsname{\color{black}}%
      \expandafter\def\csname LT4\endcsname{\color{black}}%
      \expandafter\def\csname LT5\endcsname{\color{black}}%
      \expandafter\def\csname LT6\endcsname{\color{black}}%
      \expandafter\def\csname LT7\endcsname{\color{black}}%
      \expandafter\def\csname LT8\endcsname{\color{black}}%
    \fi
  \fi
  \setlength{\unitlength}{0.0500bp}%
  \begin{picture}(7200.00,5040.00)%
    \gplgaddtomacro\gplbacktext{%
      \csname LTb\endcsname%
      \put(1650,1807){\makebox(0,0)[r]{\strut{} 1e-06}}%
      \put(1650,3006){\makebox(0,0)[r]{\strut{} 1e-05}}%
      \put(1650,4204){\makebox(0,0)[r]{\strut{} 0.0001}}%
      \put(2236,484){\makebox(0,0){\strut{} 100}}%
      \put(5164,484){\makebox(0,0){\strut{} 1000}}%
      \put(484,2740){\rotatebox{90}{\makebox(0,0){\strut{}$\log[\vert\partial^{2}_{i}\,V[t]\vert]$}}}%
      \put(4370,154){\makebox(0,0){\strut{}$\log[\langle N_{4,1}(i)\rangle]$}}%
    }%
    \gplgaddtomacro\gplfronttext{%
      \csname LTb\endcsname%
      \put(5971,4548){\makebox(0,0)[r]{\strut{}$\vert(P_{4,1})_{i,i}-2\varepsilon-k_{1}\,\partial^{2}_{i}\,\tilde{S}_{k}[t]\vert$}}%
      \csname LTb\endcsname%
      \put(5971,4218){\makebox(0,0)[r]{\strut{}$\mathrm{fit\ of}\ \vert\partial^{2}_{i}\,V[t]\vert$}}%
    }%
    \gplbacktext
    \put(0,0){\includegraphics{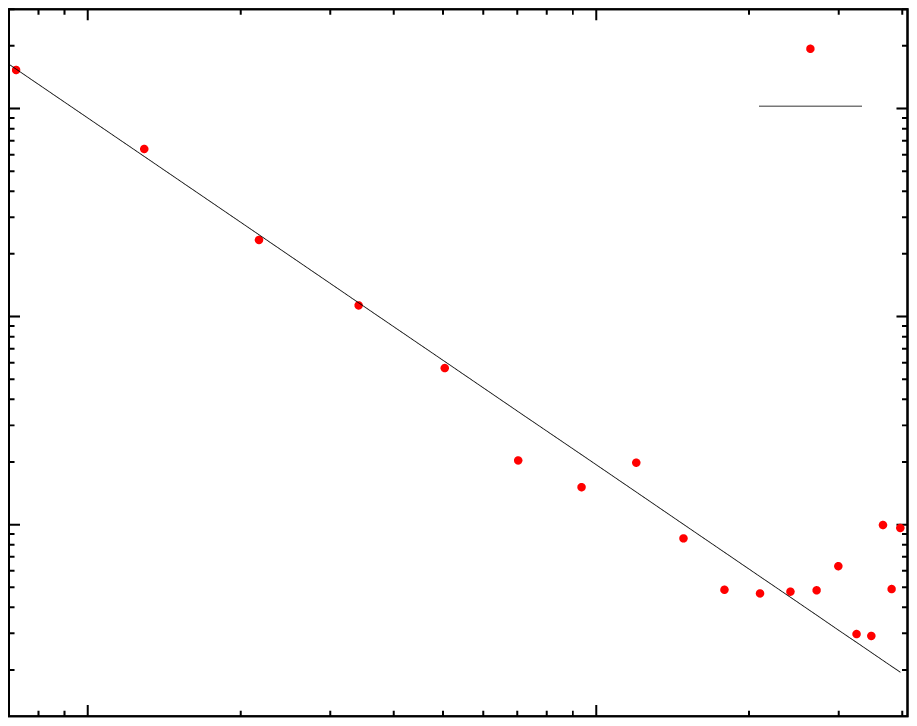}}%
    \gplfronttext
  \end{picture}%
\endgroup

%% file: avdbl.tex
\begingroup
  \makeatletter
  \providecommand\color[2][]{%
    \GenericError{(gnuplot) \space\space\space\@spaces}{%
      Package color not loaded in conjunction with
      terminal option `colourtext'%
    }{See the gnuplot documentation for explanation.%
    }{Either use 'blacktext' in gnuplot or load the package
      color.sty in LaTeX.}%
    \renewcommand\color[2][]{}%
  }%
  \providecommand\includegraphics[2][]{%
    \GenericError{(gnuplot) \space\space\space\@spaces}{%
      Package graphicx or graphics not loaded%
    }{See the gnuplot documentation for explanation.%
    }{The gnuplot epslatex terminal needs graphicx.sty or graphics.sty.}%
    \renewcommand\includegraphics[2][]{}%
  }%
  \providecommand\rotatebox[2]{#2}%
  \@ifundefined{ifGPcolor}{%
    \newif\ifGPcolor
    \GPcolortrue
  }{}%
  \@ifundefined{ifGPblacktext}{%
    \newif\ifGPblacktext
    \GPblacktexttrue
  }{}%
  \let\gplgaddtomacro\g@addto@macro
  \gdef\gplbacktext{}%
  \gdef\gplfronttext{}%
  \makeatother
  \ifGPblacktext
    \def\colorrgb#1{}%
    \def\colorgray#1{}%
  \else
    \ifGPcolor
      \def\colorrgb#1{\color[rgb]{#1}}%
      \def\colorgray#1{\color[gray]{#1}}%
      \expandafter\def\csname LTw\endcsname{\color{white}}%
      \expandafter\def\csname LTb\endcsname{\color{black}}%
      \expandafter\def\csname LTa\endcsname{\color{black}}%
      \expandafter\def\csname LT0\endcsname{\color[rgb]{1,0,0}}%
      \expandafter\def\csname LT1\endcsname{\color[rgb]{0,1,0}}%
      \expandafter\def\csname LT2\endcsname{\color[rgb]{0,0,1}}%
      \expandafter\def\csname LT3\endcsname{\color[rgb]{1,0,1}}%
      \expandafter\def\csname LT4\endcsname{\color[rgb]{0,1,1}}%
      \expandafter\def\csname LT5\endcsname{\color[rgb]{1,1,0}}%
      \expandafter\def\csname LT6\endcsname{\color[rgb]{0,0,0}}%
      \expandafter\def\csname LT7\endcsname{\color[rgb]{1,0.3,0}}%
      \expandafter\def\csname LT8\endcsname{\color[rgb]{0.5,0.5,0.5}}%
    \else
      \def\colorrgb#1{\color{black}}%
      \def\colorgray#1{\color[gray]{#1}}%
      \expandafter\def\csname LTw\endcsname{\color{white}}%
      \expandafter\def\csname LTb\endcsname{\color{black}}%
      \expandafter\def\csname LTa\endcsname{\color{black}}%
      \expandafter\def\csname LT0\endcsname{\color{black}}%
      \expandafter\def\csname LT1\endcsname{\color{black}}%
      \expandafter\def\csname LT2\endcsname{\color{black}}%
      \expandafter\def\csname LT3\endcsname{\color{black}}%
      \expandafter\def\csname LT4\endcsname{\color{black}}%
      \expandafter\def\csname LT5\endcsname{\color{black}}%
      \expandafter\def\csname LT6\endcsname{\color{black}}%
      \expandafter\def\csname LT7\endcsname{\color{black}}%
      \expandafter\def\csname LT8\endcsname{\color{black}}%
    \fi
  \fi
  \setlength{\unitlength}{0.0500bp}%
  \begin{picture}(7200.00,5040.00)%
    \gplgaddtomacro\gplbacktext{%
      \csname LTb\endcsname%
      \put(1386,704){\makebox(0,0)[r]{\strut{} 0}}%
      \put(1386,1156){\makebox(0,0)[r]{\strut{} 500}}%
      \put(1386,1609){\makebox(0,0)[r]{\strut{} 1000}}%
      \put(1386,2061){\makebox(0,0)[r]{\strut{} 1500}}%
      \put(1386,2514){\makebox(0,0)[r]{\strut{} 2000}}%
      \put(1386,2966){\makebox(0,0)[r]{\strut{} 2500}}%
      \put(1386,3419){\makebox(0,0)[r]{\strut{} 3000}}%
      \put(1386,3871){\makebox(0,0)[r]{\strut{} 3500}}%
      \put(1386,4324){\makebox(0,0)[r]{\strut{} 4000}}%
      \put(1386,4776){\makebox(0,0)[r]{\strut{} 4500}}%
      \put(1518,484){\makebox(0,0){\strut{} 0}}%
      \put(2207,484){\makebox(0,0){\strut{} 10}}%
      \put(2895,484){\makebox(0,0){\strut{} 20}}%
      \put(3584,484){\makebox(0,0){\strut{} 30}}%
      \put(4272,484){\makebox(0,0){\strut{} 40}}%
      \put(4961,484){\makebox(0,0){\strut{} 50}}%
      \put(5650,484){\makebox(0,0){\strut{} 60}}%
      \put(6338,484){\makebox(0,0){\strut{} 70}}%
      \put(484,2740){\rotatebox{90}{\makebox(0,0){\strut{}$\langle N(t)\rangle$}}}%
      \put(4238,154){\makebox(0,0){\strut{}$t$}}%
    }%
    \gplgaddtomacro\gplfronttext{%
      \csname LTb\endcsname%
      \put(5971,3888){\makebox(0,0)[r]{\strut{}\footnotesize$\langle N_{4,1}(i)\rangle$}}%
      \csname LTb\endcsname%
      \put(5971,4218){\makebox(0,0)[r]{\strut{}\footnotesize$\rho\,\langle N_{\mathit{3,2}}(i+\frac{1}{2})\rangle$}}%
      \csname LTb\endcsname%
      \put(5971,4548){\makebox(0,0)[r]{\strut{}\footnotesize$N_{0}\cos^{3}(\omega(t-t_{0}))$}}%
    }%
    \gplbacktext
    \put(0,0){\includegraphics{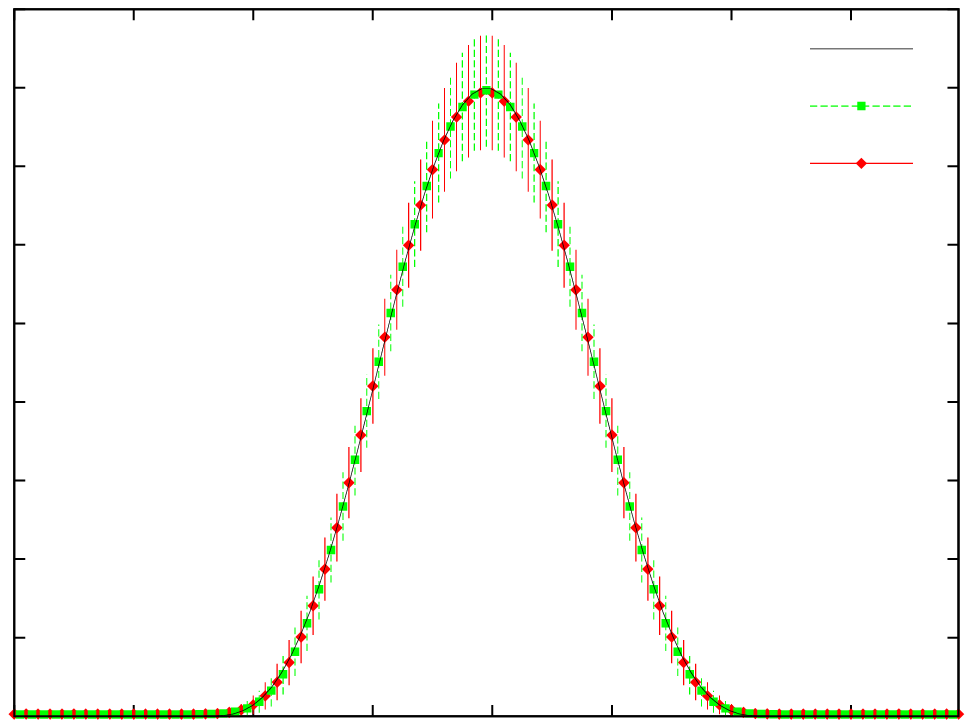}}%
    \gplfronttext
  \end{picture}%
\endgroup

%% file: ftexp41db.tex
\begingroup
  \makeatletter
  \providecommand\color[2][]{%
    \GenericError{(gnuplot) \space\space\space\@spaces}{%
      Package color not loaded in conjunction with
      terminal option `colourtext'%
    }{See the gnuplot documentation for explanation.%
    }{Either use 'blacktext' in gnuplot or load the package
      color.sty in LaTeX.}%
    \renewcommand\color[2][]{}%
  }%
  \providecommand\includegraphics[2][]{%
    \GenericError{(gnuplot) \space\space\space\@spaces}{%
      Package graphicx or graphics not loaded%
    }{See the gnuplot documentation for explanation.%
    }{The gnuplot epslatex terminal needs graphicx.sty or graphics.sty.}%
    \renewcommand\includegraphics[2][]{}%
  }%
  \providecommand\rotatebox[2]{#2}%
  \@ifundefined{ifGPcolor}{%
    \newif\ifGPcolor
    \GPcolortrue
  }{}%
  \@ifundefined{ifGPblacktext}{%
    \newif\ifGPblacktext
    \GPblacktexttrue
  }{}%
  \let\gplgaddtomacro\g@addto@macro
  \gdef\gplbacktext{}%
  \gdef\gplfronttext{}%
  \makeatother
  \ifGPblacktext
    \def\colorrgb#1{}%
    \def\colorgray#1{}%
  \else
    \ifGPcolor
      \def\colorrgb#1{\color[rgb]{#1}}%
      \def\colorgray#1{\color[gray]{#1}}%
      \expandafter\def\csname LTw\endcsname{\color{white}}%
      \expandafter\def\csname LTb\endcsname{\color{black}}%
      \expandafter\def\csname LTa\endcsname{\color{black}}%
      \expandafter\def\csname LT0\endcsname{\color[rgb]{1,0,0}}%
      \expandafter\def\csname LT1\endcsname{\color[rgb]{0,1,0}}%
      \expandafter\def\csname LT2\endcsname{\color[rgb]{0,0,1}}%
      \expandafter\def\csname LT3\endcsname{\color[rgb]{1,0,1}}%
      \expandafter\def\csname LT4\endcsname{\color[rgb]{0,1,1}}%
      \expandafter\def\csname LT5\endcsname{\color[rgb]{1,1,0}}%
      \expandafter\def\csname LT6\endcsname{\color[rgb]{0,0,0}}%
      \expandafter\def\csname LT7\endcsname{\color[rgb]{1,0.3,0}}%
      \expandafter\def\csname LT8\endcsname{\color[rgb]{0.5,0.5,0.5}}%
    \else
      \def\colorrgb#1{\color{black}}%
      \def\colorgray#1{\color[gray]{#1}}%
      \expandafter\def\csname LTw\endcsname{\color{white}}%
      \expandafter\def\csname LTb\endcsname{\color{black}}%
      \expandafter\def\csname LTa\endcsname{\color{black}}%
      \expandafter\def\csname LT0\endcsname{\color{black}}%
      \expandafter\def\csname LT1\endcsname{\color{black}}%
      \expandafter\def\csname LT2\endcsname{\color{black}}%
      \expandafter\def\csname LT3\endcsname{\color{black}}%
      \expandafter\def\csname LT4\endcsname{\color{black}}%
      \expandafter\def\csname LT5\endcsname{\color{black}}%
      \expandafter\def\csname LT6\endcsname{\color{black}}%
      \expandafter\def\csname LT7\endcsname{\color{black}}%
      \expandafter\def\csname LT8\endcsname{\color{black}}%
    \fi
  \fi
  \setlength{\unitlength}{0.0500bp}%
  \begin{picture}(7200.00,5040.00)%
    \gplgaddtomacro\gplbacktext{%
      \csname LTb\endcsname%
      \put(1254,704){\makebox(0,0)[r]{\strut{}-0.7}}%
      \put(1254,1270){\makebox(0,0)[r]{\strut{}-0.6}}%
      \put(1254,1835){\makebox(0,0)[r]{\strut{}-0.5}}%
      \put(1254,2401){\makebox(0,0)[r]{\strut{}-0.4}}%
      \put(1254,2966){\makebox(0,0)[r]{\strut{}-0.3}}%
      \put(1254,3532){\makebox(0,0)[r]{\strut{}-0.2}}%
      \put(1254,4097){\makebox(0,0)[r]{\strut{}-0.1}}%
      \put(1254,4663){\makebox(0,0)[r]{\strut{} 0}}%
      \put(1386,484){\makebox(0,0){\strut{} 0}}%
      \put(2083,484){\makebox(0,0){\strut{} 500}}%
      \put(2779,484){\makebox(0,0){\strut{} 1000}}%
      \put(3476,484){\makebox(0,0){\strut{} 1500}}%
      \put(4172,484){\makebox(0,0){\strut{} 2000}}%
      \put(4869,484){\makebox(0,0){\strut{} 2500}}%
      \put(5565,484){\makebox(0,0){\strut{} 3000}}%
      \put(6262,484){\makebox(0,0){\strut{} 3500}}%
      \put(6958,484){\makebox(0,0){\strut{} 4000}}%
      \put(484,2740){\rotatebox{90}{\makebox(0,0){\strut{}$\partial_{i}\,\tilde{S}^{(\mathrm{dbl})}_{p}[t]$}}}%
      \put(4172,154){\makebox(0,0){\strut{}$\langle N_{4,1}(i)\rangle$}}%
    }%
    \gplgaddtomacro\gplfronttext{%
      \csname LTb\endcsname%
      \put(5971,1262){\makebox(0,0)[r]{\strut{}$-\partial_{i}\,\tilde{S}^{(\mathrm{dbl})}_{k}[t]$}}%
      \csname LTb\endcsname%
      \put(5971,932){\makebox(0,0)[r]{\strut{}$\mathrm{fit\ of}\ \partial_{i}\,\tilde{S}^{(\mathrm{dbl})}_{p}[t]$}}%
    }%
    \gplbacktext
    \put(0,0){\includegraphics{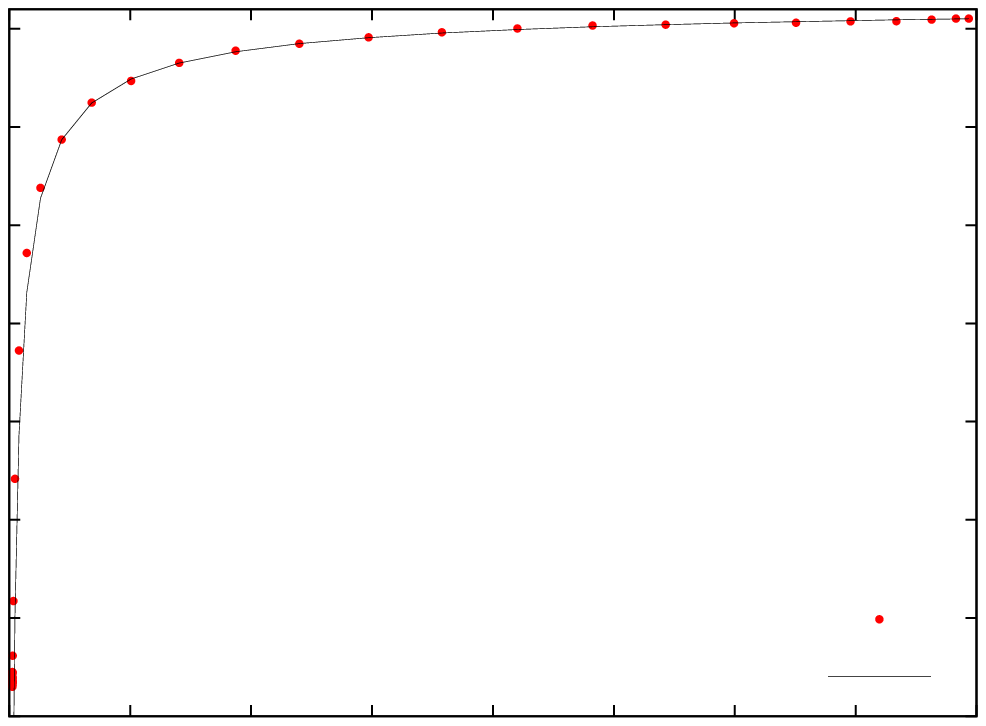}}%
    \gplfronttext
  \end{picture}%
\endgroup

%% file: ftexp32db.tex
\begingroup
  \makeatletter
  \providecommand\color[2][]{%
    \GenericError{(gnuplot) \space\space\space\@spaces}{%
      Package color not loaded in conjunction with
      terminal option `colourtext'%
    }{See the gnuplot documentation for explanation.%
    }{Either use 'blacktext' in gnuplot or load the package
      color.sty in LaTeX.}%
    \renewcommand\color[2][]{}%
  }%
  \providecommand\includegraphics[2][]{%
    \GenericError{(gnuplot) \space\space\space\@spaces}{%
      Package graphicx or graphics not loaded%
    }{See the gnuplot documentation for explanation.%
    }{The gnuplot epslatex terminal needs graphicx.sty or graphics.sty.}%
    \renewcommand\includegraphics[2][]{}%
  }%
  \providecommand\rotatebox[2]{#2}%
  \@ifundefined{ifGPcolor}{%
    \newif\ifGPcolor
    \GPcolortrue
  }{}%
  \@ifundefined{ifGPblacktext}{%
    \newif\ifGPblacktext
    \GPblacktexttrue
  }{}%
  \let\gplgaddtomacro\g@addto@macro
  \gdef\gplbacktext{}%
  \gdef\gplfronttext{}%
  \makeatother
  \ifGPblacktext
    \def\colorrgb#1{}%
    \def\colorgray#1{}%
  \else
    \ifGPcolor
      \def\colorrgb#1{\color[rgb]{#1}}%
      \def\colorgray#1{\color[gray]{#1}}%
      \expandafter\def\csname LTw\endcsname{\color{white}}%
      \expandafter\def\csname LTb\endcsname{\color{black}}%
      \expandafter\def\csname LTa\endcsname{\color{black}}%
      \expandafter\def\csname LT0\endcsname{\color[rgb]{1,0,0}}%
      \expandafter\def\csname LT1\endcsname{\color[rgb]{0,1,0}}%
      \expandafter\def\csname LT2\endcsname{\color[rgb]{0,0,1}}%
      \expandafter\def\csname LT3\endcsname{\color[rgb]{1,0,1}}%
      \expandafter\def\csname LT4\endcsname{\color[rgb]{0,1,1}}%
      \expandafter\def\csname LT5\endcsname{\color[rgb]{1,1,0}}%
      \expandafter\def\csname LT6\endcsname{\color[rgb]{0,0,0}}%
      \expandafter\def\csname LT7\endcsname{\color[rgb]{1,0.3,0}}%
      \expandafter\def\csname LT8\endcsname{\color[rgb]{0.5,0.5,0.5}}%
    \else
      \def\colorrgb#1{\color{black}}%
      \def\colorgray#1{\color[gray]{#1}}%
      \expandafter\def\csname LTw\endcsname{\color{white}}%
      \expandafter\def\csname LTb\endcsname{\color{black}}%
      \expandafter\def\csname LTa\endcsname{\color{black}}%
      \expandafter\def\csname LT0\endcsname{\color{black}}%
      \expandafter\def\csname LT1\endcsname{\color{black}}%
      \expandafter\def\csname LT2\endcsname{\color{black}}%
      \expandafter\def\csname LT3\endcsname{\color{black}}%
      \expandafter\def\csname LT4\endcsname{\color{black}}%
      \expandafter\def\csname LT5\endcsname{\color{black}}%
      \expandafter\def\csname LT6\endcsname{\color{black}}%
      \expandafter\def\csname LT7\endcsname{\color{black}}%
      \expandafter\def\csname LT8\endcsname{\color{black}}%
    \fi
  \fi
  \setlength{\unitlength}{0.0500bp}%
  \begin{picture}(7200.00,5040.00)%
    \gplgaddtomacro\gplbacktext{%
      \csname LTb\endcsname%
      \put(1386,835){\makebox(0,0)[r]{\strut{} 0}}%
      \put(1386,1492){\makebox(0,0)[r]{\strut{} 0.05}}%
      \put(1386,2149){\makebox(0,0)[r]{\strut{} 0.1}}%
      \put(1386,2806){\makebox(0,0)[r]{\strut{} 0.15}}%
      \put(1386,3462){\makebox(0,0)[r]{\strut{} 0.2}}%
      \put(1386,4119){\makebox(0,0)[r]{\strut{} 0.25}}%
      \put(1386,4776){\makebox(0,0)[r]{\strut{} 0.3}}%
      \put(1518,484){\makebox(0,0){\strut{} 0}}%
      \put(2554,484){\makebox(0,0){\strut{} 2000}}%
      \put(3590,484){\makebox(0,0){\strut{} 4000}}%
      \put(4627,484){\makebox(0,0){\strut{} 6000}}%
      \put(5663,484){\makebox(0,0){\strut{} 8000}}%
      \put(6699,484){\makebox(0,0){\strut{} 10000}}%
      \put(484,2740){\rotatebox{90}{\makebox(0,0){\strut{}$\partial_{i+\frac{1}{2}}\,\tilde{S}^{(\mathrm{dbl})}_{p}[t]$}}}%
      \put(4238,154){\makebox(0,0){\strut{}$\langle N_{\mathit{3,2}}(i+\frac{1}{2})\rangle$}}%
    }%
    \gplgaddtomacro\gplfronttext{%
      \csname LTb\endcsname%
      \put(5971,4548){\makebox(0,0)[r]{\strut{}$-\partial_{i+\frac{1}{2}}\,\tilde{S}^{(\mathrm{dbl})}_{k}[t]$}}%
      \csname LTb\endcsname%
      \put(5971,4218){\makebox(0,0)[r]{\strut{}$\mathrm{fit\ of}\ \partial_{i+\frac{1}{2}}\,\tilde{S}^{(\mathrm{dbl})}_{p}[t]$}}%
    }%
    \gplbacktext
    \put(0,0){\includegraphics{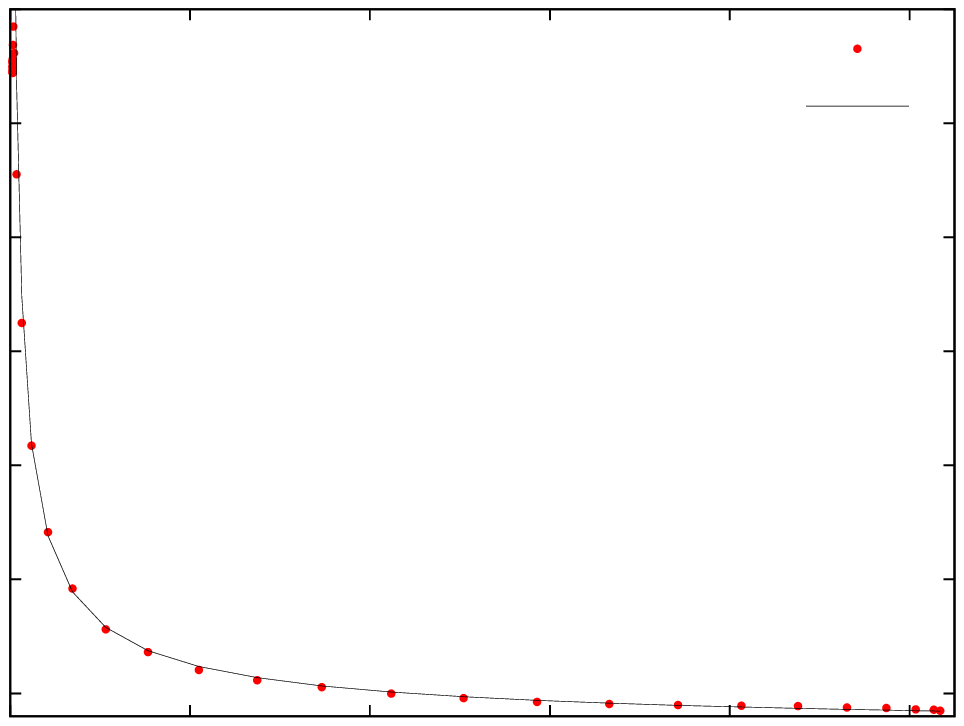}}%
    \gplfronttext
  \end{picture}%
\endgroup

%% file: smdk41db.tex
\begingroup
  \makeatletter
  \providecommand\color[2][]{%
    \GenericError{(gnuplot) \space\space\space\@spaces}{%
      Package color not loaded in conjunction with
      terminal option `colourtext'%
    }{See the gnuplot documentation for explanation.%
    }{Either use 'blacktext' in gnuplot or load the package
      color.sty in LaTeX.}%
    \renewcommand\color[2][]{}%
  }%
  \providecommand\includegraphics[2][]{%
    \GenericError{(gnuplot) \space\space\space\@spaces}{%
      Package graphicx or graphics not loaded%
    }{See the gnuplot documentation for explanation.%
    }{The gnuplot epslatex terminal needs graphicx.sty or graphics.sty.}%
    \renewcommand\includegraphics[2][]{}%
  }%
  \providecommand\rotatebox[2]{#2}%
  \@ifundefined{ifGPcolor}{%
    \newif\ifGPcolor
    \GPcolortrue
  }{}%
  \@ifundefined{ifGPblacktext}{%
    \newif\ifGPblacktext
    \GPblacktexttrue
  }{}%
  \let\gplgaddtomacro\g@addto@macro
  \gdef\gplbacktext{}%
  \gdef\gplfronttext{}%
  \makeatother
  \ifGPblacktext
    \def\colorrgb#1{}%
    \def\colorgray#1{}%
  \else
    \ifGPcolor
      \def\colorrgb#1{\color[rgb]{#1}}%
      \def\colorgray#1{\color[gray]{#1}}%
      \expandafter\def\csname LTw\endcsname{\color{white}}%
      \expandafter\def\csname LTb\endcsname{\color{black}}%
      \expandafter\def\csname LTa\endcsname{\color{black}}%
      \expandafter\def\csname LT0\endcsname{\color[rgb]{1,0,0}}%
      \expandafter\def\csname LT1\endcsname{\color[rgb]{0,1,0}}%
      \expandafter\def\csname LT2\endcsname{\color[rgb]{0,0,1}}%
      \expandafter\def\csname LT3\endcsname{\color[rgb]{1,0,1}}%
      \expandafter\def\csname LT4\endcsname{\color[rgb]{0,1,1}}%
      \expandafter\def\csname LT5\endcsname{\color[rgb]{1,1,0}}%
      \expandafter\def\csname LT6\endcsname{\color[rgb]{0,0,0}}%
      \expandafter\def\csname LT7\endcsname{\color[rgb]{1,0.3,0}}%
      \expandafter\def\csname LT8\endcsname{\color[rgb]{0.5,0.5,0.5}}%
    \else
      \def\colorrgb#1{\color{black}}%
      \def\colorgray#1{\color[gray]{#1}}%
      \expandafter\def\csname LTw\endcsname{\color{white}}%
      \expandafter\def\csname LTb\endcsname{\color{black}}%
      \expandafter\def\csname LTa\endcsname{\color{black}}%
      \expandafter\def\csname LT0\endcsname{\color{black}}%
      \expandafter\def\csname LT1\endcsname{\color{black}}%
      \expandafter\def\csname LT2\endcsname{\color{black}}%
      \expandafter\def\csname LT3\endcsname{\color{black}}%
      \expandafter\def\csname LT4\endcsname{\color{black}}%
      \expandafter\def\csname LT5\endcsname{\color{black}}%
      \expandafter\def\csname LT6\endcsname{\color{black}}%
      \expandafter\def\csname LT7\endcsname{\color{black}}%
      \expandafter\def\csname LT8\endcsname{\color{black}}%
    \fi
  \fi
  \setlength{\unitlength}{0.0500bp}%
  \begin{picture}(7200.00,5040.00)%
    \gplgaddtomacro\gplbacktext{%
      \csname LTb\endcsname%
      \put(1650,2032){\makebox(0,0)[r]{\strut{} 0.0001}}%
      \put(1650,3474){\makebox(0,0)[r]{\strut{} 0.001}}%
      \put(3663,484){\makebox(0,0){\strut{} 100}}%
      \put(5706,484){\makebox(0,0){\strut{} 1000}}%
      \put(484,2740){\rotatebox{90}{\makebox(0,0){\strut{}$\log[(P_{\mathrm{dbl}}^{11})_{i,i+1}-2\varepsilon]$}}}%
      \put(4370,154){\makebox(0,0){\strut{}$\log[\frac{1}{2}(\langle N_{4,1}(i)\rangle+\langle N_{4,1}(i+1)\rangle)]$}}%
    }%
    \gplgaddtomacro\gplfronttext{%
      \csname LTb\endcsname%
      \put(5971,4548){\makebox(0,0)[r]{\strut{}$(P_{\mathrm{dbl}}^{11})_{i,i+1}-2\varepsilon$}}%
      \csname LTb\endcsname%
      \put(5971,4218){\makebox(0,0)[r]{\strut{}$k^{(\mathrm{d})}_{1}\,\partial_{i+1}\partial_{i}\,\tilde{S}^{(\mathrm{dbl})}_{k}[t]$}}%
    }%
    \gplbacktext
    \put(0,0){\includegraphics{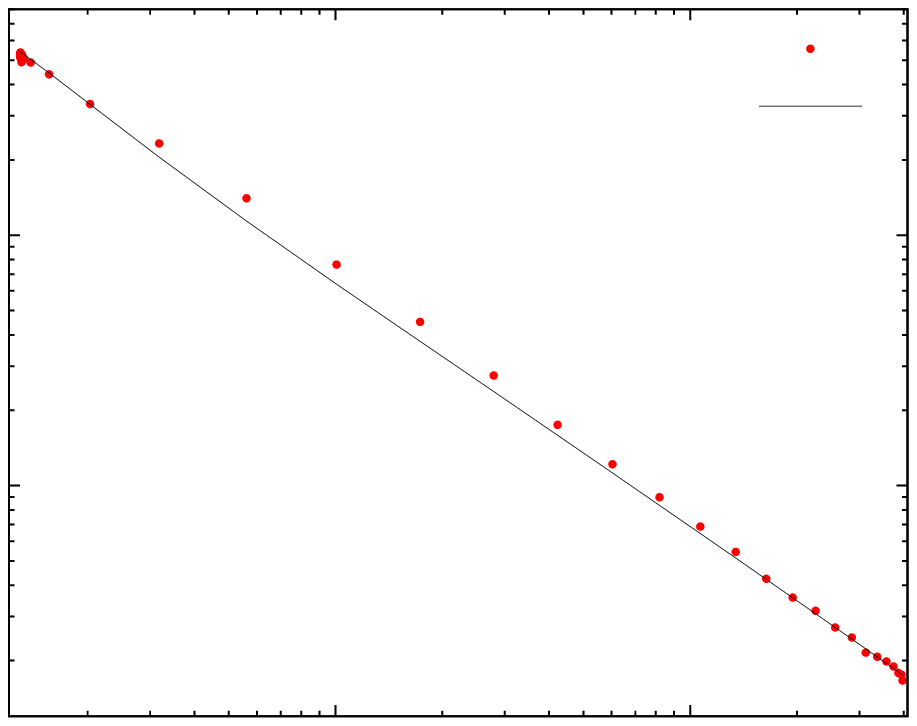}}%
    \gplfronttext
  \end{picture}%
\endgroup

%% file: smdk4132db.tex
\begingroup
  \makeatletter
  \providecommand\color[2][]{%
    \GenericError{(gnuplot) \space\space\space\@spaces}{%
      Package color not loaded in conjunction with
      terminal option `colourtext'%
    }{See the gnuplot documentation for explanation.%
    }{Either use 'blacktext' in gnuplot or load the package
      color.sty in LaTeX.}%
    \renewcommand\color[2][]{}%
  }%
  \providecommand\includegraphics[2][]{%
    \GenericError{(gnuplot) \space\space\space\@spaces}{%
      Package graphicx or graphics not loaded%
    }{See the gnuplot documentation for explanation.%
    }{The gnuplot epslatex terminal needs graphicx.sty or graphics.sty.}%
    \renewcommand\includegraphics[2][]{}%
  }%
  \providecommand\rotatebox[2]{#2}%
  \@ifundefined{ifGPcolor}{%
    \newif\ifGPcolor
    \GPcolortrue
  }{}%
  \@ifundefined{ifGPblacktext}{%
    \newif\ifGPblacktext
    \GPblacktexttrue
  }{}%
  \let\gplgaddtomacro\g@addto@macro
  \gdef\gplbacktext{}%
  \gdef\gplfronttext{}%
  \makeatother
  \ifGPblacktext
    \def\colorrgb#1{}%
    \def\colorgray#1{}%
  \else
    \ifGPcolor
      \def\colorrgb#1{\color[rgb]{#1}}%
      \def\colorgray#1{\color[gray]{#1}}%
      \expandafter\def\csname LTw\endcsname{\color{white}}%
      \expandafter\def\csname LTb\endcsname{\color{black}}%
      \expandafter\def\csname LTa\endcsname{\color{black}}%
      \expandafter\def\csname LT0\endcsname{\color[rgb]{1,0,0}}%
      \expandafter\def\csname LT1\endcsname{\color[rgb]{0,1,0}}%
      \expandafter\def\csname LT2\endcsname{\color[rgb]{0,0,1}}%
      \expandafter\def\csname LT3\endcsname{\color[rgb]{1,0,1}}%
      \expandafter\def\csname LT4\endcsname{\color[rgb]{0,1,1}}%
      \expandafter\def\csname LT5\endcsname{\color[rgb]{1,1,0}}%
      \expandafter\def\csname LT6\endcsname{\color[rgb]{0,0,0}}%
      \expandafter\def\csname LT7\endcsname{\color[rgb]{1,0.3,0}}%
      \expandafter\def\csname LT8\endcsname{\color[rgb]{0.5,0.5,0.5}}%
    \else
      \def\colorrgb#1{\color{black}}%
      \def\colorgray#1{\color[gray]{#1}}%
      \expandafter\def\csname LTw\endcsname{\color{white}}%
      \expandafter\def\csname LTb\endcsname{\color{black}}%
      \expandafter\def\csname LTa\endcsname{\color{black}}%
      \expandafter\def\csname LT0\endcsname{\color{black}}%
      \expandafter\def\csname LT1\endcsname{\color{black}}%
      \expandafter\def\csname LT2\endcsname{\color{black}}%
      \expandafter\def\csname LT3\endcsname{\color{black}}%
      \expandafter\def\csname LT4\endcsname{\color{black}}%
      \expandafter\def\csname LT5\endcsname{\color{black}}%
      \expandafter\def\csname LT6\endcsname{\color{black}}%
      \expandafter\def\csname LT7\endcsname{\color{black}}%
      \expandafter\def\csname LT8\endcsname{\color{black}}%
    \fi
  \fi
  \setlength{\unitlength}{0.0500bp}%
  \begin{picture}(7200.00,5040.00)%
    \gplgaddtomacro\gplbacktext{%
      \csname LTb\endcsname%
      \put(1650,1694){\makebox(0,0)[r]{\strut{} 0.0001}}%
      \put(1650,3110){\makebox(0,0)[r]{\strut{} 0.001}}%
      \put(1650,4527){\makebox(0,0)[r]{\strut{} 0.01}}%
      \put(3663,484){\makebox(0,0){\strut{} 100}}%
      \put(5706,484){\makebox(0,0){\strut{} 1000}}%
      \put(484,2740){\rotatebox{90}{\makebox(0,0){\strut{}$\log[\vert(P_{\mathrm{dbl}}^{12})_{i,i}\vert]$}}}%
      \put(4370,154){\makebox(0,0){\strut{}$\log[\langle N_{4,1}(i)\rangle]$}}%
    }%
    \gplgaddtomacro\gplfronttext{%
      \csname LTb\endcsname%
      \put(5971,4548){\makebox(0,0)[r]{\strut{}$\vert(P_{\mathrm{dbl}}^{12})_{i,i}\vert$}}%
      \csname LTb\endcsname%
      \put(5971,4218){\makebox(0,0)[r]{\strut{}$\vert k^{(\mathrm{d})}_{1}\,\partial_{i+\frac{1}{2}}\partial_{i}\,\tilde{S}^{(\mathrm{dbl})}_{k}[t]\vert$}}%
    }%
    \gplbacktext
    \put(0,0){\includegraphics{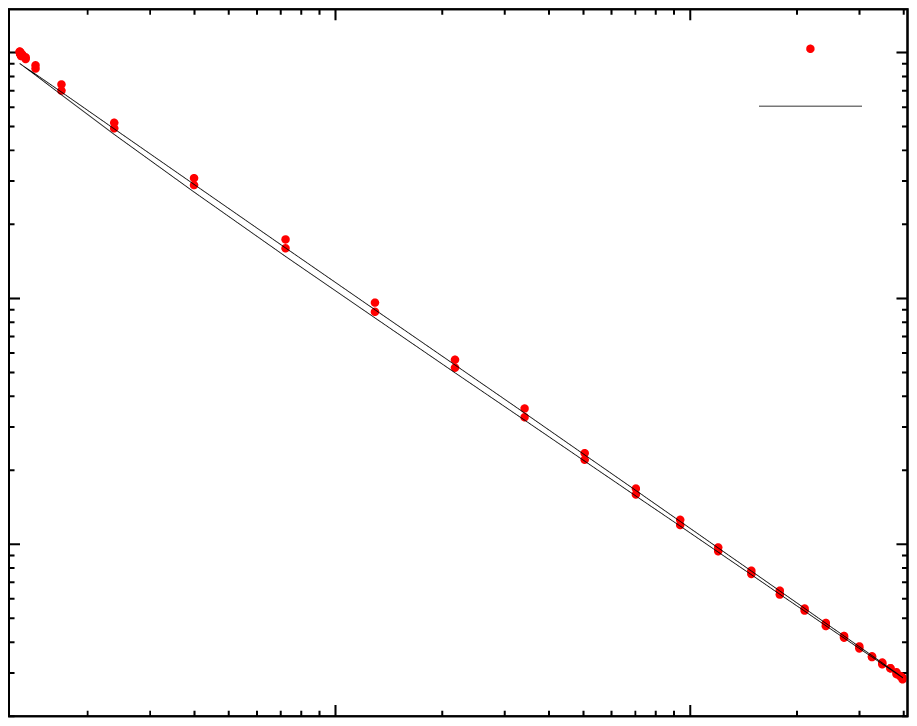}}%
    \gplfronttext
  \end{picture}%
\endgroup

%% file: shdk41db.tex
\begingroup
  \makeatletter
  \providecommand\color[2][]{%
    \GenericError{(gnuplot) \space\space\space\@spaces}{%
      Package color not loaded in conjunction with
      terminal option `colourtext'%
    }{See the gnuplot documentation for explanation.%
    }{Either use 'blacktext' in gnuplot or load the package
      color.sty in LaTeX.}%
    \renewcommand\color[2][]{}%
  }%
  \providecommand\includegraphics[2][]{%
    \GenericError{(gnuplot) \space\space\space\@spaces}{%
      Package graphicx or graphics not loaded%
    }{See the gnuplot documentation for explanation.%
    }{The gnuplot epslatex terminal needs graphicx.sty or graphics.sty.}%
    \renewcommand\includegraphics[2][]{}%
  }%
  \providecommand\rotatebox[2]{#2}%
  \@ifundefined{ifGPcolor}{%
    \newif\ifGPcolor
    \GPcolortrue
  }{}%
  \@ifundefined{ifGPblacktext}{%
    \newif\ifGPblacktext
    \GPblacktexttrue
  }{}%
  \let\gplgaddtomacro\g@addto@macro
  \gdef\gplbacktext{}%
  \gdef\gplfronttext{}%
  \makeatother
  \ifGPblacktext
    \def\colorrgb#1{}%
    \def\colorgray#1{}%
  \else
    \ifGPcolor
      \def\colorrgb#1{\color[rgb]{#1}}%
      \def\colorgray#1{\color[gray]{#1}}%
      \expandafter\def\csname LTw\endcsname{\color{white}}%
      \expandafter\def\csname LTb\endcsname{\color{black}}%
      \expandafter\def\csname LTa\endcsname{\color{black}}%
      \expandafter\def\csname LT0\endcsname{\color[rgb]{1,0,0}}%
      \expandafter\def\csname LT1\endcsname{\color[rgb]{0,1,0}}%
      \expandafter\def\csname LT2\endcsname{\color[rgb]{0,0,1}}%
      \expandafter\def\csname LT3\endcsname{\color[rgb]{1,0,1}}%
      \expandafter\def\csname LT4\endcsname{\color[rgb]{0,1,1}}%
      \expandafter\def\csname LT5\endcsname{\color[rgb]{1,1,0}}%
      \expandafter\def\csname LT6\endcsname{\color[rgb]{0,0,0}}%
      \expandafter\def\csname LT7\endcsname{\color[rgb]{1,0.3,0}}%
      \expandafter\def\csname LT8\endcsname{\color[rgb]{0.5,0.5,0.5}}%
    \else
      \def\colorrgb#1{\color{black}}%
      \def\colorgray#1{\color[gray]{#1}}%
      \expandafter\def\csname LTw\endcsname{\color{white}}%
      \expandafter\def\csname LTb\endcsname{\color{black}}%
      \expandafter\def\csname LTa\endcsname{\color{black}}%
      \expandafter\def\csname LT0\endcsname{\color{black}}%
      \expandafter\def\csname LT1\endcsname{\color{black}}%
      \expandafter\def\csname LT2\endcsname{\color{black}}%
      \expandafter\def\csname LT3\endcsname{\color{black}}%
      \expandafter\def\csname LT4\endcsname{\color{black}}%
      \expandafter\def\csname LT5\endcsname{\color{black}}%
      \expandafter\def\csname LT6\endcsname{\color{black}}%
      \expandafter\def\csname LT7\endcsname{\color{black}}%
      \expandafter\def\csname LT8\endcsname{\color{black}}%
    \fi
  \fi
  \setlength{\unitlength}{0.0500bp}%
  \begin{picture}(7200.00,5040.00)%
    \gplgaddtomacro\gplbacktext{%
      \csname LTb\endcsname%
      \put(1650,841){\makebox(0,0)[r]{\strut{} 0.0001}}%
      \put(1650,2258){\makebox(0,0)[r]{\strut{} 0.001}}%
      \put(1650,3674){\makebox(0,0)[r]{\strut{} 0.01}}%
      \put(3663,484){\makebox(0,0){\strut{} 100}}%
      \put(5706,484){\makebox(0,0){\strut{} 1000}}%
      \put(484,2740){\rotatebox{90}{\makebox(0,0){\strut{}$\log[(P_{\mathrm{dbl}}^{11})_{i,i}-2\varepsilon]$}}}%
      \put(4370,154){\makebox(0,0){\strut{}$\log[\langle N_{4,1}(i)\rangle]$}}%
    }%
    \gplgaddtomacro\gplfronttext{%
      \csname LTb\endcsname%
      \put(5971,4548){\makebox(0,0)[r]{\strut{}$(P_{\mathrm{dbl}}^{11})_{i,i}-2\varepsilon$}}%
      \csname LTb\endcsname%
      \put(5971,4218){\makebox(0,0)[r]{\strut{}$k^{(\mathrm{d})}_{1}\,\partial^{2}_{i}\,\tilde{S}^{(\mathrm{dbl})}_{k}[t]$}}%
    }%
    \gplbacktext
    \put(0,0){\includegraphics{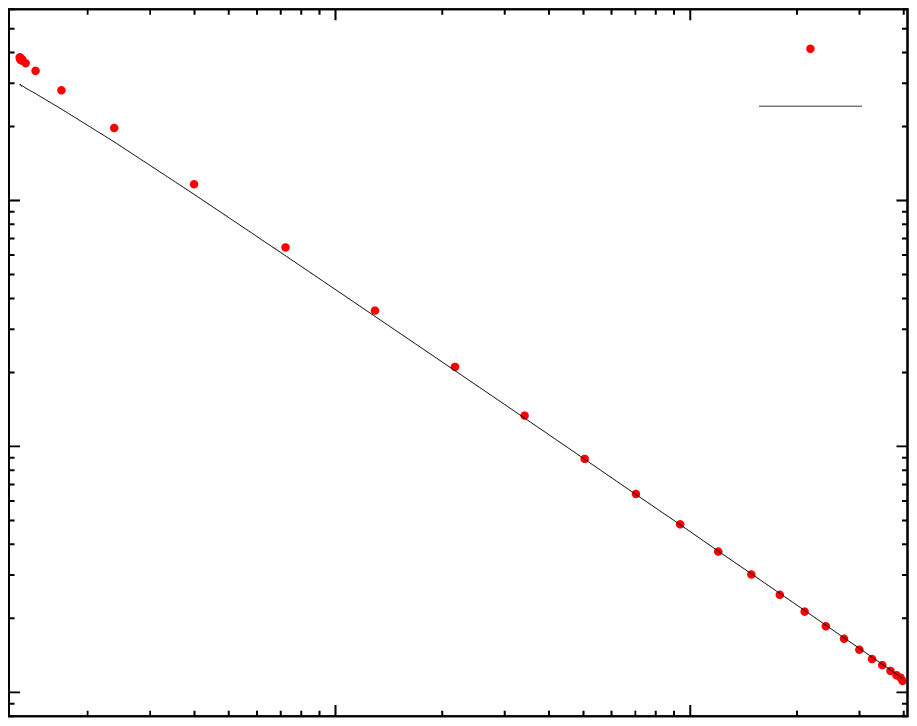}}%
    \gplfronttext
  \end{picture}%
\endgroup

%% file: shdk32db.tex
\begingroup
  \makeatletter
  \providecommand\color[2][]{%
    \GenericError{(gnuplot) \space\space\space\@spaces}{%
      Package color not loaded in conjunction with
      terminal option `colourtext'%
    }{See the gnuplot documentation for explanation.%
    }{Either use 'blacktext' in gnuplot or load the package
      color.sty in LaTeX.}%
    \renewcommand\color[2][]{}%
  }%
  \providecommand\includegraphics[2][]{%
    \GenericError{(gnuplot) \space\space\space\@spaces}{%
      Package graphicx or graphics not loaded%
    }{See the gnuplot documentation for explanation.%
    }{The gnuplot epslatex terminal needs graphicx.sty or graphics.sty.}%
    \renewcommand\includegraphics[2][]{}%
  }%
  \providecommand\rotatebox[2]{#2}%
  \@ifundefined{ifGPcolor}{%
    \newif\ifGPcolor
    \GPcolortrue
  }{}%
  \@ifundefined{ifGPblacktext}{%
    \newif\ifGPblacktext
    \GPblacktexttrue
  }{}%
  \let\gplgaddtomacro\g@addto@macro
  \gdef\gplbacktext{}%
  \gdef\gplfronttext{}%
  \makeatother
  \ifGPblacktext
    \def\colorrgb#1{}%
    \def\colorgray#1{}%
  \else
    \ifGPcolor
      \def\colorrgb#1{\color[rgb]{#1}}%
      \def\colorgray#1{\color[gray]{#1}}%
      \expandafter\def\csname LTw\endcsname{\color{white}}%
      \expandafter\def\csname LTb\endcsname{\color{black}}%
      \expandafter\def\csname LTa\endcsname{\color{black}}%
      \expandafter\def\csname LT0\endcsname{\color[rgb]{1,0,0}}%
      \expandafter\def\csname LT1\endcsname{\color[rgb]{0,1,0}}%
      \expandafter\def\csname LT2\endcsname{\color[rgb]{0,0,1}}%
      \expandafter\def\csname LT3\endcsname{\color[rgb]{1,0,1}}%
      \expandafter\def\csname LT4\endcsname{\color[rgb]{0,1,1}}%
      \expandafter\def\csname LT5\endcsname{\color[rgb]{1,1,0}}%
      \expandafter\def\csname LT6\endcsname{\color[rgb]{0,0,0}}%
      \expandafter\def\csname LT7\endcsname{\color[rgb]{1,0.3,0}}%
      \expandafter\def\csname LT8\endcsname{\color[rgb]{0.5,0.5,0.5}}%
    \else
      \def\colorrgb#1{\color{black}}%
      \def\colorgray#1{\color[gray]{#1}}%
      \expandafter\def\csname LTw\endcsname{\color{white}}%
      \expandafter\def\csname LTb\endcsname{\color{black}}%
      \expandafter\def\csname LTa\endcsname{\color{black}}%
      \expandafter\def\csname LT0\endcsname{\color{black}}%
      \expandafter\def\csname LT1\endcsname{\color{black}}%
      \expandafter\def\csname LT2\endcsname{\color{black}}%
      \expandafter\def\csname LT3\endcsname{\color{black}}%
      \expandafter\def\csname LT4\endcsname{\color{black}}%
      \expandafter\def\csname LT5\endcsname{\color{black}}%
      \expandafter\def\csname LT6\endcsname{\color{black}}%
      \expandafter\def\csname LT7\endcsname{\color{black}}%
      \expandafter\def\csname LT8\endcsname{\color{black}}%
    \fi
  \fi
  \setlength{\unitlength}{0.0500bp}%
  \begin{picture}(7200.00,5040.00)%
    \gplgaddtomacro\gplbacktext{%
      \csname LTb\endcsname%
      \put(1650,1822){\makebox(0,0)[r]{\strut{} 0.0001}}%
      \put(1650,3180){\makebox(0,0)[r]{\strut{} 0.001}}%
      \put(1650,4537){\makebox(0,0)[r]{\strut{} 0.01}}%
      \put(2894,484){\makebox(0,0){\strut{} 100}}%
      \put(4905,484){\makebox(0,0){\strut{} 1000}}%
      \put(6915,484){\makebox(0,0){\strut{} 10000}}%
      \put(484,2740){\rotatebox{90}{\makebox(0,0){\strut{}$\log[(P_{\mathrm{dbl}}^{22})_{i,i}]$}}}%
      \put(4370,154){\makebox(0,0){\strut{}$\log[\langle N_{\mathit{3,2}}(i+\frac{1}{2})\rangle]$}}%
    }%
    \gplgaddtomacro\gplfronttext{%
      \csname LTb\endcsname%
      \put(5971,4548){\makebox(0,0)[r]{\strut{}$(P_{\mathrm{dbl}}^{22})_{i,i}$}}%
      \csname LTb\endcsname%
      \put(5971,4218){\makebox(0,0)[r]{\strut{}$k^{(\mathrm{d})}_{1}\,\partial^{2}_{i+\frac{1}{2}}\,\tilde{S}^{(\mathrm{dbl})}_{k}[t]$}}%
    }%
    \gplbacktext
    \put(0,0){\includegraphics{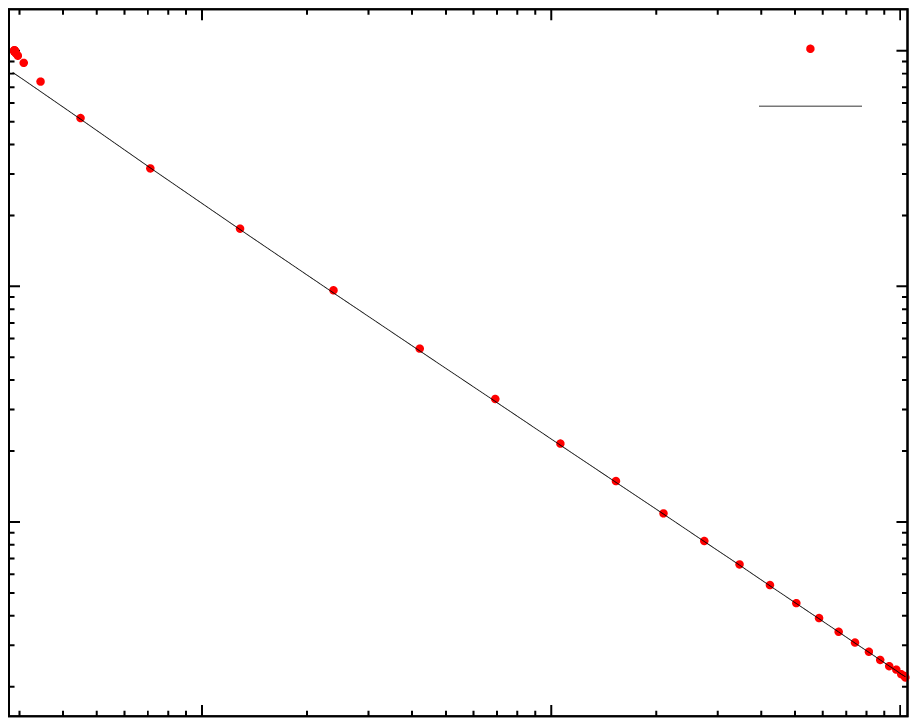}}%
    \gplfronttext
  \end{picture}%
\endgroup

%% file: avtrp.tex
\begingroup
  \makeatletter
  \providecommand\color[2][]{%
    \GenericError{(gnuplot) \space\space\space\@spaces}{%
      Package color not loaded in conjunction with
      terminal option `colourtext'%
    }{See the gnuplot documentation for explanation.%
    }{Either use 'blacktext' in gnuplot or load the package
      color.sty in LaTeX.}%
    \renewcommand\color[2][]{}%
  }%
  \providecommand\includegraphics[2][]{%
    \GenericError{(gnuplot) \space\space\space\@spaces}{%
      Package graphicx or graphics not loaded%
    }{See the gnuplot documentation for explanation.%
    }{The gnuplot epslatex terminal needs graphicx.sty or graphics.sty.}%
    \renewcommand\includegraphics[2][]{}%
  }%
  \providecommand\rotatebox[2]{#2}%
  \@ifundefined{ifGPcolor}{%
    \newif\ifGPcolor
    \GPcolortrue
  }{}%
  \@ifundefined{ifGPblacktext}{%
    \newif\ifGPblacktext
    \GPblacktexttrue
  }{}%
  \let\gplgaddtomacro\g@addto@macro
  \gdef\gplbacktext{}%
  \gdef\gplfronttext{}%
  \makeatother
  \ifGPblacktext
    \def\colorrgb#1{}%
    \def\colorgray#1{}%
  \else
    \ifGPcolor
      \def\colorrgb#1{\color[rgb]{#1}}%
      \def\colorgray#1{\color[gray]{#1}}%
      \expandafter\def\csname LTw\endcsname{\color{white}}%
      \expandafter\def\csname LTb\endcsname{\color{black}}%
      \expandafter\def\csname LTa\endcsname{\color{black}}%
      \expandafter\def\csname LT0\endcsname{\color[rgb]{1,0,0}}%
      \expandafter\def\csname LT1\endcsname{\color[rgb]{0,1,0}}%
      \expandafter\def\csname LT2\endcsname{\color[rgb]{0,0,1}}%
      \expandafter\def\csname LT3\endcsname{\color[rgb]{1,0,1}}%
      \expandafter\def\csname LT4\endcsname{\color[rgb]{0,1,1}}%
      \expandafter\def\csname LT5\endcsname{\color[rgb]{1,1,0}}%
      \expandafter\def\csname LT6\endcsname{\color[rgb]{0,0,0}}%
      \expandafter\def\csname LT7\endcsname{\color[rgb]{1,0.3,0}}%
      \expandafter\def\csname LT8\endcsname{\color[rgb]{0.5,0.5,0.5}}%
    \else
      \def\colorrgb#1{\color{black}}%
      \def\colorgray#1{\color[gray]{#1}}%
      \expandafter\def\csname LTw\endcsname{\color{white}}%
      \expandafter\def\csname LTb\endcsname{\color{black}}%
      \expandafter\def\csname LTa\endcsname{\color{black}}%
      \expandafter\def\csname LT0\endcsname{\color{black}}%
      \expandafter\def\csname LT1\endcsname{\color{black}}%
      \expandafter\def\csname LT2\endcsname{\color{black}}%
      \expandafter\def\csname LT3\endcsname{\color{black}}%
      \expandafter\def\csname LT4\endcsname{\color{black}}%
      \expandafter\def\csname LT5\endcsname{\color{black}}%
      \expandafter\def\csname LT6\endcsname{\color{black}}%
      \expandafter\def\csname LT7\endcsname{\color{black}}%
      \expandafter\def\csname LT8\endcsname{\color{black}}%
    \fi
  \fi
  \setlength{\unitlength}{0.0500bp}%
  \begin{picture}(7200.00,5040.00)%
    \gplgaddtomacro\gplbacktext{%
      \csname LTb\endcsname%
      \put(1386,704){\makebox(0,0)[r]{\strut{} 0}}%
      \put(1386,1156){\makebox(0,0)[r]{\strut{} 500}}%
      \put(1386,1609){\makebox(0,0)[r]{\strut{} 1000}}%
      \put(1386,2061){\makebox(0,0)[r]{\strut{} 1500}}%
      \put(1386,2514){\makebox(0,0)[r]{\strut{} 2000}}%
      \put(1386,2966){\makebox(0,0)[r]{\strut{} 2500}}%
      \put(1386,3419){\makebox(0,0)[r]{\strut{} 3000}}%
      \put(1386,3871){\makebox(0,0)[r]{\strut{} 3500}}%
      \put(1386,4324){\makebox(0,0)[r]{\strut{} 4000}}%
      \put(1386,4776){\makebox(0,0)[r]{\strut{} 4500}}%
      \put(1518,484){\makebox(0,0){\strut{} 0}}%
      \put(2207,484){\makebox(0,0){\strut{} 10}}%
      \put(2895,484){\makebox(0,0){\strut{} 20}}%
      \put(3584,484){\makebox(0,0){\strut{} 30}}%
      \put(4272,484){\makebox(0,0){\strut{} 40}}%
      \put(4961,484){\makebox(0,0){\strut{} 50}}%
      \put(5650,484){\makebox(0,0){\strut{} 60}}%
      \put(6338,484){\makebox(0,0){\strut{} 70}}%
      \put(484,2740){\rotatebox{90}{\makebox(0,0){\strut{}$\langle N(t)\rangle$}}}%
      \put(4238,154){\makebox(0,0){\strut{}$t$}}%
    }%
    \gplgaddtomacro\gplfronttext{%
      \csname LTb\endcsname%
      \put(5971,3558){\makebox(0,0)[r]{\strut{}\footnotesize$\langle N_{4,1}(i)\rangle$}}%
      \csname LTb\endcsname%
      \put(5971,3888){\makebox(0,0)[r]{\strut{}\footnotesize$2\rho\,\langle N_{3,2}(i+\frac{1}{3})\rangle$}}%
      \csname LTb\endcsname%
      \put(5971,4218){\makebox(0,0)[r]{\strut{}\footnotesize$2\rho\,\langle N_{2,3}(i+\frac{2}{3})\rangle$}}%
      \csname LTb\endcsname%
      \put(5971,4548){\makebox(0,0)[r]{\strut{}\footnotesize$N_{0}\cos^{3}(\omega(t-t_{0}))$}}%
    }%
    \gplbacktext
    \put(0,0){\includegraphics{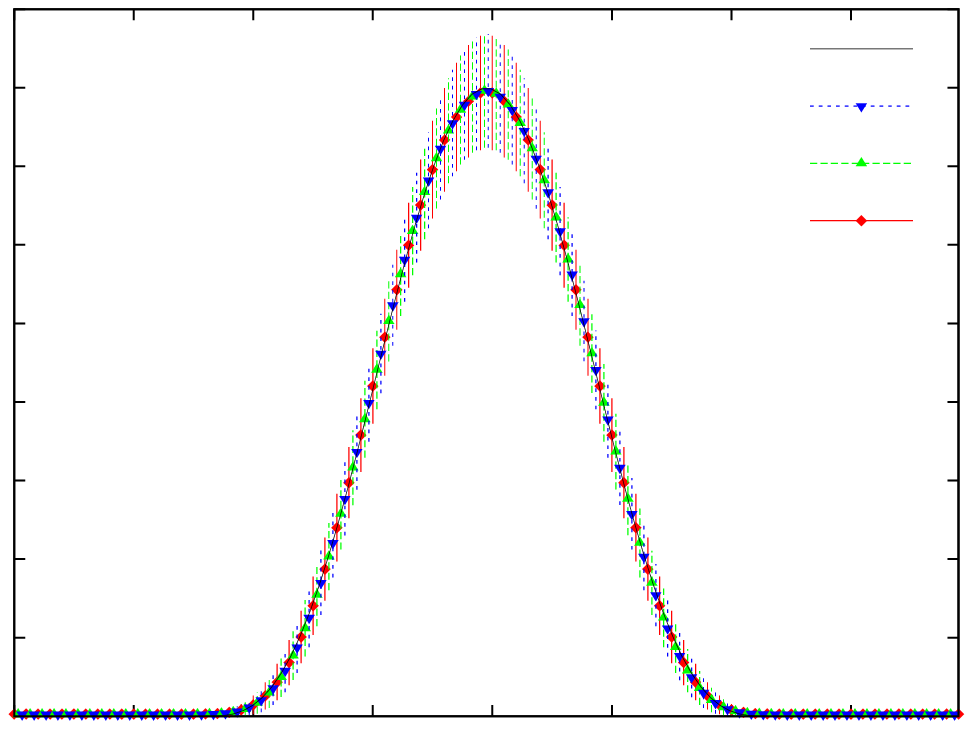}}%
    \gplfronttext
  \end{picture}%
\endgroup